\documentclass[acmsmall, screen]{acmart}
\usepackage{subcaption}
\usepackage{tcolorbox}
\usepackage{multirow} 
\usepackage{booktabs}
\usepackage{caption}

\newcommand{\ea}{\textit{et al.}}

\AtBeginDocument{%
  \providecommand\BibTeX{{%
    \normalfont B\kern-0.5em{\scshape i\kern-0.25em b}\kern-0.8em\TeX}}}

\setcopyright{acmcopyright}
\acmJournal{CSUR}
\acmYear{2023} \acmVolume{1} \acmNumber{1} \acmArticle{1} \acmMonth{10} \acmPrice{15.00}\acmDOI{10.1145/xxxxxxx              }

\acmJournal{JACM}
\acmVolume{37}
\acmNumber{4}
\acmMonth{10}

\begin{document}

\title{Pitfalls in Language Models for Code Intelligence: A Taxonomy and Survey}

\author{Xinyu She}
\email{xinyushe@hust.edu.cn}
\authornotemark[1]
\affiliation{%
  \institution{Huazhong University of Science and Technology}
  \city{Wuhan}           
  \country{China}
}
\author{Yue Liu}
\email{yue.liu1@monash.edu}
\authornote{Co-first authors who contributed equally to this work.}
\affiliation{%
  \institution{Monash University}
  \city{Melbourne}
  \country{Australia}
}
\author{Yanjie Zhao}
\email{carolzhao233@gmail.com}
\affiliation{%
  \institution{Ohio State University}
  \city{Columbus}
  \country{USA}
}
\author{Yiling He}
\email{yilinghe@zju.edu.cn}
\affiliation{%
  \institution{Zhejiang University}
  \city{Hangzhou}
  \country{China}
}
\author{Li Li}
\email{lilicoding@ieee.org}
\affiliation{%
  \institution{Beihang University}
  \city{Beijing}
  \country{China}
}
\author{Chakkrit Tantithamthavorn}
\email{chakkrit@monash.edu}
\affiliation{%
  \institution{Monash University}
  \city{Melbourne}
  \country{Australia}
}
\author{Zhan Qin}
\email{qinzhan@zju.edu.cn}
\affiliation{%
  \institution{Zhejiang University}
  \city{Hangzhou}
  \country{China}
}
\author{Haoyu Wang}
\authornote{Haoyu Wang is the corresponding author (haoyuwang@hust.edu.cn).}
\email{haoyuwang@hust.edu.cn}
\affiliation{%
  \institution{Huazhong University of Science and Technology}
  \city{Wuhan}     
  \country{China}
}
\renewcommand{\shortauthors}{ et al.}

\begin{abstract}
Modern language models (LMs) have been successfully employed in source code generation and understanding, leading to a significant increase in research focused on learning-based code intelligence, such as automated bug repair, and test case generation. 
Despite their great potential, \textbf{language models for code intelligence (LM4Code) are susceptible to potential pitfalls}, \textbf{which hinder realistic performance and further impact their reliability and applicability in real-world deployment}. 
Such challenges drive the need for a comprehensive understanding - not just identifying these issues but delving into their possible implications and existing solutions to build more reliable language models tailored to code intelligence. 
Based on a well-defined systematic research approach, we conducted an extensive literature review to uncover the pitfalls inherent in LM4Code.
Finally, 67 primary studies from top-tier venues have been identified.
After carefully examining these studies, we designed a taxonomy of pitfalls in LM4Code research and conducted a systematic study to summarize the issues, implications, current solutions, and challenges of different pitfalls for LM4Code systems. 
We developed a comprehensive classification scheme that dissects pitfalls across four crucial aspects:  data collection and labeling, system design and learning, performance evaluation, and deployment and maintenance.
Through this study, we aim to provide a roadmap for researchers and practitioners, facilitating their understanding and utilization of LM4Code in reliable and trustworthy ways.
\end{abstract}



\maketitle

\section{Introduction}
\label{sec:Introduction}
With every upgrade, language models (LMs) seem to redefine future boundaries.
Language models have achieved remarkable successes in natural language understanding and generation~\cite{vaswani2017attention, min2021recent}, underlined by the significant contributions from state-of-the-art models such as T5~\cite{wang2021codet5,fu2022VulRepair}, BERT~\cite{liu2023contrabert,gao2023keeping,gao2023two}, and GPT~\cite{xia2023automated,li2023cctest}.
Due to the format similarity between source code and natural language, language models have been widely applied in the domain of software engineering~\cite{chirkova2021empirical, niu2023empirical}.
They are now extensively researched and employed for source code understanding and generation, such as code completion~\cite{pearce_examining_2023,pearce_examining_2023,schuster2021you}, code summarization~\cite{gao2023keeping}, code generation~\cite{shen_incorporating_2022,liu_deep_2022}, code search~\cite{sun_importance_2022}, program repair~\cite{fan2023automated,xia2023automated}, and test case generation~\cite{zhang_generating_2020}. 
With powerful learning capabilities, language models have shown superior performance against traditional code intelligence approaches, such as template-based, heuristic-based, and machine learning-based approaches~\cite{jiang2023impact, openai2023gpt4, schuster2021you}.
Their superior performance stems from the fact that many LMs are trained on vast and diverse code repositories, enabling LMs to discern complex syntax, comprehend semantic context, and effectively predict code sequences~\cite{wang2022bridging}. 

However, the lack of transparency, often termed ``black-box'', poses significant challenges and concerns~\cite{DBLP:conf/kbse/Tantithamthavorn21,10109341}.
In other words, while language models for code intelligence (LM4Code) approaches offer powerful capabilities, they often lack transparency in their underlying reasoning and decision-making process.
Tantithamthavorn~\ea~also raised concerns that such a lack of transparency often leads to a lack of adoption of LM4Code in practice~
\cite{DBLP:conf/kbse/Tantithamthavorn21,DBLP:conf/msr/JiarpakdeeTG21}.
Consequently, hidden or neglected pitfalls in data or algorithms may persist, leading to unrealistic performance evaluation and unreliable code recommendations~\cite{hu2023,DBLP:conf/icse/Tantithamthavorn18}.
For example, Shi~\ea~\cite{shi2022we} found that noisy data (e.g., empty methods or duplicated code) was prevalent in widely-used benchmark datasets for code summarization, with contamination levels ranging between 31\% to 66\%.
By filtering out this noisy data, performance metrics like the BLEU-4 score witnessed a substantial increase (e.g., from 11.36\% to 16.48\%). 
Similarly, Sun~\ea~\cite{sun_importance_2022} highlighted a substantial amount of noise in user queries across various code search benchmark datasets.
Such instances underscore the hidden data noise that might undermine the trustworthiness of code produced or recommended by LMs.
What's more concerning is when these pitfalls go unnoticed, which raises significant questions about the reliability and integrity of the LM4Code systems built on them, thereby preventing the adoption of research advances in academia and industry. 

As LMs become increasingly prevalent in code intelligence despite increasing obstacles,  there emerges an urgent need for a comprehensive understanding of potential pitfalls within LM4Code systems. 
This isn't limited to pitfall identification; it demands a deeper exploration into the understanding of the implications of these pitfalls, current solutions, and possible challenges.
Although there is a growing body of research concerning or addressing pitfalls in LM4Code~\cite{sun_importance_2022, shi2022we, lo2023trustworthy, zeng_extensive_2022}, the domain lacks a comprehensive and systematic overview of these efforts.
Without such an overview, researchers, developers, and practitioners potentially overlook significant pitfalls identified in previous studies.
In this study, we conducted a systematic literature review, adhering to a well-defined approach that identifies, evaluates, and interprets the relevant literature that focuses on the pitfalls within LM4Code.
Our contributions of this paper are as follows:

\begin{itemize}
    \item \textit{Paper Collection of Pitfalls in LM4Code.} 
    Through a rigorous systematic literature review (SLR) protocol as outlined by~\cite{kitchenham_guidelines_2007,kitchenham_segress_2023} and after an in-depth analysis of the primary studies, we collected 67 primary papers (spanning 2018 to 2023) closely related to evaluating or addressing LM4Code pitfalls. Comprehensive details on our review process and the collected papers are available online~\footnote{\url{https://github.com/yueyueL/ReliableLM4Code}}.
    \item \textit{Compresensive Taxonomy.}
    We conducted a qualitative and quantitative synthesis of the collected studies. 
    We present a taxonomy of the collected studies according to the LM4Code lifecycle, including data collection and labeling, system design and learning, performance evaluation, deployment, and maintenance.
    Our synthesis investigates the pitfalls present in LM4Code, summarizes the implications of these pitfalls, investigates how these issues are addressed, and outlines future challenges in this field.
    \item \textit{Insightful Findings and Recommendations.}
    In addition to identifying and analyzing pitfalls, we distilled practical insights and recommendations for researchers and practitioners in the field of LM4Code. These findings pave the way for developing more robust and reliable language models tailored for code intelligence, mitigating potential challenges and maximizing their utility of such models in real-world applications.
    
\end{itemize}

\section{Study Design}

\begin{figure}[t]
  \centering
  \begin{minipage}{0.5\linewidth} 
    \centering
    \includegraphics[height=3.7cm, keepaspectratio]{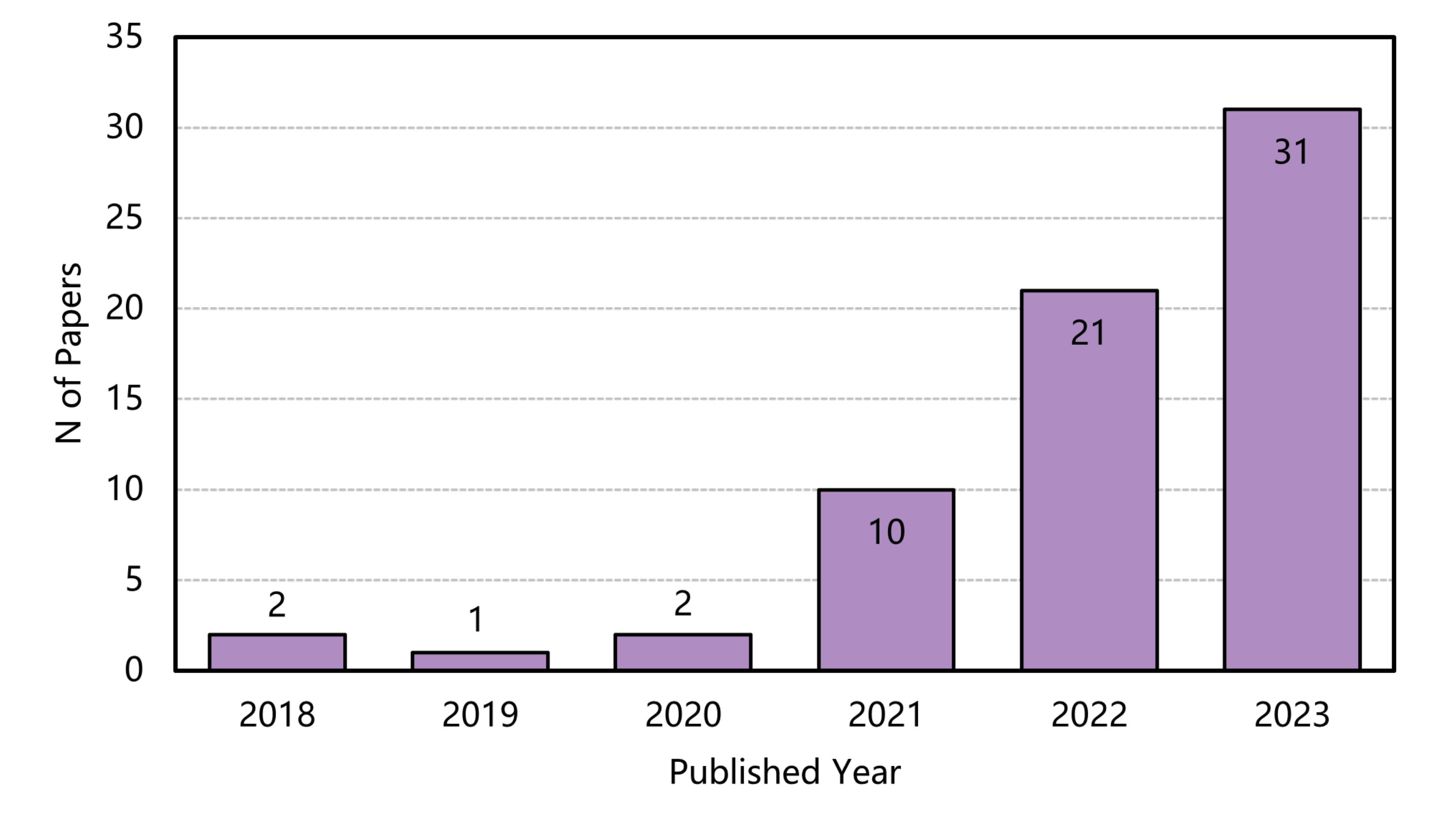} 
    \caption{Distribution of papers over years}
    \label{fig:new_annual-count}
  \end{minipage}%
  \begin{minipage}{0.5\linewidth} 
    \centering
    \includegraphics[height=3.7cm, keepaspectratio]{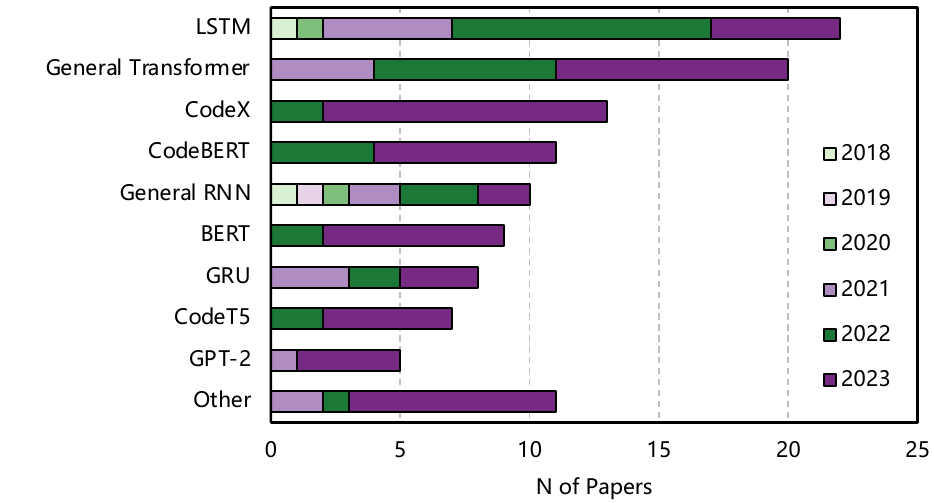} 
    \caption{Distribution of papers across LMs}
    \label{fig:new_lm_ditribution}
  \end{minipage}
\end{figure}

\subsection{Research Questions and Motivations}
In recent research, language models trained for code intelligence have shown promising performance~\cite{hou2023large, wang2023software, zeng_extensive_2022}.
However, an increasing number of literature~\cite{sun_importance_2022, shi2022we, nie2023understanding} has highlighted the existence of pitfalls in LM4Code that can skew their realistic performance, leading to either substantial overestimation or underestimation of their effectiveness.
The aim of conducting this systematic review is to gain an in-depth understanding of the pitfalls present in language models tailored for code intelligence.
Ensuring the robustness, reliability, and trustworthy deployment of LMs is important for their effective integration into the software development lifecycle.
Consequently, it is crucial to discern the nature of these pitfalls, comprehend their implications, and examine existing solutions.
Thus, we aim to answer the following research questions in this study:
\begin{itemize}
    \item \textbf{RQ1: What types of pitfalls are prevalent in language models for code intelligence?} This research question aims to identify the prevalent pitfalls in LM4Code systems, exploring how they could affect various stages of the learning-based system lifecycle.
    \item \textbf{RQ2: What are the implications of these pitfalls?} This research question investigates the implications of the identified pitfalls, specifically focusing on their impacts on the effectiveness, reliability, and ethical considerations of LM4Code systems.
    \item \textbf{RQ3: What solutions have been proposed to address these pitfalls?} This research question reviews the existing body of literature to identify proposed approaches for solving the identified pitfalls.
\end{itemize}

\subsection{Paper Collection and Selection}
To systematically identify relevant studies on pitfalls of LM4Code, we followed a rigorous methodology proposed by Kitchenham~\ea~\cite{kitchenham_guidelines_2007,kitchenham_segress_2023} and Zhang~\ea~\cite{zhang2011identifying} to perform a lightweight Systematic Literature Review (SLR).
We utilized the ``Quasi-Gold Standard'' (QGS)~\cite{zhang2011identifying} approach, combining manual and automated search strategies across major academic databases. This ensured comprehensive coverage while maintaining a focus on high-quality studies.
In total, we obtained over 100,000 papers from major academic databases including ACM Digital Library, IEEE Xplore, Springer, ScienceDirect, Web of Science, and DBLP.
Specifically, through QGS we obtained over 100,000 candidate papers from ACM Digital Library, IEEE Xplore, Springer, ScienceDirect, Web of Science, and DBLP. To filter these results, we defined robust inclusion/exclusion criteria and performed the quality assessment of full texts.
Additionally, we also conducted backward and forward snowballing~\cite{kitchenham2007guidelines} to complement the database searches and avoid excluding important works. Through snowballing, we evaluated over 1,000 additional papers.
By systematically combining these search strategies, selection criteria, and quality checks, we identified 67 high-quality studies investigating pitfalls and challenges in LM4Code.
Due to page limits, we make the review protocol details available in the supplementary report and our online repository.

Figure~\ref{fig:new_annual-count} displays the distribution of the collected research studies across the published year.
From Figure~\ref{fig:new_annual-count}, we have noted that there is a significant increase in the number of relevant research studies published annually from 2021, indicating a rising interest in investigating potential LM4Code pitfalls. 
Figure~\ref{fig:new_lm_ditribution} further presents the distribution of language modes used in the collected studies.
It is important to note that while both LSTM and GRU are types of RNN, papers that only specify the use of RNN without further detail are categorized under ``General RNN'' in this study.
Similarly, despite observing the utilization of several popular transformer-based architectures such as CodeBERT, Codex, and CodeT5, papers that merely claim the use of a self-defined or custom-designed transformer are classified as ``General Transformer'' in subsequent sections.
We find that LSTM models exhibit a higher prevalence than other types. However, over the past two years, studies utilizing transformer-based LMs, particularly pre-trained models like CodeBERT and Codex, have substantially increased.
Overall, these findings indicate growing attention toward identifying and evaluating challenges with the reliability and effectiveness of LM4Code. 
The community appears to be moving towards a comprehensive exploration of the realistic performance of LM4Code.
Our systematic collection provides an opportunity to thoroughly analyze LM4Code pitfalls.

\begin{figure}[t]
    \centering
    \includegraphics[width=\linewidth]{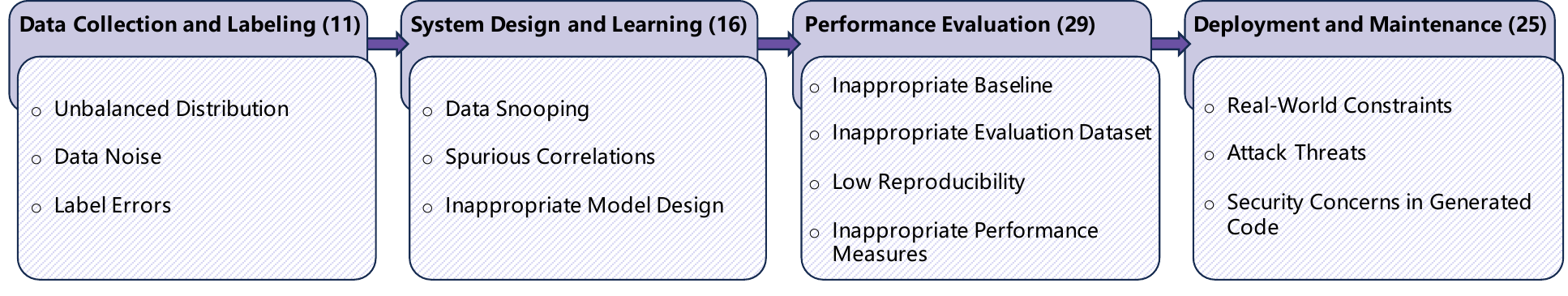}
    \caption{The overview of pitfalls of LMs for code intelligence}
    \label{fig:bias_Workflow}
\end{figure}

\subsection{Paper Organization}
Through our study collection process, we identified 67 research papers that specifically discuss or address the pitfalls in LM4Code. 
In the following content, we aim to answer our initial research questions based on these papers.
Similar to prior research~\cite{arp2022and}, our answers are organized following the typical workflow of LMs for code intelligence, which ensures the topics discussed in collected studies can be covered.
To be specific, we segment the pitfalls into four key aspects of the LM pipeline: data collection and labeling (Secton~\ref{sec:data}), system design and learning (Secton~\ref{sec:system_design}), performance evaluation (Secton~\ref{sec:performance_evaluation}), and deployment and maintenance (Secton~\ref{sec:deployment}).
This framework is depicted in Figure~\ref{fig:bias_Workflow}.
For each aspect, we first summarize the types of prevalent pitfalls discussed in the collected studies (RQ1), then introduce the implications of these pitfalls (RQ2), and finally explore potential solutions and best practices recommended in the literature (RQ3).
In Section~\ref{sec:Discussion}, we further discuss open challenges and promising research directions.
This organized structure enables a comprehensive analysis of pitfalls and considerations across the entire LM4Code pipeline. 
Our taxonomy aims to provide crucial insights for developing more robust, reliable, and practical LM systems for code intelligence tasks.

\section{Data Collection and Labeling}
\label{sec:data}
The data-hungry language models require large-scale and high-quality training datasets.
According to a survey by Hou~\ea~\cite{hou2023large}, the majority of LMs for code intelligence are trained using data from open-source platforms, with GitHub and StackOverflow being the most popular options.
However, the data in these platforms are user-contributed, varying significantly in the level of quality and reliability.
It leads to non-negligible noises, bias, and errors in the training dataset and further affects the behavior of the models, which brings significant pitfalls in LMs for code intelligence. 
In this section, we provide a brief description of related studies and discuss the implications and potential solutions during the data collection and labeling stages. 


\begin{table}[t]
  \centering
  \resizebox{1\linewidth}{!}{
    \begin{tabular}{llp{12.335em}p{22.335em}p{14.165em}}
    \toprule
    \textbf{Bias Type} & \multicolumn{1}{c}{\textbf{Paper }} & \multicolumn{1}{c}{\textbf{SE Tasks}} & \textbf{Description} & \textbf{Implications} \\
    \midrule
    \multirow{3}[6]{*}{\textbf{Unbalanced Distribution}} & \cite{chakraborty2021deep}, \cite{yang2023does}, \cite{steenhoek2023empirical} & Vulnerability Detection & The ratio of vulnerable and non-vulnerable cases in real-world projects is extremely unbalanced & F1-score drops by 73\% \\
\cmidrule{2-5}          & \cite{yedida2021value} & Defect Prediction & Many datasets are imbalanced with target classes under 30\% & Imbalanced data results in an F1 score below 0.2 \\
\cmidrule{2-5}          & \cite{li2022robust} & Bug Report Classification, Defect Prediction & Class imbalance & / \\
    \midrule
    \multirow{3}[6]{*}{\textbf{Data Noise}} & \cite{liu2018neural}, \cite{zhang2023slice} & Commit Message Generation & Commit messages mix bot-generated content with human-written trivial messages containing redundant or easily inferred information. & BLEU-4 drops from 31.92\% to 14.19\% \\
\cmidrule{2-5}          & \cite{sun_importance_2022} & Code Search & Over one-third of queries of code search datasets contain noises that make them deviate from natural user queries (e.g., HTML tags, interrogation) & MRR improves from 0.407 to 0.512 after data cleaning \\
\cmidrule{2-5}          & \cite{shi2022we} & Code Summarization & Noisy code-comment pairs, including non-literal and duplicated code, are prevalent in four benchmark datasets (31\% to 66\%) & Training three models with the cleaned datasets improves the BLEU-4 by 27\%, 21\%, and 24\% \\
    \midrule
    \multirow{3}[6]{*}{\textbf{Labeling Errors}} & \cite{li2022robust} & Bug Report Classification, Defect Prediction & Mislabelled samples - issue reports that describe defects but are not classified as such.  & / \\
\cmidrule{2-5}          & \cite{lin2022xcode} & Code Translation, Clone Detection, Code Search & Most of the collected code snippets are unlabeled & / \\
\cmidrule{2-5}          & \cite{nie2023understanding} & Vulnerability Detection & Error labels are common in many vulnerability datasets & F1-score drops by 20.7\% \\
    \bottomrule
    \end{tabular}%
}
  \caption{Summary of Common Biases in Data Collection and Labeling from Reviewed Research Studies}
  \label{tab:bias_in_data_collection}%
\end{table}%

\subsection{RQ1-Pitfalls}

From the collected papers, we identified 11 research studies focusing on pitfalls during the data collection and labeling process.
Table~\ref{tab:bias_in_data_collection} presents the statistics of literature on this topic, where the pitfalls can be grouped into three main categories.

\noindent \textbf{Unbalanced Distribution:} 
Unbalanced distribution arises when there is a lack of proper randomization in the selection of samples, leading to certain populations being underrepresented or overrepresented~\cite{shahbazi2023representation}.
In code-related scenarios, it usually refers to the gap between the sample distribution of real-world practices and training datasets.
For example, as emphasized by~\cite{chakraborty2021deep, yang2023does, steenhoek2023empirical}, vulnerable instances in vulnerability detection studies are overwhelming while neutral code instances in real-world environments considerably outnumber their vulnerable counterparts.
This imbalance extends to other code-based tasks.
In software defect prediction, where defective modules are scarce compared with non-defective cases in real-world environments~\cite {yedida2021value,li2022robust}.
Similarly, bug report classification suffers from underrepresentation of the minority bug class~\cite{li2022robust}.

\noindent \textbf{Data Noise:} 
Data noises are the samples that are meaningless or even harmful for the models to learn, such as samples with deprecated coding conventions, multi-lingual comments, and auto-generated code snippets~\cite{shi2022we}.
Such noises widely exist in code datasets.
For example, as investigated by Sun~\ea~\cite{sun_importance_2022} and Shi~\ea~\cite{shi2022we}, over one-third of the popular code dataset, CodeSearchNet~\cite{husain2019codesearchnet}, are noises that are hardly seen in neural code search.
Their analysis results show that the examined datasets contain a multitude of noise categories, including unrelated comments, non-literal characters, and issues like empty or duplicated code.
Liu~\ea~\cite{liu2018neural} and Zhang~\ea~\cite{zhang2023slice} specifically investigated data noise in commit message generations, where approximately 16\% of the commit messages of benchmark dataset by Jiang~\ea~\cite{jiang2017automatically} were identified as noises.

\noindent \textbf{Labeling Errors:}
Labeling errors arise when ground-truth labels are inaccurate, unstable, or erroneous~\cite{arp2022and, tantithamthavorn2015impact}.
In some code-related tasks, such as vulnerability detection, the raw code datasets need to be labeled by human annotators.
Nie~\ea~\cite{nie2023understanding} explored the labeling error problem in vulnerability detection wherein a vulnerable code sample is mislabeled as non-vulnerable, and vice versa.
They found that mislabeling a non-vulnerable sample as vulnerable was a more pervasive issue.
The research further assessed three prominent datasets, D2A~\cite{zheng2021d2a}, Big-Vul~\cite{fan2020ac}, and Cross-Vul~\cite{nikitopoulos2021crossvul}, discovering that in the worst cases, nearly 30\% of the labels in these datasets may cause noisy labels.
Similarly, Li~\ea~\cite{li2022robust} examined the datasets from Herzig~\ea~\cite{herzig2013s} for Bug Report Classification (BRC) and from Yatish~\ea~\cite{yatish2019mining} for Software Defect Prediction (SDP).
Their findings revealed that these datasets possess mislabel rates ranging between 2\% and 29\%.

\begin{figure}[t]
  \centering
  \begin{minipage}{0.5\linewidth}
    \centering
    \includegraphics[height=3.5cm, keepaspectratio]{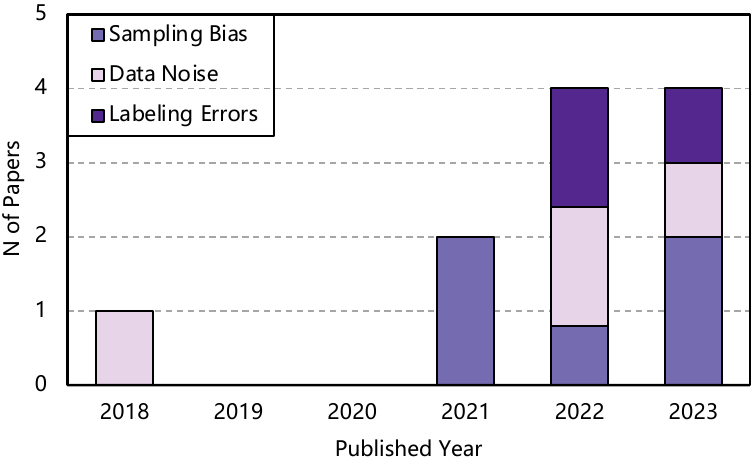} 
    \caption{Paper distribution across time (Section~\ref{sec:data}) }
    \label{fig:result_data_bias_time}
  \end{minipage}%
  \hfill
  \begin{minipage}{0.5\linewidth} 
    \centering
    \includegraphics[height=3.5cm, keepaspectratio]{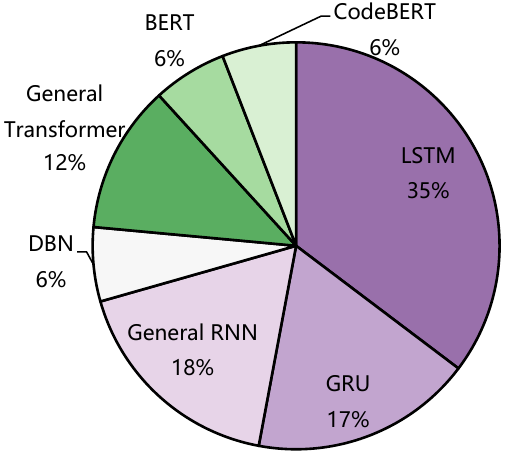}
    \caption{Distribution of LMs (Section~\ref{sec:data})}
    \label{fig:result_data_bias_models}
  \end{minipage}
\end{figure}

Overall, Figure~\ref{fig:result_data_bias_time} and Figure~\ref{fig:result_data_bias_models} present the distribution of the 11 papers concerning bias in data collection and labeling.
Figure~\ref{fig:result_data_bias_time} shows that the issue of data noise has continuously attracted the attention of researchers, and this attention has intensified significantly in recent years. 
Figure~\ref{fig:result_data_bias_models} reveals that while the emergence and widespread use of sophisticated models such as BERT and GPT~\cite{hou2023large}, the majority of the examined articles on data bias mostly concentrate on conventional language models, specifically LSTM and GRU.
This observation indicates a potential avenue for further investigation into the data biases present in modern language models.

\subsection{RQ2-Implications}
Our review results highlight the pervasive presence of pitfalls during the data collection and labeling processes in various automated code-related tasks. 
However, the deeper implications of these pitfalls remain to be fully discerned.
Thus, in this section, we aim to summarize the reviewed papers and provide insights into how these pitfalls influence the overall efficacy and performance of language models in code intelligence.

\noindent \textbf{Performance Overestimation:}
Train-test-split is a common practice for evaluating the neural models~\cite{arp2022and}.
However, derived from the same datasets, the test set contains the same bias, such as imbalanced data, as in the train set.
It leads to an overestimation of the model's performance since there is a gap between the testing dataset and real-world practices.
For example, Chakraborty~\ea~\cite{chakraborty2021deep} presented that vulnerability detection techniques are implemented on balanced datasets, and the models can achieve more than 90\% F1 scores.
However, the scenario drastically shifts when the same models are evaluated with a realistic dataset (where only 6\% of the examples are vulnerable).
In such cases, while the model could boast a recall as high as 91.24\%, this seemingly perfect metric can be deceptive.
A deeper dive reveals a mere 18.47\% F1 score, leading to a significant number of false positives.
A similar scenario arises with the introduction of data noise. 
Zhang~\ea~\cite{zhang2023slice} demonstrated this by training Transformer-based methods on a noisy dataset that included both bot-generated and trivial messages for code commit generation. 
Remarkably, the model achieved a 42.4\% BLEU-4 score under these conditions.
Yet, once the data noise was removed, the BLEU-4 score dropped sharply to 26.2\%.
Overall, pitfalls in data, whether from sampling imbalances or the presence of data noise, can lead to exaggerated performance metrics in language models for code tasks. 

\noindent \textbf{Compromised Model Efficacy:}
High-quality training data serves as the foundation for the trustworthy training of models.
When the training data introduces inherent noises and errors, language models might inadvertently learn irrelevant patterns or establish spurious correlations.
This not only distorts the model's understanding but can also undermine its performance.
For instance, Sun~\ea~\cite{sun_importance_2022} proved that code search models, when trained on carefully cleaned data without data noise, achieve a significant improvement in the number of answered queries and the rank of ground truth in search results (e.g., MRR improves from 0.407 to 0.512).
Similarly, Nie~\ea~\cite{nie2023understanding} showed that labeling errors severely compromise the performance of prevalent vulnerability detection models, with the worst instances seeing an average F1 score plummet of 20.7\%. 

\subsection{RQ3-Solutions}
Recognizing pitfalls in data collection and labeling has emphasized the need for robust solutions to address and mitigate these issues.
Several solutions have been proposed in the literature that we've reviewed. 
These solutions have been organized into distinct categories based on their underlying principles and methodologies.

\noindent \textbf{Data Cleaning/Denoising:}
Data cleaning/denoising is the process of improving a dataset by removing or correcting abnormalities, inconsistencies, and inaccuracies. 
This phase is critical to ensuring that the training data is accurate and does not contain any misleading or irrelevant information.
Many pitfalls develop as a result of noisy or erroneous data, and data cleaning is the major method for dealing with such difficulties.
Shi~\ea~\cite{shi2022we} introduced a rule-based cleaning tool, named CAT (Code-comment cleAning Tool), that employs configurable heuristics rules to automatically scan and filter out comments and code with syntactic anomalies, thereby detecting the occurrences and distribution of data noises.
Similarly, Sun~\ea~\cite{sun_importance_2022} presented a data cleaning framework tailored for code search.
It begins with a rule-based syntactic filter configured with heuristic rules to identify syntactically inconsistent comments.
This is followed by a model-based semantic filter, which focuses on comments with the fewest reconstruction discrepancies using a Variational Auto-Encoder model trained on a pre-established bootstrap query corpus. 
Their evaluation results demonstrate that this hybrid filter approach not only significantly save computational resources but also enhances model accuracy.
Nie~\ea~\cite{nie2023understanding} introduced confident learning~\cite{northcutt2021confident} and differential training~\cite{Xu2021DifferentialTA} for denoising-based noisy label detection, aiming to enhance the label quality of vulnerability datasets. They found that the effectiveness of denoising methods heavily relies on the vulnerability detection models' fitting ability to the datasets, and these denoising methods show considerable promise in boosting vulnerability detection performance.

\noindent \textbf{Real-world Benchmarks:}
To precisely evaluate the models, researchers propose to use real-world benchmarks, instead of train-test-spilt datasets.
Many benchmarks are thus constructed.
For example, many benchmarks, such as HumanEval~\cite{chen2021evaluating}, DS-1000~\cite{Lai2022DS1000AN}, and MBPP~\cite{Austin2021ProgramSW}, are constructed using human-written tasks and test cases to evaluate the code generation LMs.
Compared with the code snippets in open-source repositories, such human-written tasks are closer to the real user requests in practice, which better reflects the performance of the model.
In addition, the models can be evaluated directly using the production data accumulated during the operation of LM-based systems.
For example, Hellendoorn \ea ~\cite{hellendoorn_when_2019} and Aye \ea ~\cite{Aye2020LearningAF} adopt the production data of code completion systems as the evaluation dataset to better measure the model's performance.
Similarly, Mozannar \ea~\cite{Mozannar2022ReadingBT, Mozannar2023WhenTS} use the user behavior data to demonstrate the effectiveness of their proposed methods.
Apart from that, Lin \ea~\cite{lin2022xcode} examine their approach on a real-world dataset composed of programming exercises with multiple solutions.

\begin{tcolorbox}[title=Summary - Data Collection and Labeling, left=2pt, right=2pt,top=2pt,bottom=2pt]
Based on 11 relevant studies, our literature review reveals three prevalent pitfalls (i.e., unbalanced distribution, data noise, and labeling errors) in the data collection and labeling process. 
These pitfalls propagate, causing overestimated performance and compromised model efficacy. 
Though initial solutions like data cleaning/denoising and real-world benchmarks have been proposed, the field is far from reaching a comprehensive resolution.
The implications underscore the need for automated and scalable techniques to ensure high-quality data for LM4Code.
\end{tcolorbox}

\begin{figure}[t]
  \centering
  \begin{minipage}{0.5\linewidth}
    \centering
    \includegraphics[height=3.5cm, keepaspectratio]{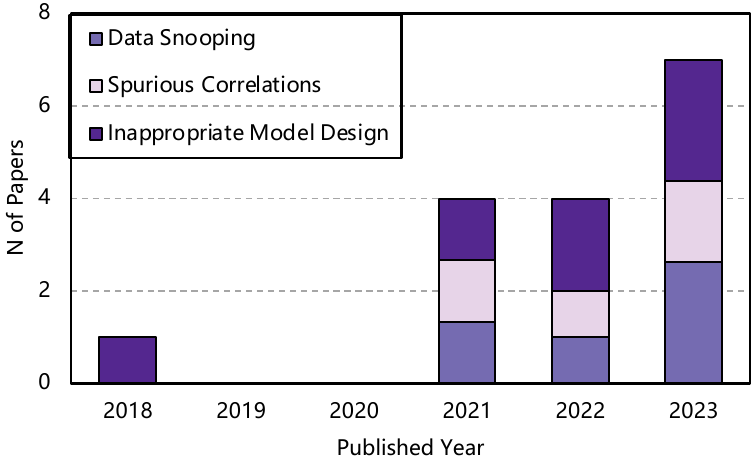} 
    \caption{Paper distribution across time (Section~\ref{sec:system_design}) }
    \label{fig:result_stage2_model_bias_time}
  \end{minipage}%
  \hfill
  \begin{minipage}{0.5\linewidth} 
    \centering
    \includegraphics[height=3.5cm, keepaspectratio]{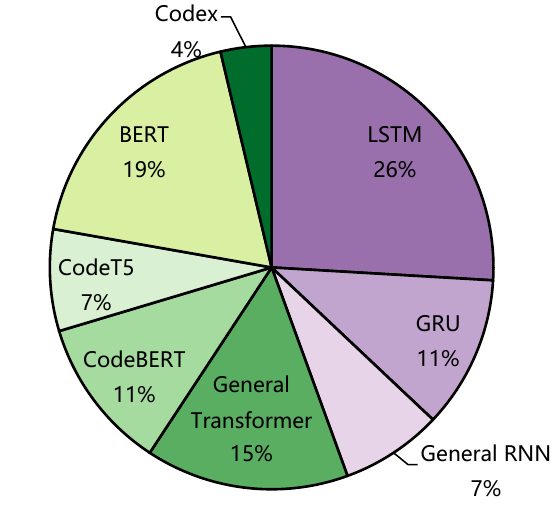}
    \caption{Distribution of LMs (Section~\ref{sec:system_design}) }
    \label{fig:result_stage2_model_bias_models}
  \end{minipage}
\end{figure}

\section{System Design and Learning}
\label{sec:system_design}
This section examines pitfalls in the system design and learning process for LM4Code.
The training of these LM4Code models directly impacts their quality and efficacy for empowering code intelligence. 
However, several challenges arise in crafting optimal model architectures, formulating strategic training-testing approaches, refining data preprocessing techniques, and selecting suitable learning algorithms. 
Each design decision risks introducing pitfalls that can undermine model robustness and effectiveness.


\subsection{RQ1-Pitfalls}
We have identified 16 research studies dedicated to the exploration of pitfalls introduced in the system design and learning process. 
These pitfalls can be broadly categorized into three categories: data snooping, spurious correlations, and inappropriate model design.
In the following, we provide comprehensive descriptions of these three pitfalls.

\noindent \textbf{Data Snooping:} 
Data snooping arises when LMs are inadvertently exposed during training to information that should be inaccessible. 
Such unintentional exposure primarily stems from improper data handling.
A common source is improper train-test split, where evaluation/testing data leaks into training.
For example, research has identified instances where training data incorporates bugs or vulnerabilities that closely mirror those encountered in testing~\cite{ding_patching_2021, wu2023effective}.
This concept of overlap is further supported by Liu~\ea~\cite{liu_can_2021} highlighting the risks associated with functionality similarities between train and test data.
Another dimension to consider is the origin of the data. 
For instance, as emphasized by Steenhoek~\ea~\cite{steenhoek2023empirical} and Wu~\ea~\cite{wu2023effective}, using samples from the same software projects for both training and testing can influence the models. 
Even data structure and representation choices enable snooping.
Liu~\ea~\cite{liu2023towards} and Shi~\ea~\cite{shi2022evaluation} emphasize that pitfalls can emerge from the data processing processes such as test prefix generation or concurrent use of same-class data.

\noindent \textbf{Spurious Correlations:}
Spurious correlations arise when language models mistakenly depend on irrelevant artifacts rather than the code's intrinsic logic or intent for decision-making, leading to misleading associations. 
The artifacts vary across SE tasks. 
For instance, in vulnerability detection, artifacts may manifest as recurring code patterns or reliance on specific function names that LMs incorrectly associate with vulnerabilities~\cite{steenhoek2023empirical, chakraborty2021deep}. 
In code summarization, models might focus more on strings or certain code structures while overlooking elements key for developers ~\cite{paltenghi_thinking_2021}. 
When generating commit messages in the context of code review, models often produce outputs that adhere to a few simple patterns, potentially failing to capture the nuances of the actual code changes~\cite{dong2023revisiting}. 
Actually, introducing advanced pre-trained models like CodeBERT has not eliminated these pitfalls. 
Specifically, if not fine-tuned appropriately for downstream tasks, these models might still overemphasize basic elements like keywords over richer code semantics~\cite{zhang_diet_2022}. 

\noindent \textbf{Inappropriate Model Design:}
Inappropriate model design in LM4Code arises when the underlying architecture fails to capture critical characteristics of code, such as hierarchy and composition.
The inability to construct robust semantic representations of code's intricate structural and logical attributes hinders model efficacy on downstream code intelligence tasks. 
Such design shortcomings can manifest in several ways. For instance, in vulnerability detection, models may exhibit a significant overlap in the feature space between classes, hindering precise vulnerability identification~\cite{chakraborty2021deep}. Code search models might lean on coarse-grained representations, capturing merely lexical or structural elements, often overlooking the true functionality of the code~\cite{wang_enhancing_2023}. Similarly, the encoder-decoder framework used in code summarization might neglect the hierarchical nature of code or struggle with sequence generation, leading to inadequate summaries~\cite{wan_improving_2018, wang_reinforcement-learning-guided_2022}. This issue is not limited to traditional architectures. Even modern program repair models, adapted from neural machine translation, face design-related challenges that affect their translation accuracy and diversity~\cite{ding_patching_2021, meng2023template}. While there are innovative attempts such as leveraging deep reinforcement learning or shared encoder-decoder architectures~\cite{lin2022xcode}, these approaches still exhibit shortcomings in addressing the diverse needs of various LM4Code applications.

To summarize, Figure~\ref{fig:result_stage2_model_bias_time} and ~\ref{fig:result_stage2_model_bias_models} display the distribution of papers that discuss LM4Code pitfalls in the system design and learning stage.
Figure~\ref{fig:result_stage2_model_bias_time} indicates that while inappropriate model design was first identified in 2018, research efforts on addressing key pitfalls have increased over the past three years. 
Among these, data snooping has garnered increasing research attention. Meanwhile, spurious correlations have become more prominent with the advent of explainable artificial intelligence~(XAI) techniques for elucidating model reasoning~\cite{10109341, 10109328,DBLP:conf/kbse/Tantithamthavorn21}. Discussions around inappropriate model design remain ongoing as new frameworks and learning strategies continue to emerge.
Contrary to our observations regarding the data collection and labeling process, Figure~\ref{fig:result_stage2_model_bias_models} reveals a greater emphasis on modern Transformer-based language models compared to conventional architectures.
Specifically, 19\% of relevant studies employ BERT, 15\% leverage general Transformer models, 11\% utilize CodeBERT, and 7\% investigate CodeT5.
This distribution highlights a shift towards examining potential pitfalls in sophisticated language models for code intelligence tasks, setting the stage for continued research focused on enhancing model transparency, interpretability, and reliability.

\subsection{RQ2-Implications} 
In the system design and learning phase, pitfalls emerge that can distort LM4Code outcomes. 
Following the pitfalls identified in the previous phase, these design-related pitfalls similarly lead to performance overestimation and compromised model efficacy. 
We will now elaborate on the specific impacts of these pitfalls and how they manifest in various software engineering tasks.

\noindent \textbf{Performance Overestimation:} 
Pitfalls in system design and learning can lead to over-optimistic performance metrics for LM4Code models.
Data snooping is a major contributor to this overestimation. 
For example, the presence of overlapping functionality between training and test sets can elevate the F1 score of a clone detection model from 0.42 to 0.96~\cite{liu_can_2021}. In vulnerability detection, a mix of projects in both training and testing phases can introduce discrepancies in F1 scores as large as 0.32~\cite{steenhoek2023empirical}. 
Spurious correlations represent a subtler and often more elusive challenge. 
These pitfalls cause models to make correct predictions, but often for the wrong reasons, leading them to rely on irrelevant code patterns or unrelated artifacts.
This not only misleads performance interpretation but also makes the models unreliable in varied scenarios~\cite{steenhoek2023empirical}. 
Models might also give undue attention to superficial code constructs, leading to inefficiencies that don't necessarily enhance outcomes~\cite{zhang_diet_2022}.

\noindent \textbf{Compromised Model Efficacy:} 
While overestimation impacts perceived performance, inappropriate model design directly compromises the efficacy of models in practical scenarios. For example, token sequence-based vulnerability detection models might fail to capture the underlying causes of vulnerabilities and instead focus on surface-level patterns, leaving a significant margin for improvement. 
Chakraborty~\ea~\cite{chakraborty2021deep} highlighted this limitation and showed that the use of gated graph neural networks can improve the baselines by 33.57\% in precision and 128.38\% in recall. 
Similarly, the design limitations in code search models lead them to generate coarse-grained representations that may overlook core functionalities~\cite{wang_enhancing_2023}. Program repair models, based on neural machine translation models, have shown slow learning curves, achieving a mere 4.5\% repair prediction accuracy even after 10 epochs~\cite{ding_patching_2021}. This indicates that the structural design of the model hinders its ability to learn and adapt efficiently.

\subsection{RQ3-Solutions}

To address the three pitfalls related to the system design and learning process, researchers have employed a variety of approaches which we describe as follows. 

\noindent \textbf{Refined Data Handling:}
To mitigate the challenges posed by data snooping, a multifaceted approach is essential.
Emphasizing rigorous data partitioning is foundational, as exemplified by Steenhoek~\ea~\cite{steenhoek2023empirical}, who advocate for cross-project validation to prevent inadvertent overlaps between training and testing sets.
Additionally, techniques like data augmentation and time-based splits can further insulate models from over-relying on specific pattern~\cite{pendlebury2019tesseract, guerra2022relativity}.
Incorporating regularization techniques, such as batch normalization, can curb overfitting and deter models from exploiting inadvertent data correlations~\cite{he2022msdroid}. 
Finally, the importance of external validation, underscored by Liu~\ea~\cite{liu_can_2021}, ensures that performance assessments are unbiased and reflective of real-world scenarios. 
By adopting these strategies, researchers can foster more reliable and generalizable outcomes.

\noindent \textbf{Model Interpretability:}
To address the biases inherent in LM4Code, especially spurious correlations, improving model interpretability has emerged as a crucial solution~\cite{10109341}. 
By examining the decision-making process of LM4Code models, researchers are better positioned to pinpoint and mitigate pitfalls, leading to more reliable predictions~\cite{he2023finer}. 
Within vulnerability detection, Fu~\ea~\cite{fu2023vulexplainer},
Li~\ea~\cite{li_vulnerability_2021}, and Zou~\ea~\cite{zou_interpreting_2021} proposed methods to enhance explanation accuracy, leveraging sophisticated visualization tools to correlate the internal dynamics of neural models with code structures, thus providing a comprehensive understanding of model reasoning. 
Cito~\ea~\cite{cito_explaining_2021} offers a distinctive perspective, centering on elucidating mispredictions. 
Their approach, which integrates neural predictions with symbolic logic, allows for precise error detection accompanied by rule-based explanations. 
Additionally, attention mechanisms to explain pre-trained models have also been analyzed.
While Shi~\ea~\cite{shi2023towards} unravels how transformer-based models allocate attention for code summarization, Wan~\ea~\cite{wan_what_2022} probes into the nuances of attention during code-to-code translation. 
These XAI approaches can serve to identify and rectify model pitfalls, ensuring the reliability of LM4Code applications.

\noindent \textbf{Model Optimization Strategies:}
In light of the pitfalls introduced by inappropriate model design, researchers have turned to model optimization strategies to address and minimize their effects. These strategies encompass several techniques designed to enhance a model's structure, training process, and generalization capabilities. 
Firstly, Model Design Adjustments involve refining the architecture to better capture data intricacies. Studies like that by Wan~\ea~\cite{wan_improving_2018} have demonstrated the benefits of introducing novel layers or structures to better understand the tree structure of code, yielding improved performance. 
Secondly, Model Ensembling is gaining traction, where the strengths of multiple models are leveraged to offset individual biases, as seen in the work by Zhang~\ea~\cite{zhang_diet_2022} which employs multiple views of the same data for more robust predictions. 
Lastly, Regularization and Fine-tuning techniques play a pivotal role. Regularization, such as dropout or L2 regularization, helps in preventing overfitting, while fine-tuning allows pre-trained models, like CodeBERT, to adapt to specific dataset nuances, as demonstrated by Fang~\ea~~\cite{fang2023representthemall}. 
By integrating these strategies, models can be better positioned to achieve superior outcomes.

\begin{tcolorbox}[title=Summary - System Design and Learning, left=2pt, right=2pt,top=2pt,bottom=2pt]
In this study, we uncover 16 research studies related to pitfalls in system design and learning. 
These pitfalls can be categorized into three main categories: data snooping, spurious correlations, and inappropriate model design.
These pitfalls lead to overestimated performance and compromised efficacy of LMs. 
Proposed solutions encompass refined data handling, model explainability, and optimization strategies like architecture adjustments, ensembling, and regularization. 

\end{tcolorbox}

\section{Performance Evaluation}
\label{sec:performance_evaluation}

The performance evaluation stage focuses on precisely assessing and analyzing the model's performance using predefined test sets and evaluation metrics.
Additionally, comparative performance evaluation against benchmarks provides insights into a model's strengths and weaknesses on specific code-related tasks.
However, potential pitfalls can emerge from factors such as improper baselines, test sets, and performance metrics.
These challenges must be thoroughly examined and addressed to ensure that the evaluation is unbiased, comprehensive, and representative of a model's true capabilities.
Thus, this section provides a brief description of related studies and discusses the implications and potential solutions during the performance evaluation stages.


\begin{figure}[t]
  \centering
  \begin{minipage}{0.5\linewidth}
    \centering
    \includegraphics[height=3.5cm, keepaspectratio]{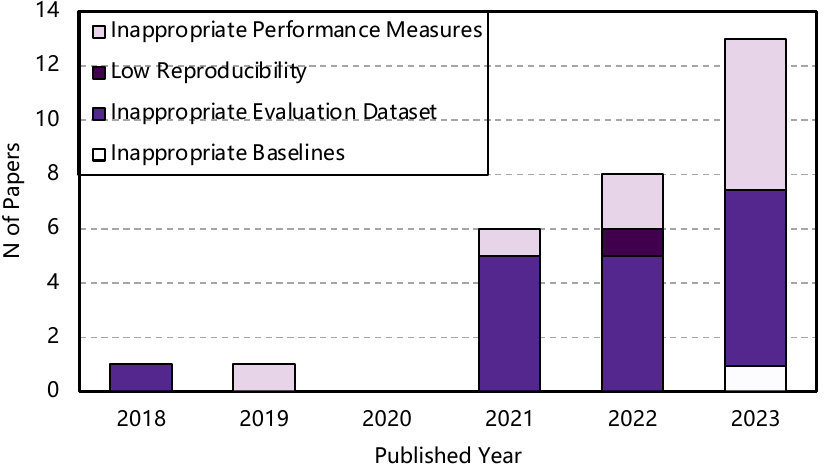} 
    \caption{Paper distribution across time (Section~\ref{sec:performance_evaluation}) }
    \label{fig:result_stage3_deploy_bias_time}
  \end{minipage}%
  \hfill
  \begin{minipage}{0.5\linewidth} 
    \centering
    \includegraphics[height=3.5cm, keepaspectratio]{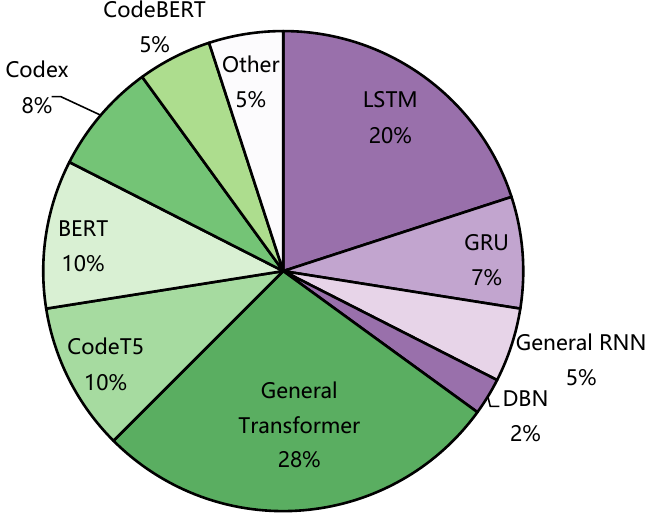}
    \caption{Distribution of LMs (Section~\ref{sec:performance_evaluation}) }
    \label{fig:result_stage3_deploy_bias_models}
  \end{minipage}
\end{figure}

\subsection{RQ1-Pitfalls}
From the collected studies, we identified 29 research studies focusing on pitfalls during the performance evaluation phase.
We have methodically categorized the collected literature into four categories, as shown in Figure~\ref{fig:result_stage3_deploy_bias_time}.

\noindent \textbf{Inappropriate Baseline:}
Inappropriate baselines arise when the performance evaluation for LM4Code is carried out without, with limited, or with skewed baseline approaches.
Such poor comparisons fail to convincingly demonstrate the improvements or strengths of newly proposed LM4Code approaches, potentially leading to misguided experimental findings or exaggerated efficacy claims.
For instance, comparing only against basic rule-based approaches would inflate capabilities, while limiting comparisons to other advanced LM-based approaches masks potential weaknesses.
Liu~\ea~\cite{liu2023towards} highlight this issue in state-of-the-art Neural Test Oracle Generation (NTOG) evaluation, where the lack of a straightforward baseline causes considerable gaps between reported and real-world performance.

\noindent \textbf{Inappropriate Evaluation Datasets:} 
Inappropriate evaluation datasets refer to the use of unsuitable, non-representative, or limited test sets that fail to adequately represent the true complexities of the studied tasks.
Such test sets can provide misleading results, and offer an obscured, over-optimistic, or pessimistic view of a model's realistic capabilities. 
Liu \ea~\cite{liu_deep_2022} highlight the frequent use of small, non-representive, and non-diverse datasets in code generation evaluations that fail to capture real-world software complexities. 
Specifically, the widely-used HS dataset~\cite{ling2016latent} comprises code from a single project, exhibiting poor diversity. 
The average code length in the Django dataset~\cite{oda2015learning} is a mere 33 characters, indicating limited program complexity.
The CoNaLa dataset~\cite{yin2018learning} utilizes automatically extracted Stack Overflow questions instead of real software requirements.
In short, common benchmarks poorly approximate the complexities of realistic code generation scenarios.
In vulnerability detection, Liu \ea~\cite{liu_cd-vuld_2022} highlight that most evaluations make the assumption that training and test datasets are drawn from the same distribution. However, they argue this assumption overlooks the continuous evolution of software vulnerabilities and projects, leading to the Cross-Domain issue where test sets should contain novel vulnerabilities or projects.
Furthermore, Nong \ea~\cite{nong_generating_2022} further reveal that vulnerability datasets like SARD~\cite{black2017sard} comprise unrealistic synthetic examples, which exhibit smaller vocabulary, smaller program length, and higher pattern frequency compared to real-world code.
Zeng \ea~\cite{zeng_deep_2021} claim that the state-of-the-art Just-in-Time (JIT) defect prediction tool, CC2Vec~\cite{hoang2020cc2vec}, was only evaluated on a limited dataset with marginal improvements to demonstrate generalizability and scalability. 
Overall, common benchmarks in code-based research utilize inappropriate test sets that fail to capture real-world complexities, leading to unrealistic performance evaluation.

\noindent \textbf{Low Reproducibility:}
Reproducibility is a fundamental requirement of scientific research to verify the validity and generalizability of findings through consistent replication across settings~\cite{ivie2018reproducibility, li2023trustworthy}.
However, LM-based research introduces specific challenges around efficiently storing big data, communication between distributed clusters, algorithm implementation, and appropriate software/hardware environments~\cite{li2023trustworthy, jiang2023empirical}.
Thus, low reproducibility in LM4Code significantly impacts performance evaluation, affecting the generalizability and trustworthiness of findings.
Zeng \ea~\cite{zeng_extensive_2022} conducted an extensive empirical study revealing reproducibility issues with pre-trained LMs for program understanding and generation.
As described in their study, multiple performance comparison results can be reversed compared to original publications. 
For example, CodeGPT incurred the largest variance across 5 clone detection runs under the benchmark BigCloneBench dataset, able to outperform or underperform other models depending on the run. 
Although PLBART reported a 0.7 F1 advantage over CodeBERT in the original paper, its min-max F1 variations of 0.84 make this finding questionable. 
Additionally, CodeBERT and PLBART's defect detection accuracy reversed between the original PLBART paper and the subsequent research by Zeng~\ea~\cite{zeng_extensive_2022}.
Such statistically significant variations could invert performance comparisons, thereby undermining LM4Code's trustworthiness.


\noindent \textbf{Inappropriate Performance Measures:} 
Inappropriate performance measures arise from a mismatch between standardized metrics and the distinct challenges inherent to software engineering tasks. 
Prior survey studies~\cite{wang_machinedeep_2023,hou2023large} have shown widespread use of generic metrics like accuracy, precision, BLEU, and Pass@k in LM4Code research.
However, these metrics often provide an incomplete view of model capabilities on code-based tasks, owing to the complex and multifaceted nature of software engineering scenarios. 
For instance, Roy~\ea~\cite{roy2021reassessing} conducted an empirical study with 226 human annotators and showed that popular metrics like BLEU and ROUGE are poor reliable indicators of human judgment. 
This demonstrates how common metrics may not fully align with code quality assessments.
Additionally, when evaluating classification tasks like vulnerability detection, reporting solely accuracy is insufficient, as true-positive and false-positive decisions are not observable~\cite{liu2023towards, chakraborty2021deep}.
Moreover, popular metrics like BLEU derived from natural language processing overlook critical attributes of code. 
As a textual similarity metric, BLEU calculates n-gram precision between generated and reference sentences. However, programming languages contain many ``trivially shared n-grams'', rendering BLEU ineffective at distinguishing actually similar code from coincidental similarities~\cite{eghbali2022crystalbleu}.
As we discussed before, code-based tasks exhibit multiple distinct challenges such as data imbalance, data snooping, execution correctness, and exception handling.
For example, despite high BLEU scores, CodeT5 generated code with only 6.4\% compilation rate, indicating poor execution correctness~\cite{wu2023effective}.
Overall, it is critical to utilize appropriate evaluation metrics in LM4Code research to provide a comprehensive and accurate understanding of model performance.

Figure~\ref{fig:result_stage3_deploy_bias_time} presents the distribution of relevant literature across years, while Figure~\ref{fig:result_stage3_deploy_bias_models} summarizes the distribution across different LMs.
Between 2018 and 2023, there was a noticeable rise in the number of papers, suggesting a growing focus by researchers on pitfalls in LM4Code during the performance evaluation phase. 
Among the identified studies, inappropriate evaluation datasets have the most mentions, with 19 studies, showing the difficulty in establishing representative datasets of LM4Code for reliable model performance. 
Additionally, inappropriate performance measures also garnered considerable research attention, with 6 identified studies.
From Figure~\ref{fig:result_stage3_deploy_bias_models}, we can observe that Transformer-based models were most discussed, evidencing their increasing adoption for SE tasks. 
In summary, these observations emphasize critical areas for improvement to enable robust evaluation and unbiased analysis of capabilities as LM4Code evolves. 
Careful consideration of evaluation datasets, metrics, and baselines will be integral to advancing progress in the field.

\subsection{RQ2-Implications}
Through a thorough examination of 29 relevant studies, our review has identified several pitfalls present in current approaches to evaluating the performance of LM4Code.
If left unaddressed, these pitfalls can produce misleading or unrealistic results, skewing perceptions of how these models might perform in real-world settings. In this section, we discuss the broader implications of the identified evaluation pitfalls and their potential impacts on LM4Code research and practical applications.

\noindent \textbf{Performance Overestimation:}
Pitfalls during performance evaluation can lead to a concerning overestimation of model capabilities, creating a misleading gap between reported metrics and real-world performance.
Key contributors include the use of inappropriate and unrealistic evaluation datasets and improper performance measures misaligned with practical objectives. 
Small and synthetic evaluation datasets often fail to adequately represent the true complexities that models face in real-world code intelligence tasks. 
Yet these limited datasets yield seemingly “perfect” metrics during evaluation. 
As Liu~\ea~\cite{liu_deep_2022} empirically showed, the capabilities of code generation models are often overestimated due to the use of limited datasets. When evaluated on the small Django~\cite{oda2015learning} and HS~\cite{ling2016latent} datasets, code generation approaches yielded strong BLEU scores of 0.811 and 0.646 respectively. However, simply switching to a new and more practical dataset led to a drastic BLEU score drop to 0.167, revealing how exaggerated initial metrics can be. 
Similarly, performance metrics misaligned with practical objectives tend to provide an unrealistic overestimate of capabilities. These inflated metrics collapse when models face the true complexities of real-world deployment.
As Ahmed~\ea~\cite{ahmed_synshine_2023} showed,  as the length of input tokens increases for program repair tasks, the accuracy of language models can decrease substantially from 82.88\% to 55\%. While overall accuracy may seem high during evaluation, metrics fail to reflect significant drops in certain practical scenarios.
Furthermore, popular metrics like BLEU can fail to accurately assess model capabilities, as demonstrated by Eghbali~\ea~\cite{eghbali2022crystalbleu}. 
Their analysis showed BLEU's shortcomings in distinguishing between a neural code generation model and a dummy model that simply exploited common n-grams. Despite the dummy model producing low-quality code, BLEU scored it equivalent to the neural model. 
This indicates BLEU's inability to differentiate between models genuinely solving complex tasks and those exploiting superficial patterns.
Overall, inappropriate evaluation practices systematically and significantly overestimate the capabilities of LM4Code models, obscuring major gaps between reported metrics and real-world effectiveness. More rigorous and realistic benchmarking is critical.

\noindent \textbf{Compromised Reproducibility:}
The reproducibility of research findings through independent verification is a fundamental requirement of the scientific process that allows reported improvements to be validated and incrementally built upon~\cite{ivie2018reproducibility}.
However, our analysis reveals that performance evaluation in LM4Code research may lack sufficient reproducibility.
The proposed approaches frequently bypass consistent evaluation protocols, datasets, and implementation details, making it difficult for others to replicate the experiments described in the literature and validate their accuracy~\cite{zeng_extensive_2022}.
Furthermore, inconsistent performance across different runs or implementations obscures the reliability of findings, as initial results that appear promising frequently fail to fully replicate in subsequent studies~\cite{zeng_extensive_2022}.
This phenomenon indicates that minor variances in experimental conditions, which are rarely comprehensively reported, can significantly influence outcomes.
Limited methodological transparency through selective reporting of details further inhibits reproducibility~\cite{jorgensen2016incorrect}.


\noindent \textbf{Misleading Benchmarks:}
Misleading benchmarks and exaggerated claims lead research communities astray and substantially hinder actual progress~\cite{van2019sok}. 
For instance, evaluating models on limited datasets or with improper metrics often amplifies perceived improvements beyond actual capabilities.
However, identifying possible pitfalls remains an open challenge for future LM4Code researchers to address.
As Liu~\ea~\cite{liu2023towards} highlighted, the lack of appropriate baselines frequently causes considerable gaps between reported and real-world performance. 
Promoting such flawed evaluations through publications and conferences propagates unreliable techniques built on shaky foundations. 
This questionable practice squanders precious community resources as researchers may pursue exaggerated claims rather than meaningful progress. 
More alarmingly, it obscures the path forward for meaningful innovation that solves real-world needs, as progress is measured on inflated claims instead~\cite{van2019sok}. 
Ultimately, systemic misleading benchmarks threaten to steer LM4Code research astray, leading the community away from impactful advancements. 
Establishing rigorous and peer-validated standards for benchmarking to guide productive research directions is an urgent need.

\subsection{RQ3-Solutions}
Given the pitfalls identified during the performance evaluation phase, it is essential to propose solutions to address these challenges and optimize the evaluation process.
We summarize the existing solutions as follows.

\noindent \textbf{Standardized and Realistic Benchmarks:}
Establishing robust and realistic benchmarks is critical for rigorous performance evaluation in LM4Code research.
Researchers should carefully select appropriate baselines based on thorough literature analysis, opting for well-established approaches tailored to specific code intelligence tasks.
This enables comprehensive capability assessment. Moreover, the community must collaboratively institute standardized benchmarks and protocols, fostering consistent evaluation. 
For example, Liu \ea~\cite{liu2023towards} highlight the need for establishing straightforward and realistic baselines to ensure truthful evaluation outcomes. 
For specific tasks, standardized task-specific benchmarks have emerged, such as VJBench proposed by Wu~\ea~\cite{wu2023effective} for automated program repair. 
Such domain-specific benchmarks facilitate targeted capability assessment.
Furthermore, diverse benchmark datasets play a significant role in comprehensively evaluating and advancing LM4Code.
For instance, Lu~\ea~\cite{lu2021codexglue} introduced CodeXGLUE, a comprehensive collection that encompasses 10 SE tasks across 14 datasets. It provides a platform for structured model evaluation and comparison across tasks using standardized baselines.

\noindent \textbf{Enhancing Reproducibility:}
Inconsistent replication remains a key pitfall during performance evaluation in LM4Code~\cite{jiang2023empirical}. 
Some solutions to enhance reproducibility have been proposed.
At its core, it's critical that code, datasets, and evaluation scripts are made publicly accessible, echoing the call for open-source initiative and providing continuously maintained links to high-quality replication packages~\cite{liu2021Reproducibility}.
Furthermore, reporting key statistical measures like variance, confidence intervals, and significance quantifies stability across runs.
Such transparency fosters community collaboration, facilitating the validation of research outcomes and paving the way for future scholarly advancements.
Carefully controlling and reporting experimental conditions like hardware specifications, software versions, random seeds, and tuning details significantly influences outcomes~\cite{hutson2018artificial}.
Chen~\ea~\cite{chen2022towards} introduced a systematic approach to train reproducible AI systems, with general criteria to evaluate reproducibility, and a unified framework to decrease randomness.

\noindent \textbf{Refined Performance Measures:}
Appropriate and comprehensive performance metrics are critical for accurately evaluating model performance on complex code-related tasks.
To address the limitations of common generic metrics like BLEU and accuracy, researchers have proposed more refined measures tailored to LM4Code challenges.
Ren~\ea~\cite{ren2020codebleu} introduced CodeBLEU to improve upon BLEU for assessing deeper semantic similarities in code generation tasks, while Eghbali~\ea~\cite{eghbali2022crystalbleu} introduced CrystalBLEU for minimizing the noise from commonly shared n-grams in programming languages.
Given the runtime nature of programming languages, metrics like compilation rate and execution accuracy on test cases have also been used~\cite{wu2023effective}.
For specific LM4Code tasks, domain-specific metrics can prove more reliable.
For example, for test oracle generation, Liu~\ea~\cite{liu2023towards} presented an evaluation metric named Found@K, which counts how many bugs can be found if developers only check the top-K recommended test cases per bug.
To better capture complex code quality attributes, direct human assessment can complement automated metrics, evaluating aspects like readability, conciseness, and correctness ~\cite{paltenghi_thinking_2021, dakhel2023github}.
Overall, more refined, code-aware, and multifaceted performance measures aligned with practical goals provide a more accurate and reliable model capability assessment.

\begin{tcolorbox}[title=Summary - Performance Evaluation, left=2pt, right=2pt,top=2pt,bottom=2pt]
In the performance evaluation of LM4Code, we identify four pitfalls related to performance evaluation, including inappropriate baseline, inappropriate evaluation datasets, low reproducibility, and inappropriate performance measures. 
From 29 relevant research studies, inappropriate datasets and metrics receive much attention. 
Such pitfalls can lead to overestimated evaluation and compromised reproducibility, misleading benchmarks for the future. 
To address this, tailored solutions like standardized benchmarks, transparency, realistic assessments, and community coordination are needed.
\end{tcolorbox}
\section{Deployment and Maintenance}
\label{sec:deployment}
Systems based on advanced LM4Code, such as GitHub Copilot~\cite{github2021copilot}, Amazon CodeWhisperer~\cite{amazon2022codewhisperer}, and Microsoft IntelliCode~\cite{svyatkovskiy2020intellicode}, have already been deployed in real-world IDEs and have garnered a large number of users.
There exist various challenges when such LM4Code systems are deployed in practice, like security threats and how LM4Code systems should be updated to adapt the rapidly changing software practices.
This section discusses pitfalls, implications, and current solutions when deploying and maintaining LM4Code.

\begin{figure}[t]
  \centering
  \begin{minipage}{0.5\linewidth}
    \centering
    \includegraphics[height=3.5cm, keepaspectratio]{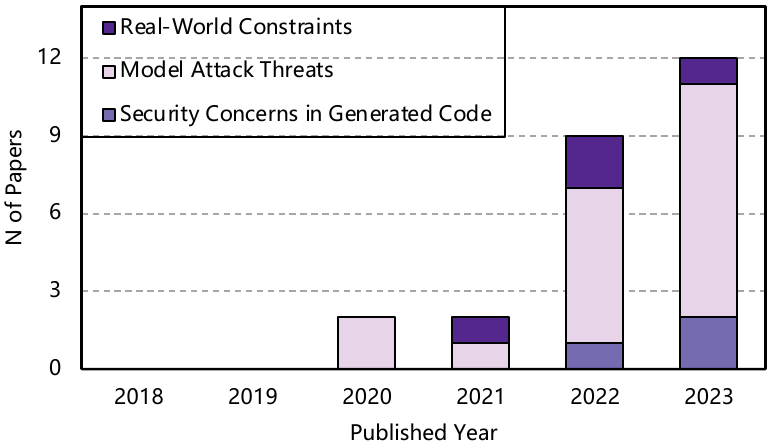} 
    \caption{Paper distribution across time (Section~\ref{sec:deployment}) }
    \label{fig:result_deploy_bias_time}
  \end{minipage}%
  \hfill
  \begin{minipage}{0.5\linewidth} 
    \centering
    \includegraphics[height=3.5cm, keepaspectratio]{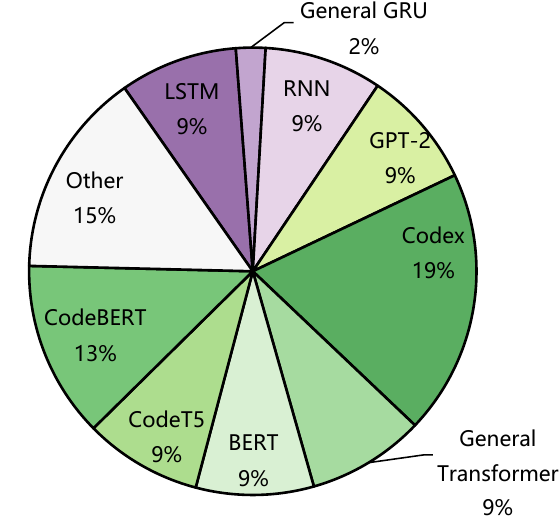}
    \caption{Distribution of LMs (Section~\ref{sec:deployment}) }
    \label{fig:result_deploy_bias_models}
  \end{minipage}
\end{figure}

\subsection{RQ1-Pitfalls}
In our literature review, we identified 25 research papers involving the pitfalls of LM4Code deployment and maintenance.
This number of identified publications surpasses that related to other stages, highlighting the fact that deploying and maintaining LM4Code presents a complex set of challenges that have gained considerable attention from the research community.

Figure~\ref{fig:result_deploy_bias_time} and Figure~\ref{fig:result_deploy_bias_models} present the distribution of these 25 research studies over time and different types of LLM4Code.
From Figure~\ref{fig:result_deploy_bias_time}, we can observe that there has been a significant increase in the number of research papers focusing on the pitfalls in the deployment and maintenance of LM4Code, especially in the past two years. 
The data presented in Figure~\ref{fig:result_deploy_bias_models} demonstrates a significant preference for utilizing the Codex model for analyzing biases related to deployment and maintenance, which may be due to the fact that the widely used Github Copilot is based on Codex.
Moreover, large pre-trained models such as CodeT5, BERT, Codex, and GPT-2 appear to surpass traditional LMs such as RNN and LSTM in terms of widespread use and popularity.
These observed patterns indicate a shift in research focus towards investigating the complexities of advanced models when deployed in real-world contexts. 
In the subsequent sections, we will examine the particular biases and issues that arise in regard to the development and maintenance of LM4Code.

\vspace{0.2cm}
\noindent \textbf{Real-World Constraints:} 
Evaluating LM4Code systems solely in controlled lab settings often overlooks practical constraints and complexities of real-world deployment~\cite{arp2022and}.
Although controlled evaluations provide useful insights into the effectiveness of a model within specific settings, they sometimes overlook the complexity and diversity of real-world deployments.
It is important to consider runtime and storage constraints when deploying an LM4Code system in the wild.
Svyatkovskiy~\ea~\cite{svyatkovskiy2021fast} mentioned that a reasonable upper bound of model size for an IDE plugin is 50 MB.
However, the size of LM4Code keeps increasing.
For example, the popular LM4Code, CodeBERT~\cite{feng2020codebert}, has 125 million parameters and is 425 MB in size, which is way larger than the suggestion by Svyatkovskiy~\ea~\cite{svyatkovskiy2021fast}.
Recently proposed models are even larger.
Considering LLaMa, developed by Meta, its smallest version has 6.7B parameters~\cite{lu2023llama}.
Models that appear to have good performance in a controlled setting may be resource-intensive or excessively large for practical applications, especially in environments with limited computational power or storage capacity~\cite{feng_performance-sensitive_2021}.

\vspace{0.2cm}
\noindent 
\textbf{Attack Threats:}
LM4Code systems face various threats from malicious attackers.
Despite the breakthrough capabilities exhibited by LMs, prior works have noted that the state-of-the-art LMs are vulnerable to a variety of attack threats such as evasion attacks~\cite{zhou_adversarial_2022, wang2023robustness, mastropaolo2023robustness}, backdoor attacks, privacy attacks, etc.
These threats can appear across the entire lifecycle of the model, from data collection to model deployment.
We identified a total of 17 research papers that specifically focus on the analysis of security threats within the LM4Code system.
Figure~\ref{fig:result_deploy_bias_time} shows that the number of these studies has increased over the recent years.
Our study employs the classification scheme mentioned in Battita \ea's work~\cite{biggio2018wild}, which has identified a diverse range of attack types in machine learning.

\begin{enumerate}
    \item \textit{Evasion Attacks:} Evasion attacks (a.k.a. adversarial attacks) leverage adversarial samples~\cite{liu2023contrabert, zhou_adversarial_2022}, which are carefully perturbed instances while appearing as regular and benign inputs to the human observer.
    Yet, these small perturbations can mislead the trained LM4Code model to produce incorrect predictions.
    As shown in Table~\ref{tab:summary_model_attacks}, evasion attacks are a major concern, with 9 out of the 17 examined papers investigating them.
    Zeng~\ea~\cite{zeng_extensive_2022} thoroughly evaluated eight different adversarial attack approaches (i.e., word importance rank, genetic algorithm, random substitution) against widely-used LM4Code, including CodeBERT, GraphCodeBERT, and CodeT5.
    Their findings emphasized the vulnerability of these models to semantics-preserving adversarial samples.
    Interestingly, even simple random attack techniques showed significant effectiveness in degrading pre-trained LM4Code.
    
    \item \textit{Poisoning Attacks:} Poisoning attacks inject malicious examples into the training data, aiming to manipulate the model's behavior~\cite{biggio2018wild, sun2022coprotector}. 
    This can lead the model to produce incorrect or attacker-chosen outputs when certain triggers appear in the inputs~\cite{schuster2021you}.
    Poisoning attacks can be categorized into two main classes: untargeted poisoning attack and targeted poisoning attack~\cite{tian2022comprehensive}.
    One special case of the target poisoning attack is the backdoor attack, where adversaries carefully insert a distinct pattern into a subset of training samples to embed a backdoor.
    When the pattern appears, the model produces pre-defined outputs (e.g., recommending vulnerable APIs).
    Otherwise, the model behaves normally.
    Table~\ref{tab:summary_model_attacks} lists six papers related to poisoning attacks, with all of them specifically focusing on backdoor attacks.
    Among various code scenarios, three studies~\cite{wan_you_2022, sun2023backdooring, sun2022coprotector} target LMs-based code search models, demonstrating that backdoor samples within code search tasks closely resemble clean code and are not easily differentiated. 
    Four studies~\cite{schuster2021you, li2023multi, sun2022coprotector, peng2022security} focus on code generation scenarios, demonstrating that attackers can manipulate code recommendations.

    \item \textit{Privacy Attacks:}  
    Privacy attacks refer to attacks aiming to infer the private information of LM4Code, e.g., parameters of models that are hosted remotely and the model training data.
    One such example is model stealing (a.k.a. model extraction) involves extracting knowledge (e.g., hyperparameters, model architecture, and training data) from a trained model without direct access to its parameters or training data~\cite{carlini2021extracting, oliynyk2023know}. 
    Adversaries use this technique to `copy' the functionality of a model, often by querying it repeatedly and analyzing the outputs~\cite{oliynyk2023know}.
    For example, Lukas~\ea~\cite{lukas2023analyzing} delved into the risk of language models, like GPT-2, leaking personally identifiable information. 
    Yang et al.~\cite{yang2023code} found that simply extracting 20,000 outputs (each having 512 tokens) from CodeParrot~\cite{codeparrot} can produce over 40,125 code snippets that are memorized from its training data.
    In addition, Niu~\ea~\cite{niu2023codexleaks} introduced and evaluated a semi-automated pipeline that employs a membership inference approach on various code generation models like CodeParrot~\cite{codeparrot} and Polycoder~\cite{xu2022systematic}. 
    By leveraging the GitHub Search API's hit rate as a distinguishing heuristic and incorporating human-in-the-loop evaluations, they found that approximately 8\% (43) of the prompts in the Codex model, used in GitHub Copilot, resulted in privacy leaks.

\end{enumerate}

\begin{table}[t]
  \centering
    \resizebox{1\linewidth}{!}{
    \begin{tabular}{llrl}
    \toprule
          & \multicolumn{1}{c}{\textbf{Category}} & \multicolumn{1}{c}{\textbf{Percentage}} & \multicolumn{1}{c}{\textbf{References \& Details}} \\
    \midrule
    \multicolumn{1}{l}{\multirow{3}[2]{*}{\textbf{Attack Objectives}}} & Evasion attacks & 59\%  &  \cite{zhou_adversarial_2022, zeng_extensive_2022, zhang_generating_2020, henkel_semantic_2022, zhang_towards_2022, liu2023contrabert, gao2023two, wang2022recode, yang2022natural}  \\
          & Poisoning attacks & 35\%  &  \cite{schuster2021you, wan_you_2022, li2023multi, sun2023backdooring, sun2022coprotector, peng2022security}  \\
          & Privacy/confidential attacks & 6\%   &  \cite{lukas2023analyzing, niu2023codexleaks}  \\
    \midrule
    \multirow{3}[2]{*}{\textbf{Attack Models}} & RNN-based (e.g., LSTM) & 41\%  &  \cite{schuster2021you, zhou_adversarial_2022, zhang_generating_2020, henkel_semantic_2022, zhang_towards_2022, wan_you_2022, gao2023two}  \\
          & General Transformer & 29\%  &  \cite{zhou_adversarial_2022, zeng_extensive_2022, wan_you_2022, gao2023two, wang2022recode}  \\
          & LLM (e.g., Codex and T5) & 41\%  &  \cite{schuster2021you, zeng_extensive_2022, lukas2023analyzing, niu2023codexleaks, li2023multi, sun2022coprotector, peng2022security, sun2023backdooring}  \\
    \midrule
    \multirow{2}[2]{*}{\textbf{Countermeasure-inclusive}} & Include & 71\%  &  \cite{schuster2021you, zhou_adversarial_2022, zhang_generating_2020, henkel_semantic_2022, zhang_towards_2022, wan_you_2022, liu2023contrabert, gao2023two, lukas2023analyzing, sun2023backdooring, yang2022natural, sun2022coprotector}  \\
          & Non-include & 29\%  &  \cite{zeng_extensive_2022, niu2023codexleaks, li2023multi, wang2022recode, peng2022security}  \\
    \bottomrule
    \end{tabular}%
  }
  \caption{Summary of Attack Threats in Reviewed LM4Code Papers}
  \label{tab:summary_model_attacks}%
\end{table}%

\vspace{0.2cm}
\noindent 
\textbf{Security Concerns in Generated Code:} 
The outputs from LM4Code, i.e., generated code, will be further used in other software systems. 
Consequently, the safety and robustness of the generated code come under scrutiny. 
Several recent investigations have shown that LM-generated code can contain vulnerabilities, emphasizing the need for rigorous validation and enhancement of their outputs.
Pearce~\ea~\cite{pearce_asleep_2022} analyzed the generated code of GitHub Copilot and identified security vulnerabilities. 
The authors produced 89 different scenarios for Copilot to complete, resulting in 1,689 programs. Alarmingly, they found approximately 40\% of these to be vulnerable. 
Such vulnerabilities arise because code often contains bugs, and given the vast quantity of unvetted code that Copilot processes, the language model is prone to learning from exploitable and buggy code.

\subsection{RQ2-Implications}
According to the reviewed papers, the identified pitfalls and security concerns related to the deployment and maintenance of LM4Code have significant implications for both researchers and practitioners in artificial intelligence and software engineering.
Next, we will describe the implications in detail.

\noindent 
\textbf{Untrustworthy Results:} 
Both the pitfalls from external attackers and internal drawbacks can lead to untrustworthy outputs, requiring the developers to carefully review and test the generated code.
External attacks like evasion, poisoning, and backdoor attacks, can induce LM4Code systems to demonstrate manipulated behavior, i.e., generating outputs specifically chosen by attackers.
For example, Schuster \ea~\cite{schuster2021you} demonstrate that code completion models can be manipulated to use unsafe encryption algorithms and deprecated security protocol so that the system built based on these suggestions is vulnerable to further attacks.
Aghakhani \ea~\cite{aghababaeyan_black-box_2023} reveal that the backdoors in code generation models can inject various insecure code, including Cross-Site Scripting (CWE-79), Path Traversal (CWE-22), and Deserialization of Untrusted Data (CWE-502).
Wan \ea~\cite{wan_you_2022} can mislead the code search models to retrieve the code snippets containing malicious actions, such as deleting specific files.
Attackers can also provoke systems into producing unexpected predictions so that the application behaves in an unintended way.
For example, the adversarial attacks~\cite{yang2022natural, zhang_challenging_2023} against LM4Code Systems can lead to completely wrong predictions.
Even without external attacks, the model itself may also naturally generate untrustworthy results, which has been largely reported in various studies.
For instance, Pearce~\ea~\cite{pearce_asleep_2022} revealed vulnerabilities frequently present in GitHub Copilot's generated code.
Furthermore, Lukas~\ea~\cite{lukas2023analyzing} suggested LMs may inadvertently retain or even leak portions of their training data, which raises concerns regarding unintentional memorization of sensitive information.
Niu~\ea~~\cite{niu2023codexleaks} investigated code completion models like the popular Codex model leaking private information (e.g., passwords and addresses) through generated outputs.


\noindent 
\textbf{Copyright Infringement:}
Pitfalls in LM4Code deployment raise significant concerns regarding copyright infringement and loss of intellectual property.
Attackers might resort to model stealing, effectively replicating the functionality of proprietary models without authorization~\cite{johnson2023ru, li2023feasibility}.
Li~\ea~\cite{li2023feasibility} demonstrate that imitation models can even exceed the performance of the victim model.
Additionally, when models are deployed on user clients, they face the potential threat of reverse engineering.
Zhou~\ea~\cite{zhou2023modelobfuscator} highlight that the on-device models may leak their confidential information, such as hyper-parameters and weights.
The risk of copyright infringement and intellectual property theft not only undermines incentives to develop innovative models, but also threatens the commercialization prospects and trustworthiness of the LM4Code industry.


\noindent 
\textbf{Compromised Model Efficacy:} 
In real-world practice, many constraints, such as latency, device memory, and computational costs, need to be satisfied.
On the one hand, it prevents some powerful models from being deployed.
For instance, Feng~\ea~\cite{feng_performance-sensitive_2021} noted multiple malware detection models are infeasible on mobile devices due to limitations in computational power, memory, and energy consumption. 
Zhang~\ea~\cite{zhang_diet_2022} highlighted that even standard CodeBERT requires extensive resources for pre-training and fine-tuning before use, which may not be affordable for individual developers.
On the other hand, the performance of deployed systems may be compromised as a trade-off. 
For example, to be deployed on the user devices, the models have to be pruned or distilled, where considerable performance degradation can be observed~\cite{shi2022compressing, shi2023smaller}.
Furthermore, Ganesh~\ea~\cite{ganesh2021compressing} showed that while quantization and unstructured pruning can reduce model size, these techniques alone do not improve runtime inference speed or memory usage during execution, unless paired with specialized hardware or processing libraries.
Therefore, while constraints exist in real-world deployment that compels simpler models, such compromises often degrade model performance, highlighting the need for techniques that can provide comparable accuracy with greater efficiency.

\subsection{RQ3-Solutions}
\subsubsection{Improving Model Robustness} 
Improving model robustness is essential to defend against evolving security threats targeting LM4Code systems. 
Here, we summarize some of the available solutions that have been proposed to enhance the robustness of LM4Code models:

\noindent \textbf{Adversarial Training:}
Adversarial training incorporates adversarial examples into training data to improve model robustness against adversarial attacks.
Regarding code comment generation, Zhou~\ea~\cite{zhou_adversarial_2022} demonstrated masked adversarial training can significantly enhance robustness while retaining performance on normal test data. 
Zhang~\ea~\cite{zhang_generating_2020} illustrated that adversarially trained language models exhibit markedly reduced attack success rates (approximately 30\%-40\%).
Moreover, Henkel~\ea~\cite{henkel_semantic_2022} implemented robust optimization employing a semantics-preserving adversary, and this approach outperformed standard data augmentation, optimally balancing accuracy on clean samples with robustness to perturbations.

\noindent \textbf{Inference with Self-repair:}
Self-repair is to let the model introspect and correct mistakes or vulnerabilities in its own code.
Delving into this, Olausson~\ea~\cite{olausson2023demystifying} investigated to what extent the GPT models can provide accurate feedback on the causes of code errors, validating the efficacy of GPT-4's self-repair capabilities. 
Moreover, Chen~\ea~\cite{chen2023improving} enhanced the self-repair process by incorporating knowledge from human-written natural language feedback.

\noindent \textbf{Domain Knowledge:}
Beyond the internal capability of the model, external aids with domain knowledge can also be employed to address the challenges.
For instance, Jain~\ea~\cite{jain2022jigsaw} enhances these LLMs by utilizing a post-processing step based on program analysis and synthesis techniques, thereby improving the correctness of the syntax and semantics in code. Wei~\ea~\cite{wei2023copiloting} introduced Repilot, a framework designed to further assist AI ``copilots'' (i.e., LLMs) by synthesizing more valid patches during the repair process.
Moreover, Johnson~\ea~\cite{johnson2023ru} proposed Random Utility-Driven Synthesis Under Uncertain Regions (R-U-SURE), a method that builds uncertainty-aware suggestions based on a decision theory model of objective conditional utility, using random samples from generative models as proxies for unobserved potential intents of end-users.
Poesia \ea~\cite{poesia2022synchromesh} propose a framework to apply constraints on partial outputs to produce complete correct programs without re-training or fine-tuning of the language model.


\noindent \textbf{Train with Real-world Dataset:}
Directly learning from the real-world datasets helps the model to better align with the user intention.
For instance, Aye et al.~\cite{aye2021learning} observed a significant decline in the accuracy of Transformer sequence models when tested using real-world data from production logs.
Moreover, they demonstrated that training on real-world examples yields a more robust model.
However, there is a significant shortage of large-scale, real-world datasets, especially for vulnerability.
Nong~\ea~\cite{nong_generating_2022} suggest a promising alternative solution: using deep learning to generate real-world samples.

\subsubsection{Enhancing Computational Efficiency}
To make full use of the limited resources, various measures have been proposed.

\noindent \textbf{Model Compression and Specialization:}
To enable the deployment of large language models in resource-constrained environments, model compression techniques have been widely explored~\cite{polino2018model, gupta2022compression}. 
Shi~\ea~\cite{shi2022compressing} introduced Compressor, which uses a genetic algorithm to guide the simplification of pre-trained code models. 
This compressed models to significantly smaller sizes with acceptable accuracy loss.
In addition to compression, methods to specialize large language models (LLMs) for specific tasks show promise. 
Parameter-Efficient Fine-Tuning (PEFT) effectively adapts LLMs using limited task data. 
As Weyssow~\ea~\cite{weyssow2023exploring} demonstrated, PEFT outperforms approaches like Incremental Curriculum Learning in reducing computational overhead and boosting performance when specializing broad LLMs.



\noindent \textbf{Efficient Inference:}
Numerous inference optimization strategies have also been proposed. 
For instance, Zhang~\ea~\cite{zhang_diet_2022} adopted input simplification strategies like word dropout and frequency filtering to simplify input programs.
By focusing on the most informative tokens, computational costs are significantly reduced.
Additionally, Chirkova~\ea~\cite{chirkova2023codebpe} introduced a tokenization method that reduces the average token length by 17\% without any downstream performance loss.
They further demonstrated that a properly chosen tokenization can even enhance the model's performance by 0.5-2\%. 
Furthermore, Svyatkovskiy~\ea~\cite{svyatkovskiy2021fast} unveiled an innovative neural completion model by combining static analysis with granular token encoding. This model boasts a lean memory footprint, consuming just 6 MB of RAM — a significant 19x reduction compared to previous models.
It can generate a single piece of code completion in a mere 8 ms and delivers an impressive 90\% accuracy rate for its top five suggestions.


\subsubsection{Privacy and Copyright Protection}
The ability of large language models to memorize and reproduce training data raises critical privacy and copyright concerns.
Thus, developing effective protection for privacy and copyright has been widely investigated.

\noindent \textbf{Privacy-preserving Techniques:}
The memorization and regeneration capabilities of large language models raise critical privacy concerns that demand research attention.
Some state-of-the-art models like StarCoder employed human annotators to mask any personal information such as keys and addresses present in the training data, in an effort to mitigate privacy risks~\cite{li2023starcoder}.
In addition, differential privacy techniques have emerged as promising strategies for mitigating privacy risks.
Lukas~\ea~\cite{lukas2023analyzing} introduce a sentence-level differential privacy approach, which provides guarantees under the assumption that records are unlikely to be duplicated.
Their results show that while helpful in reducing privacy leakage to a large extent, differential privacy alone cannot completely eliminate risks.
Further research into complementary privacy-preserving mechanisms is needed to develop LM4Code that generates high-quality outputs while provably protecting user privacy.

\noindent \textbf{Model Obfuscation:}
Obfuscating models by hiding their structure and parameters has been proposed as a technique to protect against extraction attacks.
Prior Research has shown that attackers can easily craft white-box-like attacks against models on devices, even to the extent of reversing their training data~\cite{huang2021robustness}.
For example, Zhou~\ea~\cite{zhou2023modelobfuscator} developed ModelObfuscator to apply techniques like model file obfuscation and model structure obfuscation. 
In model file obfuscation, they utilized renaming, parameter encapsulation, and neural structure obfuscation approaches, effectively obfuscating the data and structures of on-device models.
model structure obfuscation utilizes shortcut and extra layer injection, making reverse engineering harder.
Although model obfuscation has shown promise by Zhou~\ea~\cite{zhou2023modelobfuscator}, it can increase library size, introducing extra memory overhead and computation time.
Further research is needed to balance security, efficiency, and accuracy as model obfuscation techniques are applied.



\noindent \textbf{Watermarking:} 
Watermarking techniques can help protect model intellectual property.
For instance, Sun~\ea~\cite{sun2022coprotector} introduced CoProtector, which uses data poisoning to embed watermarks into source code repositories.
This ensures open-source code can't be exploited by models while providing a way to reveal watermark backdoors.
However, CoProtector notably diminishes model performance, making it hard to adopt broadly.
In neural code completion, Sun~\ea~\cite{sun2023codemark} proposed CodeMark which embeds imperceptible user-defined watermarks into code.
This traces code utilization while meeting key watermark properties like harmlessness, verifiability, and robustness.

\begin{tcolorbox}[title=Summary - Deployment and Maintenance, left=2pt, right=2pt,top=2pt,bottom=2pt]
Based on 25 relevant studies, our literature review reveals three main pitfalls in deploying and maintaining LM4Code systems: real-world constraints, attack threats, and generated code containing security concerns.
These pitfalls can lead to untrustworthy results, copyright infringement, and reduced model efficacy.
Proposed solutions involve improving robustness through adversarial training and error detection, enhancing efficiency via compression and optimized inference, and protecting privacy and copyright through obfuscation and watermarking.
However, there remains a need for robust evaluation frameworks and techniques that balance security, efficiency, and accuracy. The implications highlight that real-world deployment introduces complex challenges for LM4Code systems.
\end{tcolorbox}

\section{Discussion}
\label{sec:Discussion}
\subsection{Recommendations for LM4Code Research}
Our systematic literature review reveals numerous pitfalls that can undermine the realistic performance and real-world effectiveness of LM4Code systems. 
These pitfalls span the data, models, evaluation, and deployment phases of the LM4Code lifecycle. 
LM4Code has become an increasingly prominent research area, evidenced by the rapid increase in publications. 
A prior survey by Hou~\ea~\cite{hou2023large} uncovered 229 papers on large language models for software engineering between 2020-2023, while Wang~\ea~\cite{wang_machinedeep_2023} uncovered 350 papers on deep learning for software engineering between 2015-2020. 
However, our focused study on LM4Code pitfalls only identified 67 relevant papers. 
This indicates that research attention to pitfalls in LM4Code is still insufficient, compared to the overall research volume.
Thus, future LM4Code research must not overlook the pitfalls when applying LM4Code models to software engineering tasks.

\noindent \textbf{\underline{Recognizing Existing Pitfalls.}}
In this study, we summarized multiple common pitfalls associated with LM4Code. 
These pitfalls,  each with distinct implications, highlight the inherent complexities in applying LM4Code to real-world software engineering problems. 
As our review results demonstrate, pitfalls can introduce unrealistic performance evaluation, compromise model efficacy, and raise security concerns. 
Thus, it becomes important for future LM4Code research to recognize and avoid potential pitfalls when building LM4Code systems for SE tasks.
To ensure trustworthy findings of LM4Code research, it is essential to demonstrate effectiveness through a rigorous and reliable experimental design that reflects real-world scenarios.
Furthermore, although our review has summarized common pitfalls, as previous surveys like Hou~\ea~\cite{hou2023large} present, there exist more than 50 specific large language models tailored to over 55 software engineering scenarios. 
So while we report general implications in some common settings using prevalent models, further investigation is required to discern more specific implications in specific or unconventional scenarios.

\noindent \textbf{\underline{Addressing Existing Pitfalls.}}
Addressing the identified pitfalls is vital for advancing robust and reliable LM4Code techniques.
As our study reveals, solutions like data cleaning, model explainability, optimized model design, and rigorous benchmarking have shown promise in mitigating certain pitfalls. 
However, these solutions, although effective in specific contexts, may not be universally applicable due to the complexity and ever-evolving nature of the software engineering landscape.
A coordinated effort by the community is required to establish guidelines and best practices that enable mitigating pitfalls in data construction, model design, performance evaluation, and deployment.

\noindent \textbf{\underline{Uncovering New Pitfalls.}}
The dynamic nature of the LM4Code field means novel pitfalls will likely emerge as techniques rapidly evolve. 
Specifically, the ever-evolving software engineering landscape, increasingly complex codebases, and new LM4Code techniques provide fertile ground for novel pitfalls to emerge.
Thus, identifying emerging pitfalls is critical.
The community needs to continuously analyze model reasoning, behaviors, and performance under realistic experimental settings to uncover new pitfalls in a timely manner. 
For example, testing on diverse unexplored situations or probing model decisions via XAI techniques yield valuable insights.
Through periodically updating benchmarks, refining evaluation approaches, and incorporating real-world deployment scenarios, we can ensure that we not only keep pace with the ever-evolving landscape, but also recognize and mitigate new challenges.


\subsection{Open Challenges and Research Directions}
\subsubsection{Improving Data Quality for LM4Code}
The processes of data collection and labeling play critical roles in the model construction, determining their efficiency, trustworthiness, and overall performance~\cite{lo2023trustworthy}.
With the emergence of large language models such as GPT-4~\cite{openai2023gpt4}, the reliance on massive data has increased significantly. 
However, as we have reviewed, there are numerous obstacles and unanswered questions associated with these phases. 

\noindent \textbf{\underline{Volume vs. Quality.}}
The increasing number and widespread use of large language models such as GPT-4,  underscore the inherent conflict between the quantity and quality of data.
These models, with their vast number of parameters, heavily rely on extensive datasets to achieve their remarkable performance.
For example, Codex~\cite{chen2021evaluating}, which is a variant of the GPT-3.5 framework introduced in 2021, conducted training using a dataset that was sourced from 54 million publicly available software repositories on the GitHub platform, resulting in a total data size of 159 GB.
Yet, collecting vast amounts of data, a process that is both time-consuming and labor-intensive, is not a ``silver bullet''; data quality is also important.
Even with advanced large models, ``garbage in - garbage out''~\cite{liu2023refining, pearce_asleep_2022}.
Notably, even widely-used benchmark datasets like CodeSearchNet~\cite{husain2019codesearchnet} have been observed to contain considerable noise and errors, as highlighted in prior research~\cite{sun_importance_2022,shi2022we}.
Over-reliance on volume can lead to models that are prone to biases, noise, and even adversarial attacks. 
On the contrary, an excessive focus on data quality can inadvertently decrease the diversity and richness of the dataset. 
Finding the ideal compromise between these conflicting requirements offers an appealing opportunity for future investigation.

\noindent \textbf{\underline{Automated Data Quality Assurance.}}
As the implications of low-quality data, it is essential to utilize high-quality datasets to develop LM4Code approaches.
In the dynamic evolution environment of software and language models, manually searching and examining extensive datasets for data noise or errors is neither practical and efficient.
As a result, there is an urgent need for automated tools and frameworks that can assure and maintain data trustworthiness and quality, particularly for LM4Code models.
We can systematically discover and correct data noise, labeling errors, and other anomalies, allowing models to train on robust and reliable datasets and ensuring their reliability and efficiency in real-world applications.

\subsubsection{Strengthening Robustness and Trustworthiness in LM4Code}
LM4Code models are becoming increasingly integrated into the software development lifecycle, influencing everything from code generation to vulnerability detection. 
Ensuring that these models are robust and trustworthy is essential. This does not just relate to their prediction accuracy but extends to the reliability, interpretability, and generalization capacity of the model, especially in diverse and evolving coding environments~\cite{niu2023crosscodebench,wang2022machine}.

\noindent \textbf{\underline{Building Interpretable LM4Code.}}
The black-box nature of language models has been a long-term concern, especially when LM4Code applications directly influence software development outcomes~\cite{wang2022machine, watson2022systematic,10109341, 10109328}. 
Transparency in LM4Code requires an in-depth examination of the correlations and reasoning processes that models depend on, instead of just knowing the model's predictions. 
Our review results show that pitfalls can exist throughout the entire LM4Code lifecycle, potentially resulting in spurious correlations.
These misleading correlations are based on wrong artifacts for generating predictions, presenting significant challenges for practical real-world applications.
Although prior studies~\cite{cito_explaining_2021, shi2023towards, wan_what_2022, zou_interpreting_2021} have introduced various explainable AI (XAI) techniques into LM4Code, the current solutions lag behind the rapid evolution of language models~\cite{zini2022explainability, madsen2022post}.
Existing XAI techniques for LMs, in particular, only provide explanations as either the contribution of individual words to the decision or the layer/neuron at which syntax or semantics are encoded~\cite{zini2022explainability}.
Although helpful, these explanations only offer a fragmented picture of the model's decision-making process, ignoring a considerable amount of its intricate reasoning.
As the complexity of language models grows, there is an imperative need to develop more comprehensive and accessible XAI approaches for LM4Code.

\noindent \textbf{\underline{Improving Robustness Against Errors and Threats.}}
Recent literature~\cite{niu2023codexleaks, sun2022coprotector, peng2022security} reveals that state-of-the-art LM4Code models like Codex~\cite{chen2021evaluating}, GPT-3, and Starcoder~\cite{li2023starcoder} are susceptible to inadvertent data errors and malicious threats. 
Such pitfalls not only degrade model performance, but raise concerns about the security and trustworthiness of LM4Code systems.
Improving model robustness is therefore an urgent need to enable reliable LM4Code adoption.
While prior studies~\cite{sun_importance_2022, sun2022coprotector, sun2023backdooring, zhang_towards_2022} have proven that techniques like adversarial training and data augmentation can enhance robustness, more efforts should be spent by our research community to holistically defend against newly emerging issues and threats. 
Specifically, possible solutions like domain-specific preprocessing and learning, continuous evaluation, hybrid models, and anomaly detection should be explored. 
Overall, a multilayered defense-in-depth strategy is essential to ensure LM4Code reliability and trustworthiness against growing pitfalls or issues.


\noindent \textbf{\underline{Adapting to Ever-evolving Code Environments.}}
Within the ever-evolving field of software development, new programming languages emerge, old ones get updated, and coding techniques and habits are constantly transforming.
Against this backdrop,  a primary challenge for LM4Code is ensuring the broad applicability and robust generalizability of LM4Code systems.
Prior studies~\cite{nong_generating_2022, thongtanunam_autotransform_2022, rabin2021generalizability, wang2022machine} have highlighted the superior generalizability of advanced LMs over traditional ML/DL techniques,  particularly when it comes to previously unseen distributions.
However, there remain significant research gaps. 
Specifically, while the improvements in models like the advanced Transformer are promising, they may not always translate to practical efficiency. 
For instance, Thongtanunam~\ea~\cite{thongtanunam_autotransform_2022} indicate that while the Transformer demonstrates an improvement of 490\% to 567\% on new tokens, its accuracy remains around 10\%, indicating a significant room for improvement.
Thus, the incorporation of domain-specific knowledge, continuous model updating, and feedback loops with developers could pave the way for more adaptable and reliable LM4Code solutions in an ever-evolving coding landscape.

\subsubsection{Towards Reliable Performance Evaluation of LM4Code}
Performance evaluation plays a key role in demonstrating the efficiency and capability of the proposed LM4Code approaches.
However, as discussed in Section~\ref{sec:performance_evaluation}, there are several pitfalls that can undermine realistic performance, including inappropriate baselines, inappropriate test sets, reproducibility issues, and inappropriate performance measures.
Therefore, it is important to focus future research efforts on solid benchmarks and rigorous evaluation methodology. 

\noindent \textbf{\underline{Towards Reliable and Standardized Benchmarks.}}
It is important to establish a rigorous evaluation methodology supported by trustworthy and standardized benchmarks.
These benchmarks provide consistent frameworks for comparing different LM4Code approaches and set baselines for new techniques~\cite{beyer2019reliable}.
However, our review results reveal that multiple pitfalls related to performance evaluation like inappropriate baselines, limited test sets, and reproducibility issues distort understanding of actual LM4Code capabilities, questioning research correctness~\cite{liu2023towards, zeng_extensive_2022, liu_deep_2022}. 
For example, Laaber~\ea~\cite{laaber2021predicting} emphasize the inherent benchmark instability, stressing the importance of identifying and rectifying these instabilities to ensure accurate evaluations. 
Without addressing pitfalls, perceived model performance can become inflated, leading to potentially misleading conclusions. 
Related benchmarking challenges have been noted across various research domains like computer security~\cite{van2019sok}, and cloud computing~\cite{schwarzkopf2012seven, cooper2010benchmarking}.
Hence, there is an urgent need to prioritize the development of standardized, stable, and reliable benchmarks in LM4Code research. 
Future research should focus on creating benchmarks that are both comprehensive and representative of real-world scenarios.
To achieve this, the community should coordinate on curating representative test sets, establishing strong baselines, quantifying uncertainty, and promoting reproducible experiments, 
ensuring that advancements in the field of LM4Code are grounded in robust and reliable evaluations. 
This will not only foster trustworthiness within the research community but also drive meaningful progress in LM4Code.

\noindent \textbf{\underline{Towards Reliable Evaluation Metrics.}}
As language models like GPT-4 grow in complexity, accurately assessing their capabilities and limitations becomes imperative.
However, as discussed, traditional evaluation metrics, often borrowed from the NLP domain, may fail to capture the intricacies and details that are crucial for code generation and understanding tasks~\cite{scalabrino2017automatically, ren2020codebleu}.
For example, metrics like accuracy and BLEU focusing solely on syntactic correctness fail to account for potential semantic errors with practical consequences, as noted by Fan~\ea~\cite{fan2023automated}. 
In addition, most studies rely heavily on automated metrics, but human evaluation remains indispensable for assessing the quality and utility of generated code~\cite{dakhel2023github,paltenghi_thinking_2021}.
Additionally, the variability and uncertainty inherent in generative models like GPT-4 need to be measured and analyzed in order to understand reliability in real-world settings~\cite{openai2023gpt4}.
Therefore, developing holistic and standardized metrics tailored to code intelligence tasks is urgently needed. 
These should measure syntactic validity, semantic consistency, coherence, human quality judgments, and variability.
For LM4Code to move from impressing lab-only evaluation to transforming real-world applications, reliable evaluation metrics are crucial.


\subsubsection{Optimizing LM4Code Deployment for Real-World Scenarios}
Realizing LM4Code's potential requires addressing deployment challenges in transitioning from controlled research to practice. 
The integration, security, and scalability of LM4Code techniques become important considerations.

\noindent \textbf{\underline{Integrating LM4Code into Developer Workflows and Tools.}}
While modern LM4Code models like GitHub Copilot show potential for integrating AI-driven code suggestions into developers' workflows, realizing this in practice presents unique challenges.
First, it is crucial to ensure the correctness and human comprehensibility of suggestions, especially since the generated code can contain errors and vulnerabilities that may be problematic for real-world usage~\cite{nguyen2022empirical,peng2022security, pearce_asleep_2022}.
Additionally, customization and personalization are also important, since every developer has a unique coding style and preferences.
Modern LM4Code tools should be adaptable and learn from individual developer behaviors to provide personalized code suggestions.
Finally, integration with developer tools like version control and debugging is critical for comprehensive functionality. 
Future research should focus on addressing these challenges, ensuring that the integration of LM4Code models into developer workflows is smooth, efficient, and beneficial.

\noindent \textbf{\underline{Towards LM4Code Security.}}
As discussed in Section~\ref{sec:deployment}, LM4Code faces several security concerns that need to be addressed before responsible and ethical deployment can be achieved. 
These include risks of evasion attacks, data poisoning, privacy leakage, and generated code potentially containing vulnerabilities. 
Implementing comprehensive solutions is thus critical to safeguard models, data, and users.
Advancing the security of LM4Code requires continued research across multiple domains:
(1) Adversarial training techniques~\cite{zhou_adversarial_2022} can potentially increase model robustness against adversarial inputs; 
(2) Employing formal verification approaches to evaluate the security and correctness of generated code~\cite{pearce_asleep_2022}; 
(3) Enhanced data construction and quality assurance processes are needed to systematically identify and eliminate data poisoning, preventing the propagation of risks~\cite{schuster2021you};
(4) Approaches such as differential privacy~\cite{lukas2023analyzing} and federated learning~\cite{li2020federated} may strengthen privacy preservation in model training;
(5) Watermarking and provenance tracking mechanisms can enable authentication of model ownership and detect plagiarism or unauthorized use~\cite{kirchenbauer2023watermark, sun2022coprotector};
(6) Manual examination approaches including code reviews and human-AI collaborative interactions are important.
Substantial multidisciplinary efforts are critical to ensure the secure and ethical development and deployment of powerful generative AI systems like LM4Code. 
This remains an open research challenge requiring continued progress across communities.

\noindent \textbf{\underline{Scalability and Latency Concerns.}}
While powerful general large language models like GPT-4~\cite{openai2023gpt4}, PaLM~\cite{google2023palm}, and Claude~\cite{anthropic2023claude} offer public API access, privately deployed specialized LM4Code models can face greater scalability challenges.
Developing private models is important to customize LM4Code systems to specific domains and tasks. 
However, large language models often comprise billions of parameters, requiring massive computational resources~\cite{hou2023large}.
This poses challenges for real-time efficient integration into developer workflows. 
Reducing inference latency through methods like knowledge distillation~\cite{shi2022compressing} and efficient attention~\cite{kaddour2023challenges} is crucial for reasonable responsiveness.
Additionally, scaling up throughput for concurrent users via cloud-native architectures is also essential. 
Optimizing memory utilization via compression and caching helps in deploying large LM4Code models~\cite{shi2022compressing}. 
Energy-efficient deployment through quantization and lightweight model distillation is likewise critical~\cite{guo2019empirical}.
Tackling these scalability and latency issues will be critical to realize LM4Code's full potential through deployments that smoothly integrate LM4Code into practical developer environments in a sustainable and user-friendly manner.

\subsection{Threats to Validity}
\label{sec:threats}

This systematic literature review was conducted according to the established guidelines~\cite{kitchenham_guidelines_2007, zhang2011identifying} to mitigate potential threats to validity. 
However, there are still certain limitations primarily associated with our search strategy and the data extraction process used for constructing our paper taxonomy.

One primary threat is selection bias, wherein the selection process may miss some relevant research studies.
First, one possible cause may be that some search engines may provide some irrelevant studies or overlook some studies.
Another reason could be that our keywords don't cover all the relevant research studies.
With an emerging field like LM4Code, important ongoing work may not yet be indexed in the primary digital libraries we searched. 
In addition, as we stated before, pitfalls do not have consistent keywords or terminology, so our manual checking of the pitfalls of LM4Code papers also has the potential for omissions.
To minimize this risk, we systematically performed the paper searching across six major digital libraries in computer science, manually searched top venues, iteratively refined search strings based on a quasi-gold standard approach defined by Zhang~\ea~\cite{zhang2011identifying}, and conducted backward/forward snowballing.
Moreover, each manuscript that an individual author expressed uncertainty about regarding its including/excluding underwent thorough discussion between the first two authors before making a final decision.

Another threat is internal validity in constructing our taxonomy of LM4Code pitfalls.
A significant contribution of this paper is developing a taxonomy to categorize and synthesize key pitfalls across the LM4Code field. 
To mitigate subjectivity in our taxonomy, we adapted a framework from Arp~\ea~\cite{arp2022and} previously applied across computer security, which was collaboratively validated by the first six authors with LM4Code expertise.
At the same time, to enhance accuracy, each primary study classification was reviewed by at least three authors, with disagreements resolved through discussion.
Despite the fact that multiple evaluators reduce the possibility of bias, subjective factors persist.
To improve the integrity of our taxonomy and offer transparency into our workings, all of our collected research studies and their classification are available in our online repository.

\section{Conclusion}
\label{sec:conclusion}
In this research study, we conducted a comprehensive and rigorous systematic literature review to examine the pitfalls present in LM4Code.
We utilized a well-defined systematic literature review approach and finally obtained 67 relevant research studies from top-tier venues.
We first provided a taxonomy and we classified the existing pitfalls in LM4Code based on the various stages of the LM4Code lifecycle, including data collection and labeling, system design and learning, performance evaluation, and deployment and maintenance.
For each stage, we provided a thorough review of the relevant studies based on the pitfall types, implications, and existing solutions.
Finally, we described the current challenges and discussed the open opportunities that demand more study in this area.
We hope that our work will motivate other researchers, making language models enhanced for code intelligence more reliable and trustworthy, thereby ensuring their effective deployment into real-world applications.

\section*{Acknowledgements}
We sincerely thank Zhou Yang and Zhensu Sun, PhD students at Singapore Management University, for their invaluable insights, feedback, and assistance in clearly defining the taxonomy of pitfalls in language models for code. Their contributions greatly helped shape and improve this work.


%





\bibliographystyle{sample-format/ACM-Reference-Format}
\bibliography{main}


\begin{thebibliography}{187}


\ifx \showCODEN    \undefined \def \showCODEN     #1{\unskip}     \fi
\ifx \showDOI      \undefined \def \showDOI       #1{#1}\fi
\ifx \showISBNx    \undefined \def \showISBNx     #1{\unskip}     \fi
\ifx \showISBNxiii \undefined \def \showISBNxiii  #1{\unskip}     \fi
\ifx \showISSN     \undefined \def \showISSN      #1{\unskip}     \fi
\ifx \showLCCN     \undefined \def \showLCCN      #1{\unskip}     \fi
\ifx \shownote     \undefined \def \shownote      #1{#1}          \fi
\ifx \showarticletitle \undefined \def \showarticletitle #1{#1}   \fi
\ifx \showURL      \undefined \def \showURL       {\relax}        \fi
\providecommand\bibfield[2]{#2}
\providecommand\bibinfo[2]{#2}
\providecommand\natexlab[1]{#1}
\providecommand\showeprint[2][]{arXiv:#2}

\bibitem[\protect\citeauthoryear{huggingface.co}{cod}{[n.d.]}]%
        {codeparrot}
 \bibinfo{year}{[n.d.]}\natexlab{}.
\newblock \bibinfo{title}{codeparrot (CodeParrot)}.
\newblock
\newblock
\urldef\tempurl%
\url{https://huggingface.co/codeparrot}
\showURL{%
\tempurl}


\bibitem[\protect\citeauthoryear{Aghababaeyan, Abdellatif, Briand, S, and
  Bagherzadeh}{Aghababaeyan et~al\mbox{.}}{2023}]%
        {aghababaeyan_black-box_2023}
\bibfield{author}{\bibinfo{person}{Zohreh Aghababaeyan}, \bibinfo{person}{Manel
  Abdellatif}, \bibinfo{person}{Lionel Briand}, \bibinfo{person}{Ramesh S},
  {and} \bibinfo{person}{Mojtaba Bagherzadeh}.}
  \bibinfo{year}{2023}\natexlab{}.
\newblock \showarticletitle{Black-{Box} {Testing} of {Deep} {Neural} {Networks}
  through {Test} {Case} {Diversity}}.
\newblock \bibinfo{journal}{\emph{IIEEE Trans. Software Eng.}}
  \bibinfo{volume}{49}, \bibinfo{number}{5} (\bibinfo{date}{May}
  \bibinfo{year}{2023}), \bibinfo{pages}{3182--3204}.
\newblock
\showISSN{0098-5589, 1939-3520, 2326-3881}
\urldef\tempurl%
\url{https://doi.org/10.1109/TSE.2023.3243522}
\showDOI{\tempurl}


\bibitem[\protect\citeauthoryear{Ahmed, Ledesma, and Devanbu}{Ahmed
  et~al\mbox{.}}{2023}]%
        {ahmed_synshine_2023}
\bibfield{author}{\bibinfo{person}{Toufique Ahmed}, \bibinfo{person}{Noah~Rose
  Ledesma}, {and} \bibinfo{person}{Premkumar Devanbu}.}
  \bibinfo{year}{2023}\natexlab{}.
\newblock \showarticletitle{{SynShine}: {Improved} {Fixing} of {Syntax}
  {Errors}}.
\newblock \bibinfo{journal}{\emph{IIEEE Trans. Software Eng.}}
  \bibinfo{volume}{49}, \bibinfo{number}{4} (\bibinfo{date}{April}
  \bibinfo{year}{2023}), \bibinfo{pages}{2169--2181}.
\newblock
\showISSN{0098-5589, 1939-3520, 2326-3881}
\urldef\tempurl%
\url{https://doi.org/10.1109/TSE.2022.3212635}
\showDOI{\tempurl}


\bibitem[\protect\citeauthoryear{Amazon}{Amazon}{2023}]%
        {amazon2022codewhisperer}
\bibfield{author}{\bibinfo{person}{Amazon}.} \bibinfo{year}{2023}\natexlab{}.
\newblock \bibinfo{title}{Amazon CodeWhisperer}.
\newblock \bibinfo{howpublished}{\url{ https://aws.amazon.com/codewhisperer/}}.
\newblock


\bibitem[\protect\citeauthoryear{Anthropic}{Anthropic}{2023}]%
        {anthropic2023claude}
\bibfield{author}{\bibinfo{person}{Anthropic}.}
  \bibinfo{year}{2023}\natexlab{}.
\newblock \bibinfo{title}{Getting started with Claude}.
\newblock \bibinfo{howpublished}{\url{
  https://docs.anthropic.com/claude/docs}}.
\newblock


\bibitem[\protect\citeauthoryear{Arp, Quiring, Pendlebury, Warnecke, Pierazzi,
  Wressnegger, Cavallaro, and Rieck}{Arp et~al\mbox{.}}{2022}]%
        {arp2022and}
\bibfield{author}{\bibinfo{person}{Daniel Arp}, \bibinfo{person}{Erwin
  Quiring}, \bibinfo{person}{Feargus Pendlebury}, \bibinfo{person}{Alexander
  Warnecke}, \bibinfo{person}{Fabio Pierazzi}, \bibinfo{person}{Christian
  Wressnegger}, \bibinfo{person}{Lorenzo Cavallaro}, {and}
  \bibinfo{person}{Konrad Rieck}.} \bibinfo{year}{2022}\natexlab{}.
\newblock \showarticletitle{Dos and don'ts of machine learning in computer
  security}. In \bibinfo{booktitle}{\emph{31st USENIX Security Symposium
  (USENIX Security 22)}}. \bibinfo{pages}{3971--3988}.
\newblock


\bibitem[\protect\citeauthoryear{Austin, Odena, Nye, Bosma, Michalewski, Dohan,
  Jiang, Cai, Terry, Le, et~al\mbox{.}}{Austin et~al\mbox{.}}{2021}]%
        {Austin2021ProgramSW}
\bibfield{author}{\bibinfo{person}{Jacob Austin}, \bibinfo{person}{Augustus
  Odena}, \bibinfo{person}{Maxwell Nye}, \bibinfo{person}{Maarten Bosma},
  \bibinfo{person}{Henryk Michalewski}, \bibinfo{person}{David Dohan},
  \bibinfo{person}{Ellen Jiang}, \bibinfo{person}{Carrie Cai},
  \bibinfo{person}{Michael Terry}, \bibinfo{person}{Quoc Le}, {et~al\mbox{.}}}
  \bibinfo{year}{2021}\natexlab{}.
\newblock \showarticletitle{Program synthesis with large language models}.
\newblock \bibinfo{journal}{\emph{arXiv preprint arXiv:2108.07732}}
  (\bibinfo{year}{2021}).
\newblock


\bibitem[\protect\citeauthoryear{Aye, Kim, and Li}{Aye et~al\mbox{.}}{2021a}]%
        {Aye2020LearningAF}
\bibfield{author}{\bibinfo{person}{Gareth~Ari Aye}, \bibinfo{person}{Seohyun
  Kim}, {and} \bibinfo{person}{Hongyu Li}.} \bibinfo{year}{2021}\natexlab{a}.
\newblock \showarticletitle{Learning autocompletion from real-world datasets}.
\newblock  (\bibinfo{year}{2021}), \bibinfo{pages}{131--139}.
\newblock


\bibitem[\protect\citeauthoryear{Aye, Kim, and Li}{Aye et~al\mbox{.}}{2021b}]%
        {aye2021learning}
\bibfield{author}{\bibinfo{person}{Gareth~Ari Aye}, \bibinfo{person}{Seohyun
  Kim}, {and} \bibinfo{person}{Hongyu Li}.} \bibinfo{year}{2021}\natexlab{b}.
\newblock \showarticletitle{Learning autocompletion from real-world datasets}.
  In \bibinfo{booktitle}{\emph{2021 IEEE/ACM 43rd International Conference on
  Software Engineering: Software Engineering in Practice (ICSE-SEIP)}}. IEEE,
  \bibinfo{pages}{131--139}.
\newblock


\bibitem[\protect\citeauthoryear{Beyer, L{\"o}we, and Wendler}{Beyer
  et~al\mbox{.}}{2019}]%
        {beyer2019reliable}
\bibfield{author}{\bibinfo{person}{Dirk Beyer}, \bibinfo{person}{Stefan
  L{\"o}we}, {and} \bibinfo{person}{Philipp Wendler}.}
  \bibinfo{year}{2019}\natexlab{}.
\newblock \showarticletitle{Reliable benchmarking: requirements and solutions}.
\newblock \bibinfo{journal}{\emph{International Journal on Software Tools for
  Technology Transfer}}  \bibinfo{volume}{21} (\bibinfo{year}{2019}),
  \bibinfo{pages}{1--29}.
\newblock


\bibitem[\protect\citeauthoryear{Biggio and Roli}{Biggio and Roli}{2018}]%
        {biggio2018wild}
\bibfield{author}{\bibinfo{person}{Battista Biggio} {and}
  \bibinfo{person}{Fabio Roli}.} \bibinfo{year}{2018}\natexlab{}.
\newblock \showarticletitle{Wild patterns: Ten years after the rise of
  adversarial machine learning}. In \bibinfo{booktitle}{\emph{Proceedings of
  the 2018 ACM SIGSAC Conference on Computer and Communications Security}}.
  \bibinfo{pages}{2154--2156}.
\newblock


\bibitem[\protect\citeauthoryear{Black}{Black}{2017}]%
        {black2017sard}
\bibfield{author}{\bibinfo{person}{Paul~E Black}.}
  \bibinfo{year}{2017}\natexlab{}.
\newblock \showarticletitle{Sard: A software assurance reference dataset}.
\newblock  (\bibinfo{year}{2017}).
\newblock


\bibitem[\protect\citeauthoryear{Carlini, Tramer, Wallace, Jagielski,
  Herbert-Voss, Lee, Roberts, Brown, Song, Erlingsson, et~al\mbox{.}}{Carlini
  et~al\mbox{.}}{2021}]%
        {carlini2021extracting}
\bibfield{author}{\bibinfo{person}{Nicholas Carlini}, \bibinfo{person}{Florian
  Tramer}, \bibinfo{person}{Eric Wallace}, \bibinfo{person}{Matthew Jagielski},
  \bibinfo{person}{Ariel Herbert-Voss}, \bibinfo{person}{Katherine Lee},
  \bibinfo{person}{Adam Roberts}, \bibinfo{person}{Tom Brown},
  \bibinfo{person}{Dawn Song}, \bibinfo{person}{Ulfar Erlingsson},
  {et~al\mbox{.}}} \bibinfo{year}{2021}\natexlab{}.
\newblock \showarticletitle{Extracting training data from large language
  models}. In \bibinfo{booktitle}{\emph{30th USENIX Security Symposium (USENIX
  Security 21)}}. \bibinfo{pages}{2633--2650}.
\newblock


\bibitem[\protect\citeauthoryear{Chakraborty, Krishna, Ding, and
  Ray}{Chakraborty et~al\mbox{.}}{2021}]%
        {chakraborty2021deep}
\bibfield{author}{\bibinfo{person}{Saikat Chakraborty}, \bibinfo{person}{Rahul
  Krishna}, \bibinfo{person}{Yangruibo Ding}, {and} \bibinfo{person}{Baishakhi
  Ray}.} \bibinfo{year}{2021}\natexlab{}.
\newblock \showarticletitle{Deep learning based vulnerability detection: Are we
  there yet}.
\newblock \bibinfo{journal}{\emph{IEEE Transactions on Software Engineering}}
  (\bibinfo{year}{2021}).
\newblock


\bibitem[\protect\citeauthoryear{Chen, Scheurer, Korbak, Campos, Chan, Bowman,
  Cho, and Perez}{Chen et~al\mbox{.}}{2023}]%
        {chen2023improving}
\bibfield{author}{\bibinfo{person}{Angelica Chen},
  \bibinfo{person}{J{\'e}r{\'e}my Scheurer}, \bibinfo{person}{Tomasz Korbak},
  \bibinfo{person}{Jon~Ander Campos}, \bibinfo{person}{Jun~Shern Chan},
  \bibinfo{person}{Samuel~R Bowman}, \bibinfo{person}{Kyunghyun Cho}, {and}
  \bibinfo{person}{Ethan Perez}.} \bibinfo{year}{2023}\natexlab{}.
\newblock \showarticletitle{Improving code generation by training with natural
  language feedback}.
\newblock \bibinfo{journal}{\emph{arXiv preprint arXiv:2303.16749}}
  (\bibinfo{year}{2023}).
\newblock


\bibitem[\protect\citeauthoryear{Chen, Wen, Shi, Lin, Rajbahadur, and
  Jiang}{Chen et~al\mbox{.}}{2022}]%
        {chen2022towards}
\bibfield{author}{\bibinfo{person}{Boyuan Chen}, \bibinfo{person}{Mingzhi Wen},
  \bibinfo{person}{Yong Shi}, \bibinfo{person}{Dayi Lin},
  \bibinfo{person}{Gopi~Krishnan Rajbahadur}, {and} \bibinfo{person}{Zhen~Ming
  Jiang}.} \bibinfo{year}{2022}\natexlab{}.
\newblock \showarticletitle{Towards training reproducible deep learning
  models}. In \bibinfo{booktitle}{\emph{Proceedings of the 44th International
  Conference on Software Engineering}}. \bibinfo{pages}{2202--2214}.
\newblock


\bibitem[\protect\citeauthoryear{Chen, Tworek, Jun, Yuan, Pinto, Kaplan,
  Edwards, Burda, Joseph, Brockman, et~al\mbox{.}}{Chen et~al\mbox{.}}{2021}]%
        {chen2021evaluating}
\bibfield{author}{\bibinfo{person}{Mark Chen}, \bibinfo{person}{Jerry Tworek},
  \bibinfo{person}{Heewoo Jun}, \bibinfo{person}{Qiming Yuan},
  \bibinfo{person}{Henrique Ponde de~Oliveira Pinto}, \bibinfo{person}{Jared
  Kaplan}, \bibinfo{person}{Harri Edwards}, \bibinfo{person}{Yuri Burda},
  \bibinfo{person}{Nicholas Joseph}, \bibinfo{person}{Greg Brockman},
  {et~al\mbox{.}}} \bibinfo{year}{2021}\natexlab{}.
\newblock \showarticletitle{Evaluating large language models trained on code}.
\newblock \bibinfo{journal}{\emph{arXiv preprint arXiv:2107.03374}}
  (\bibinfo{year}{2021}).
\newblock


\bibitem[\protect\citeauthoryear{Chirkova and Troshin}{Chirkova and
  Troshin}{2021}]%
        {chirkova2021empirical}
\bibfield{author}{\bibinfo{person}{Nadezhda Chirkova} {and}
  \bibinfo{person}{Sergey Troshin}.} \bibinfo{year}{2021}\natexlab{}.
\newblock \showarticletitle{Empirical study of transformers for source code}.
  In \bibinfo{booktitle}{\emph{Proceedings of the 29th ACM joint meeting on
  European software engineering conference and symposium on the foundations of
  software engineering}}. \bibinfo{pages}{703--715}.
\newblock


\bibitem[\protect\citeauthoryear{Chirkova and Troshin}{Chirkova and
  Troshin}{2023}]%
        {chirkova2023codebpe}
\bibfield{author}{\bibinfo{person}{Nadezhda Chirkova} {and}
  \bibinfo{person}{Sergey Troshin}.} \bibinfo{year}{2023}\natexlab{}.
\newblock \showarticletitle{CodeBPE: Investigating Subtokenization Options for
  Large Language Model Pretraining on Source Code}.
\newblock \bibinfo{journal}{\emph{arXiv preprint arXiv:2308.00683}}
  (\bibinfo{year}{2023}).
\newblock


\bibitem[\protect\citeauthoryear{Cito, Chandra, Tantithamthavorn, and
  Hemmati}{Cito et~al\mbox{.}}{2023}]%
        {10109328}
\bibfield{author}{\bibinfo{person}{Jürgen Cito}, \bibinfo{person}{Satish
  Chandra}, \bibinfo{person}{Chakkrit Tantithamthavorn}, {and}
  \bibinfo{person}{Hadi Hemmati}.} \bibinfo{year}{2023}\natexlab{}.
\newblock \showarticletitle{Expert Perspectives on Explainability}.
\newblock \bibinfo{journal}{\emph{IEEE Software}} \bibinfo{volume}{40},
  \bibinfo{number}{3} (\bibinfo{year}{2023}), \bibinfo{pages}{84--88}.
\newblock
\urldef\tempurl%
\url{https://doi.org/10.1109/MS.2023.3255663}
\showDOI{\tempurl}


\bibitem[\protect\citeauthoryear{Cito, Dillig, Kim, Murali, and Chandra}{Cito
  et~al\mbox{.}}{2021}]%
        {cito_explaining_2021}
\bibfield{author}{\bibinfo{person}{J{\"u}rgen Cito}, \bibinfo{person}{Isil
  Dillig}, \bibinfo{person}{Seohyun Kim}, \bibinfo{person}{Vijayaraghavan
  Murali}, {and} \bibinfo{person}{Satish Chandra}.}
  \bibinfo{year}{2021}\natexlab{}.
\newblock \showarticletitle{Explaining mispredictions of machine learning
  models using rule induction}. In \bibinfo{booktitle}{\emph{Proceedings of the
  29th {ACM} {Joint} {Meeting} on {European} {Software} {Engineering}
  {Conference} and {Symposium} on the {Foundations} of {Software}
  {Engineering}}} \emph{(\bibinfo{series}{{ESEC}/{FSE} 2021})}.
  \bibinfo{publisher}{Association for Computing Machinery},
  \bibinfo{address}{New York, NY, USA}, \bibinfo{pages}{716--727}.
\newblock
\showISBNx{978-1-4503-8562-6}
\urldef\tempurl%
\url{https://doi.org/10.1145/3468264.3468614}
\showDOI{\tempurl}


\bibitem[\protect\citeauthoryear{Cooper, Silberstein, Tam, Ramakrishnan, and
  Sears}{Cooper et~al\mbox{.}}{2010}]%
        {cooper2010benchmarking}
\bibfield{author}{\bibinfo{person}{Brian~F Cooper}, \bibinfo{person}{Adam
  Silberstein}, \bibinfo{person}{Erwin Tam}, \bibinfo{person}{Raghu
  Ramakrishnan}, {and} \bibinfo{person}{Russell Sears}.}
  \bibinfo{year}{2010}\natexlab{}.
\newblock \showarticletitle{Benchmarking cloud serving systems with YCSB}. In
  \bibinfo{booktitle}{\emph{Proceedings of the 1st ACM symposium on Cloud
  computing}}. \bibinfo{pages}{143--154}.
\newblock


\bibitem[\protect\citeauthoryear{Croft, Xie, and Babar}{Croft
  et~al\mbox{.}}{2022}]%
        {croft2022data}
\bibfield{author}{\bibinfo{person}{Roland Croft}, \bibinfo{person}{Yongzheng
  Xie}, {and} \bibinfo{person}{Muhammad~Ali Babar}.}
  \bibinfo{year}{2022}\natexlab{}.
\newblock \showarticletitle{Data preparation for software vulnerability
  prediction: A systematic literature review}.
\newblock \bibinfo{journal}{\emph{IEEE Transactions on Software Engineering}}
  \bibinfo{volume}{49}, \bibinfo{number}{3} (\bibinfo{year}{2022}),
  \bibinfo{pages}{1044--1063}.
\newblock


\bibitem[\protect\citeauthoryear{Dakhel, Majdinasab, Nikanjam, Khomh,
  Desmarais, and Jiang}{Dakhel et~al\mbox{.}}{2023}]%
        {dakhel2023github}
\bibfield{author}{\bibinfo{person}{Arghavan~Moradi Dakhel},
  \bibinfo{person}{Vahid Majdinasab}, \bibinfo{person}{Amin Nikanjam},
  \bibinfo{person}{Foutse Khomh}, \bibinfo{person}{Michel~C Desmarais}, {and}
  \bibinfo{person}{Zhen Ming~Jack Jiang}.} \bibinfo{year}{2023}\natexlab{}.
\newblock \showarticletitle{Github copilot ai pair programmer: Asset or
  liability?}
\newblock \bibinfo{journal}{\emph{Journal of Systems and Software}}
  \bibinfo{volume}{203} (\bibinfo{year}{2023}), \bibinfo{pages}{111734}.
\newblock


\bibitem[\protect\citeauthoryear{Ding, Ray, Devanbu, and Hellendoorn}{Ding
  et~al\mbox{.}}{2021}]%
        {ding_patching_2021}
\bibfield{author}{\bibinfo{person}{Yangruibo Ding}, \bibinfo{person}{Baishakhi
  Ray}, \bibinfo{person}{Premkumar Devanbu}, {and} \bibinfo{person}{Vincent~J.
  Hellendoorn}.} \bibinfo{year}{2021}\natexlab{}.
\newblock \showarticletitle{Patching as translation: the data and the
  metaphor}. In \bibinfo{booktitle}{\emph{Proceedings of the 35th {IEEE}/{ACM}
  {International} {Conference} on {Automated} {Software} {Engineering}}}
  \emph{(\bibinfo{series}{{ASE} '20})}. \bibinfo{publisher}{Association for
  Computing Machinery}, \bibinfo{address}{Virtual Event Australia},
  \bibinfo{pages}{275--286}.
\newblock
\showISBNx{978-1-4503-6768-4}
\urldef\tempurl%
\url{https://doi.org/10.1145/3324884.3416587}
\showDOI{\tempurl}


\bibitem[\protect\citeauthoryear{Dong, Lou, Hao, and Tan}{Dong
  et~al\mbox{.}}{2023}]%
        {dong2023revisiting}
\bibfield{author}{\bibinfo{person}{Jinhao Dong}, \bibinfo{person}{Yiling Lou},
  \bibinfo{person}{Dan Hao}, {and} \bibinfo{person}{Lin Tan}.}
  \bibinfo{year}{2023}\natexlab{}.
\newblock \showarticletitle{Revisiting Learning-based Commit Message
  Generation}. In \bibinfo{booktitle}{\emph{2023 IEEE/ACM 45th International
  Conference on Software Engineering (ICSE)}}. IEEE, \bibinfo{pages}{794--805}.
\newblock


\bibitem[\protect\citeauthoryear{Durelli, Durelli, Borges, Endo, Eler, Dias,
  and Guimar{\~a}es}{Durelli et~al\mbox{.}}{2019}]%
        {durelli2019machine}
\bibfield{author}{\bibinfo{person}{Vinicius~HS Durelli},
  \bibinfo{person}{Rafael~S Durelli}, \bibinfo{person}{Simone~S Borges},
  \bibinfo{person}{Andre~T Endo}, \bibinfo{person}{Marcelo~M Eler},
  \bibinfo{person}{Diego~RC Dias}, {and} \bibinfo{person}{Marcelo~P
  Guimar{\~a}es}.} \bibinfo{year}{2019}\natexlab{}.
\newblock \showarticletitle{Machine learning applied to software testing: A
  systematic mapping study}.
\newblock \bibinfo{journal}{\emph{IEEE Transactions on Reliability}}
  \bibinfo{volume}{68}, \bibinfo{number}{3} (\bibinfo{year}{2019}),
  \bibinfo{pages}{1189--1212}.
\newblock


\bibitem[\protect\citeauthoryear{Eghbali and Pradel}{Eghbali and
  Pradel}{2022}]%
        {eghbali2022crystalbleu}
\bibfield{author}{\bibinfo{person}{Aryaz Eghbali} {and}
  \bibinfo{person}{Michael Pradel}.} \bibinfo{year}{2022}\natexlab{}.
\newblock \showarticletitle{CrystalBLEU: precisely and efficiently measuring
  the similarity of code}. In \bibinfo{booktitle}{\emph{Proceedings of the 37th
  IEEE/ACM International Conference on Automated Software Engineering}}.
  \bibinfo{pages}{1--12}.
\newblock


\bibitem[\protect\citeauthoryear{Fan, Li, Wang, and Nguyen}{Fan
  et~al\mbox{.}}{2020}]%
        {fan2020ac}
\bibfield{author}{\bibinfo{person}{Jiahao Fan}, \bibinfo{person}{Yi Li},
  \bibinfo{person}{Shaohua Wang}, {and} \bibinfo{person}{Tien~N Nguyen}.}
  \bibinfo{year}{2020}\natexlab{}.
\newblock \showarticletitle{AC/C++ code vulnerability dataset with code changes
  and CVE summaries}. In \bibinfo{booktitle}{\emph{Proceedings of the 17th
  International Conference on Mining Software Repositories}}.
  \bibinfo{pages}{508--512}.
\newblock


\bibitem[\protect\citeauthoryear{Fan, Gao, Mirchev, Roychoudhury, and Tan}{Fan
  et~al\mbox{.}}{2023}]%
        {fan2023automated}
\bibfield{author}{\bibinfo{person}{Zhiyu Fan}, \bibinfo{person}{Xiang Gao},
  \bibinfo{person}{Martin Mirchev}, \bibinfo{person}{Abhik Roychoudhury}, {and}
  \bibinfo{person}{Shin~Hwei Tan}.} \bibinfo{year}{2023}\natexlab{}.
\newblock \showarticletitle{Automated repair of programs from large language
  models}. In \bibinfo{booktitle}{\emph{2023 IEEE/ACM 45th International
  Conference on Software Engineering (ICSE)}}. IEEE,
  \bibinfo{pages}{1469--1481}.
\newblock


\bibitem[\protect\citeauthoryear{Fang, Zhang, Tan, Jiang, Xia, and Sun}{Fang
  et~al\mbox{.}}{2023}]%
        {fang2023representthemall}
\bibfield{author}{\bibinfo{person}{Sen Fang}, \bibinfo{person}{Tao Zhang},
  \bibinfo{person}{Youshuai Tan}, \bibinfo{person}{He Jiang},
  \bibinfo{person}{Xin Xia}, {and} \bibinfo{person}{Xiaobing Sun}.}
  \bibinfo{year}{2023}\natexlab{}.
\newblock \showarticletitle{RepresentThemAll: A Universal Learning
  Representation of Bug Reports}. In \bibinfo{booktitle}{\emph{2023 IEEE/ACM
  45th International Conference on Software Engineering (ICSE)}}. IEEE,
  \bibinfo{pages}{602--614}.
\newblock


\bibitem[\protect\citeauthoryear{Federation}{Federation}{2023}]%
        {ccf2023ranking}
\bibfield{author}{\bibinfo{person}{China~Computer Federation}.}
  \bibinfo{year}{2023}\natexlab{}.
\newblock \bibinfo{title}{CCF Recommended List of International Conferences and
  Periodicals}.
\newblock \bibinfo{howpublished}{\url{
  https://www.ccf.org.cn/en/Bulletin/2019-05-13/663884.shtml}}.
\newblock


\bibitem[\protect\citeauthoryear{Feng, Chen, Xie, Meng, Lin, and Liu}{Feng
  et~al\mbox{.}}{2021}]%
        {feng_performance-sensitive_2021}
\bibfield{author}{\bibinfo{person}{Ruitao Feng}, \bibinfo{person}{Sen Chen},
  \bibinfo{person}{Xiaofei Xie}, \bibinfo{person}{Guozhu Meng},
  \bibinfo{person}{Shang-Wei Lin}, {and} \bibinfo{person}{Yang Liu}.}
  \bibinfo{year}{2021}\natexlab{}.
\newblock \showarticletitle{A {Performance}-{Sensitive} {Malware} {Detection}
  {System} {Using} {Deep} {Learning} on {Mobile} {Devices}}.
\newblock \bibinfo{journal}{\emph{IEEE Trans.Inform.Forensic Secur.}}
  \bibinfo{volume}{16} (\bibinfo{year}{2021}), \bibinfo{pages}{1563--1578}.
\newblock
\showISSN{1556-6013, 1556-6021}
\urldef\tempurl%
\url{https://doi.org/10.1109/TIFS.2020.3025436}
\showDOI{\tempurl}


\bibitem[\protect\citeauthoryear{Feng, Guo, Tang, Duan, Feng, Gong, Shou, Qin,
  Liu, Jiang, et~al\mbox{.}}{Feng et~al\mbox{.}}{2020}]%
        {feng2020codebert}
\bibfield{author}{\bibinfo{person}{Zhangyin Feng}, \bibinfo{person}{Daya Guo},
  \bibinfo{person}{Duyu Tang}, \bibinfo{person}{Nan Duan},
  \bibinfo{person}{Xiaocheng Feng}, \bibinfo{person}{Ming Gong},
  \bibinfo{person}{Linjun Shou}, \bibinfo{person}{Bing Qin},
  \bibinfo{person}{Ting Liu}, \bibinfo{person}{Daxin Jiang}, {et~al\mbox{.}}}
  \bibinfo{year}{2020}\natexlab{}.
\newblock \showarticletitle{Codebert: A pre-trained model for programming and
  natural languages}.
\newblock \bibinfo{journal}{\emph{arXiv preprint arXiv:2002.08155}}
  (\bibinfo{year}{2020}).
\newblock


\bibitem[\protect\citeauthoryear{Fu, Tantithamthavorn, Le, Nguyen, and
  Phung}{Fu et~al\mbox{.}}{2022}]%
        {fu2022VulRepair}
\bibfield{author}{\bibinfo{person}{Michael Fu}, \bibinfo{person}{Chakkrit
  Tantithamthavorn}, \bibinfo{person}{Trung Le}, \bibinfo{person}{Van Nguyen},
  {and} \bibinfo{person}{Dinh Phung}.} \bibinfo{year}{2022}\natexlab{}.
\newblock \showarticletitle{VulRepair: A T5-Based Automated Software
  Vulnerability Repair} \emph{(\bibinfo{series}{ESEC/FSE 2022})}.
  \bibinfo{publisher}{Association for Computing Machinery},
  \bibinfo{address}{New York, NY, USA}, \bibinfo{pages}{935–947}.
\newblock
\showISBNx{9781450394130}
\urldef\tempurl%
\url{https://doi.org/10.1145/3540250.3549098}
\showDOI{\tempurl}


\bibitem[\protect\citeauthoryear{Fu, Van~Nguyen, Le, and Phung}{Fu
  et~al\mbox{.}}{2023}]%
        {fu2023vulexplainer}
\bibfield{author}{\bibinfo{person}{Michael Fu},
  \bibinfo{person}{Chakkrit~Tantithamthavorn Van~Nguyen},
  \bibinfo{person}{Trung Le}, {and} \bibinfo{person}{Dinh Phung}.}
  \bibinfo{year}{2023}\natexlab{}.
\newblock \showarticletitle{VulExplainer: A Transformer-based Hierarchical
  Distillation for Explaining Vulnerability Types}.
\newblock \bibinfo{journal}{\emph{IEEE Transactions on Software Engineering
  (TSE)}} (\bibinfo{year}{2023}).
\newblock


\bibitem[\protect\citeauthoryear{Ganesh, Chen, Lou, Khan, Yang, Sajjad, Nakov,
  Chen, and Winslett}{Ganesh et~al\mbox{.}}{2021}]%
        {ganesh2021compressing}
\bibfield{author}{\bibinfo{person}{Prakhar Ganesh}, \bibinfo{person}{Yao Chen},
  \bibinfo{person}{Xin Lou}, \bibinfo{person}{Mohammad~Ali Khan},
  \bibinfo{person}{Yin Yang}, \bibinfo{person}{Hassan Sajjad},
  \bibinfo{person}{Preslav Nakov}, \bibinfo{person}{Deming Chen}, {and}
  \bibinfo{person}{Marianne Winslett}.} \bibinfo{year}{2021}\natexlab{}.
\newblock \showarticletitle{Compressing large-scale transformer-based models: A
  case study on bert}.
\newblock \bibinfo{journal}{\emph{Transactions of the Association for
  Computational Linguistics}}  \bibinfo{volume}{9} (\bibinfo{year}{2021}),
  \bibinfo{pages}{1061--1080}.
\newblock


\bibitem[\protect\citeauthoryear{Gao, Gao, Wang, Sun, Lo, and Yu}{Gao
  et~al\mbox{.}}{2023a}]%
        {gao2023two}
\bibfield{author}{\bibinfo{person}{Shuzheng Gao}, \bibinfo{person}{Cuiyun Gao},
  \bibinfo{person}{Chaozheng Wang}, \bibinfo{person}{Jun Sun},
  \bibinfo{person}{David Lo}, {and} \bibinfo{person}{Yue Yu}.}
  \bibinfo{year}{2023}\natexlab{a}.
\newblock \showarticletitle{Two sides of the same coin: Exploiting the impact
  of identifiers in neural code comprehension}. In
  \bibinfo{booktitle}{\emph{2023 IEEE/ACM 45th International Conference on
  Software Engineering (ICSE)}}. IEEE, \bibinfo{pages}{1933--1945}.
\newblock


\bibitem[\protect\citeauthoryear{Gao, Zhang, Gao, and Wang}{Gao
  et~al\mbox{.}}{2023b}]%
        {gao2023keeping}
\bibfield{author}{\bibinfo{person}{Shuzheng Gao}, \bibinfo{person}{Hongyu
  Zhang}, \bibinfo{person}{Cuiyun Gao}, {and} \bibinfo{person}{Chaozheng
  Wang}.} \bibinfo{year}{2023}\natexlab{b}.
\newblock \showarticletitle{Keeping Pace with Ever-Increasing Data: Towards
  Continual Learning of Code Intelligence Models}.
\newblock \bibinfo{journal}{\emph{arXiv preprint arXiv:2302.03482}}
  (\bibinfo{year}{2023}).
\newblock


\bibitem[\protect\citeauthoryear{GitHub}{GitHub}{2023}]%
        {github2021copilot}
\bibfield{author}{\bibinfo{person}{GitHub}.} \bibinfo{year}{2023}\natexlab{}.
\newblock \bibinfo{title}{Github copilot}.
\newblock \bibinfo{howpublished}{\url{ https://copilot.github.com}}.
\newblock


\bibitem[\protect\citeauthoryear{Google}{Google}{2023}]%
        {google2023palm}
\bibfield{author}{\bibinfo{person}{Google}.} \bibinfo{year}{2023}\natexlab{}.
\newblock \bibinfo{title}{Build generative AI applications with Google}.
\newblock \bibinfo{howpublished}{\url{
  https://developers.generativeai.google/}}.
\newblock


\bibitem[\protect\citeauthoryear{Guerra-Manzanares and Bahsi}{Guerra-Manzanares
  and Bahsi}{2022}]%
        {guerra2022relativity}
\bibfield{author}{\bibinfo{person}{Alejandro Guerra-Manzanares} {and}
  \bibinfo{person}{Hayretdin Bahsi}.} \bibinfo{year}{2022}\natexlab{}.
\newblock \showarticletitle{On the relativity of time: Implications and
  challenges of data drift on long-term effective android malware detection}.
\newblock \bibinfo{journal}{\emph{Computers \& Security}}
  \bibinfo{volume}{122} (\bibinfo{year}{2022}), \bibinfo{pages}{102835}.
\newblock


\bibitem[\protect\citeauthoryear{Guo, Chen, Xie, Ma, Hu, Liu, Liu, Zhao, and
  Li}{Guo et~al\mbox{.}}{2019}]%
        {guo2019empirical}
\bibfield{author}{\bibinfo{person}{Qianyu Guo}, \bibinfo{person}{Sen Chen},
  \bibinfo{person}{Xiaofei Xie}, \bibinfo{person}{Lei Ma},
  \bibinfo{person}{Qiang Hu}, \bibinfo{person}{Hongtao Liu},
  \bibinfo{person}{Yang Liu}, \bibinfo{person}{Jianjun Zhao}, {and}
  \bibinfo{person}{Xiaohong Li}.} \bibinfo{year}{2019}\natexlab{}.
\newblock \showarticletitle{An empirical study towards characterizing deep
  learning development and deployment across different frameworks and
  platforms}. In \bibinfo{booktitle}{\emph{2019 34th IEEE/ACM International
  Conference on Automated Software Engineering (ASE)}}. IEEE,
  \bibinfo{pages}{810--822}.
\newblock


\bibitem[\protect\citeauthoryear{Gupta and Agrawal}{Gupta and Agrawal}{2022}]%
        {gupta2022compression}
\bibfield{author}{\bibinfo{person}{Manish Gupta} {and} \bibinfo{person}{Puneet
  Agrawal}.} \bibinfo{year}{2022}\natexlab{}.
\newblock \showarticletitle{Compression of deep learning models for text: A
  survey}.
\newblock \bibinfo{journal}{\emph{ACM Transactions on Knowledge Discovery from
  Data (TKDD)}} \bibinfo{volume}{16}, \bibinfo{number}{4}
  (\bibinfo{year}{2022}), \bibinfo{pages}{1--55}.
\newblock


\bibitem[\protect\citeauthoryear{He, Liu, Wu, Yang, Ren, and Qin}{He
  et~al\mbox{.}}{2022}]%
        {he2022msdroid}
\bibfield{author}{\bibinfo{person}{Yiling He}, \bibinfo{person}{Yiping Liu},
  \bibinfo{person}{Lei Wu}, \bibinfo{person}{Ziqi Yang}, \bibinfo{person}{Kui
  Ren}, {and} \bibinfo{person}{Zhan Qin}.} \bibinfo{year}{2022}\natexlab{}.
\newblock \showarticletitle{MsDroid: Identifying malicious snippets for android
  malware detection}.
\newblock \bibinfo{journal}{\emph{IEEE Transactions on Dependable and Secure
  Computing}} (\bibinfo{year}{2022}).
\newblock


\bibitem[\protect\citeauthoryear{He, Lou, Qin, and Ren}{He
  et~al\mbox{.}}{2023}]%
        {he2023finer}
\bibfield{author}{\bibinfo{person}{Yiling He}, \bibinfo{person}{Jian Lou},
  \bibinfo{person}{Zhan Qin}, {and} \bibinfo{person}{Kui Ren}.}
  \bibinfo{year}{2023}\natexlab{}.
\newblock \showarticletitle{FINER: Enhancing State-of-the-art Classifiers with
  Feature Attribution to Facilitate Security Analysis}.
\newblock \bibinfo{journal}{\emph{arXiv preprint arXiv:2308.05362}}
  (\bibinfo{year}{2023}).
\newblock


\bibitem[\protect\citeauthoryear{Hellendoorn, Proksch, Gall, and
  Bacchelli}{Hellendoorn et~al\mbox{.}}{2019}]%
        {hellendoorn_when_2019}
\bibfield{author}{\bibinfo{person}{Vincent~J. Hellendoorn},
  \bibinfo{person}{Sebastian Proksch}, \bibinfo{person}{Harald~C. Gall}, {and}
  \bibinfo{person}{Alberto Bacchelli}.} \bibinfo{year}{2019}\natexlab{}.
\newblock \showarticletitle{When {Code} {Completion} {Fails}: {A} {Case}
  {Study} on {Real}-{World} {Completions}}. In \bibinfo{booktitle}{\emph{2019
  {IEEE}/{ACM} 41st {International} {Conference} on {Software} {Engineering}
  ({ICSE})}}. \bibinfo{publisher}{IEEE}, \bibinfo{address}{Montreal, QC,
  Canada}, \bibinfo{pages}{960--970}.
\newblock
\showISBNx{978-1-72810-869-8}
\urldef\tempurl%
\url{https://doi.org/10.1109/ICSE.2019.00101}
\showDOI{\tempurl}


\bibitem[\protect\citeauthoryear{Henkel, Ramakrishnan, Wang, Albarghouthi, Jha,
  and Reps}{Henkel et~al\mbox{.}}{2022}]%
        {henkel_semantic_2022}
\bibfield{author}{\bibinfo{person}{Jordan Henkel}, \bibinfo{person}{Goutham
  Ramakrishnan}, \bibinfo{person}{Zi Wang}, \bibinfo{person}{Aws Albarghouthi},
  \bibinfo{person}{Somesh Jha}, {and} \bibinfo{person}{Thomas Reps}.}
  \bibinfo{year}{2022}\natexlab{}.
\newblock \showarticletitle{Semantic {Robustness} of {Models} of {Source}
  {Code}}. In \bibinfo{booktitle}{\emph{2022 {IEEE} {International}
  {Conference} on {Software} {Analysis}, {Evolution} and {Reengineering}
  ({SANER})}}. \bibinfo{publisher}{IEEE}, \bibinfo{address}{Honolulu, HI, USA},
  \bibinfo{pages}{526--537}.
\newblock
\showISBNx{978-1-66543-786-8}
\urldef\tempurl%
\url{https://doi.org/10.1109/SANER53432.2022.00070}
\showDOI{\tempurl}


\bibitem[\protect\citeauthoryear{Herzig, Just, and Zeller}{Herzig
  et~al\mbox{.}}{2013}]%
        {herzig2013s}
\bibfield{author}{\bibinfo{person}{Kim Herzig}, \bibinfo{person}{Sascha Just},
  {and} \bibinfo{person}{Andreas Zeller}.} \bibinfo{year}{2013}\natexlab{}.
\newblock \showarticletitle{It's not a bug, it's a feature: how
  misclassification impacts bug prediction}. In \bibinfo{booktitle}{\emph{2013
  35th international conference on software engineering (ICSE)}}. IEEE,
  \bibinfo{pages}{392--401}.
\newblock


\bibitem[\protect\citeauthoryear{Hoang, Kang, Lo, and Lawall}{Hoang
  et~al\mbox{.}}{2020}]%
        {hoang2020cc2vec}
\bibfield{author}{\bibinfo{person}{Thong Hoang}, \bibinfo{person}{Hong~Jin
  Kang}, \bibinfo{person}{David Lo}, {and} \bibinfo{person}{Julia Lawall}.}
  \bibinfo{year}{2020}\natexlab{}.
\newblock \showarticletitle{Cc2vec: Distributed representations of code
  changes}. In \bibinfo{booktitle}{\emph{Proceedings of the ACM/IEEE 42nd
  International Conference on Software Engineering}}.
  \bibinfo{pages}{518--529}.
\newblock


\bibitem[\protect\citeauthoryear{Hou, Zhao, Liu, Yang, Wang, Li, Luo, Lo,
  Grundy, and Wang}{Hou et~al\mbox{.}}{2023}]%
        {hou2023large}
\bibfield{author}{\bibinfo{person}{Xinyi Hou}, \bibinfo{person}{Yanjie Zhao},
  \bibinfo{person}{Yue Liu}, \bibinfo{person}{Zhou Yang},
  \bibinfo{person}{Kailong Wang}, \bibinfo{person}{Li Li},
  \bibinfo{person}{Xiapu Luo}, \bibinfo{person}{David Lo},
  \bibinfo{person}{John Grundy}, {and} \bibinfo{person}{Haoyu Wang}.}
  \bibinfo{year}{2023}\natexlab{}.
\newblock \showarticletitle{Large Language Models for Software Engineering: A
  Systematic Literature Review}.
\newblock \bibinfo{journal}{\emph{arXiv preprint arXiv:2308.10620}}
  (\bibinfo{year}{2023}).
\newblock


\bibitem[\protect\citeauthoryear{Hu, Liu, Zhao, Liu, Sun, Tantithamthavorn, and
  Li}{Hu et~al\mbox{.}}{2023}]%
        {hu2023}
\bibfield{author}{\bibinfo{person}{Haonan Hu}, \bibinfo{person}{Yue Liu},
  \bibinfo{person}{Yanjie Zhao}, \bibinfo{person}{Yonghui Liu},
  \bibinfo{person}{Xiaoyu Sun}, \bibinfo{person}{Chakkrit Tantithamthavorn},
  {and} \bibinfo{person}{Li Li}.} \bibinfo{year}{2023}\natexlab{}.
\newblock \showarticletitle{Detecting Temporal Inconsistency in Biased Datasets
  for Android Malware Detection}. In \bibinfo{booktitle}{\emph{The 6th
  International Workshop on Advances in Mobile App Analysis (A-Mobile)}}.
  \bibinfo{pages}{0--0}.
\newblock


\bibitem[\protect\citeauthoryear{Huang, Hu, and Chen}{Huang
  et~al\mbox{.}}{2021}]%
        {huang2021robustness}
\bibfield{author}{\bibinfo{person}{Yujin Huang}, \bibinfo{person}{Han Hu},
  {and} \bibinfo{person}{Chunyang Chen}.} \bibinfo{year}{2021}\natexlab{}.
\newblock \showarticletitle{Robustness of on-device models: Adversarial attack
  to deep learning models on android apps}. In \bibinfo{booktitle}{\emph{2021
  IEEE/ACM 43rd International Conference on Software Engineering: Software
  Engineering in Practice (ICSE-SEIP)}}. IEEE, \bibinfo{pages}{101--110}.
\newblock


\bibitem[\protect\citeauthoryear{Husain, Wu, Gazit, Allamanis, and
  Brockschmidt}{Husain et~al\mbox{.}}{2019}]%
        {husain2019codesearchnet}
\bibfield{author}{\bibinfo{person}{Hamel Husain}, \bibinfo{person}{Ho-Hsiang
  Wu}, \bibinfo{person}{Tiferet Gazit}, \bibinfo{person}{Miltiadis Allamanis},
  {and} \bibinfo{person}{Marc Brockschmidt}.} \bibinfo{year}{2019}\natexlab{}.
\newblock \showarticletitle{Codesearchnet challenge: Evaluating the state of
  semantic code search}.
\newblock \bibinfo{journal}{\emph{arXiv preprint arXiv:1909.09436}}
  (\bibinfo{year}{2019}).
\newblock


\bibitem[\protect\citeauthoryear{Hutson}{Hutson}{2018}]%
        {hutson2018artificial}
\bibfield{author}{\bibinfo{person}{Matthew Hutson}.}
  \bibinfo{year}{2018}\natexlab{}.
\newblock \bibinfo{title}{Artificial intelligence faces reproducibility
  crisis}.
\newblock
\newblock


\bibitem[\protect\citeauthoryear{Ivie and Thain}{Ivie and Thain}{2018}]%
        {ivie2018reproducibility}
\bibfield{author}{\bibinfo{person}{Peter Ivie} {and} \bibinfo{person}{Douglas
  Thain}.} \bibinfo{year}{2018}\natexlab{}.
\newblock \showarticletitle{Reproducibility in scientific computing}.
\newblock \bibinfo{journal}{\emph{ACM Computing Surveys (CSUR)}}
  \bibinfo{volume}{51}, \bibinfo{number}{3} (\bibinfo{year}{2018}),
  \bibinfo{pages}{1--36}.
\newblock


\bibitem[\protect\citeauthoryear{Jain, Vaidyanath, Iyer, Natarajan,
  Parthasarathy, Rajamani, and Sharma}{Jain et~al\mbox{.}}{2022}]%
        {jain2022jigsaw}
\bibfield{author}{\bibinfo{person}{Naman Jain}, \bibinfo{person}{Skanda
  Vaidyanath}, \bibinfo{person}{Arun Iyer}, \bibinfo{person}{Nagarajan
  Natarajan}, \bibinfo{person}{Suresh Parthasarathy}, \bibinfo{person}{Sriram
  Rajamani}, {and} \bibinfo{person}{Rahul Sharma}.}
  \bibinfo{year}{2022}\natexlab{}.
\newblock \showarticletitle{Jigsaw: Large language models meet program
  synthesis}. In \bibinfo{booktitle}{\emph{Proceedings of the 44th
  International Conference on Software Engineering}}.
  \bibinfo{pages}{1219--1231}.
\newblock


\bibitem[\protect\citeauthoryear{Jiang, Liu, Lutellier, and Tan}{Jiang
  et~al\mbox{.}}{2023a}]%
        {jiang2023impact}
\bibfield{author}{\bibinfo{person}{Nan Jiang}, \bibinfo{person}{Kevin Liu},
  \bibinfo{person}{Thibaud Lutellier}, {and} \bibinfo{person}{Lin Tan}.}
  \bibinfo{year}{2023}\natexlab{a}.
\newblock \showarticletitle{Impact of code language models on automated program
  repair}.
\newblock \bibinfo{journal}{\emph{Proceedings of the 45th International
  Conference on Software Engineering}} (\bibinfo{year}{2023}).
\newblock


\bibitem[\protect\citeauthoryear{Jiang, Armaly, and McMillan}{Jiang
  et~al\mbox{.}}{2017}]%
        {jiang2017automatically}
\bibfield{author}{\bibinfo{person}{Siyuan Jiang}, \bibinfo{person}{Ameer
  Armaly}, {and} \bibinfo{person}{Collin McMillan}.}
  \bibinfo{year}{2017}\natexlab{}.
\newblock \showarticletitle{Automatically generating commit messages from diffs
  using neural machine translation}. In \bibinfo{booktitle}{\emph{2017 32nd
  IEEE/ACM International Conference on Automated Software Engineering (ASE)}}.
  IEEE, \bibinfo{pages}{135--146}.
\newblock


\bibitem[\protect\citeauthoryear{Jiang, Synovic, Hyatt, Schorlemmer, Sethi, Lu,
  Thiruvathukal, and Davis}{Jiang et~al\mbox{.}}{2023b}]%
        {jiang2023empirical}
\bibfield{author}{\bibinfo{person}{Wenxin Jiang}, \bibinfo{person}{Nicholas
  Synovic}, \bibinfo{person}{Matt Hyatt}, \bibinfo{person}{Taylor~R.
  Schorlemmer}, \bibinfo{person}{Rohan Sethi}, \bibinfo{person}{Yung-Hsiang
  Lu}, \bibinfo{person}{George~K. Thiruvathukal}, {and}
  \bibinfo{person}{James~C. Davis}.} \bibinfo{year}{2023}\natexlab{b}.
\newblock \showarticletitle{An Empirical Study of Pre-Trained Model Reuse in
  the Hugging Face Deep Learning Model Registry}.
\newblock  (\bibinfo{year}{2023}), \bibinfo{pages}{2463--2475}.
\newblock
\urldef\tempurl%
\url{https://doi.org/10.1109/ICSE48619.2023.00206}
\showDOI{\tempurl}


\bibitem[\protect\citeauthoryear{Jiarpakdee, Tantithamthavorn, and
  Grundy}{Jiarpakdee et~al\mbox{.}}{2021}]%
        {DBLP:conf/msr/JiarpakdeeTG21}
\bibfield{author}{\bibinfo{person}{Jirayus Jiarpakdee},
  \bibinfo{person}{Chakkrit Tantithamthavorn}, {and} \bibinfo{person}{John~C.
  Grundy}.} \bibinfo{year}{2021}\natexlab{}.
\newblock \showarticletitle{Practitioners' Perceptions of the Goals and Visual
  Explanations of Defect Prediction Models}. In \bibinfo{booktitle}{\emph{18th
  {IEEE/ACM} International Conference on Mining Software Repositories, {MSR}
  2021, Madrid, Spain, May 17-19, 2021}}. \bibinfo{publisher}{{IEEE}},
  \bibinfo{pages}{432--443}.
\newblock
\urldef\tempurl%
\url{https://doi.org/10.1109/MSR52588.2021.00055}
\showDOI{\tempurl}


\bibitem[\protect\citeauthoryear{Johnson, Tarlow, and Walder}{Johnson
  et~al\mbox{.}}{2023}]%
        {johnson2023ru}
\bibfield{author}{\bibinfo{person}{Daniel~D Johnson}, \bibinfo{person}{Daniel
  Tarlow}, {and} \bibinfo{person}{Christian Walder}.}
  \bibinfo{year}{2023}\natexlab{}.
\newblock \showarticletitle{RU-SURE? Uncertainty-Aware Code Suggestions By
  Maximizing Utility Across Random User Intents}.
\newblock \bibinfo{journal}{\emph{arXiv preprint arXiv:2303.00732}}
  (\bibinfo{year}{2023}).
\newblock


\bibitem[\protect\citeauthoryear{J{\o}rgensen, Dyb{\aa}, Liest{\o}l, and
  Sj{\o}berg}{J{\o}rgensen et~al\mbox{.}}{2016}]%
        {jorgensen2016incorrect}
\bibfield{author}{\bibinfo{person}{Magne J{\o}rgensen}, \bibinfo{person}{Tore
  Dyb{\aa}}, \bibinfo{person}{Knut Liest{\o}l}, {and} \bibinfo{person}{Dag~IK
  Sj{\o}berg}.} \bibinfo{year}{2016}\natexlab{}.
\newblock \showarticletitle{Incorrect results in software engineering
  experiments: How to improve research practices}.
\newblock \bibinfo{journal}{\emph{Journal of Systems and Software}}
  \bibinfo{volume}{116} (\bibinfo{year}{2016}), \bibinfo{pages}{133--145}.
\newblock


\bibitem[\protect\citeauthoryear{Kaddour, Harris, Mozes, Bradley, Raileanu, and
  McHardy}{Kaddour et~al\mbox{.}}{2023}]%
        {kaddour2023challenges}
\bibfield{author}{\bibinfo{person}{Jean Kaddour}, \bibinfo{person}{Joshua
  Harris}, \bibinfo{person}{Maximilian Mozes}, \bibinfo{person}{Herbie
  Bradley}, \bibinfo{person}{Roberta Raileanu}, {and} \bibinfo{person}{Robert
  McHardy}.} \bibinfo{year}{2023}\natexlab{}.
\newblock \showarticletitle{Challenges and applications of large language
  models}.
\newblock \bibinfo{journal}{\emph{arXiv preprint arXiv:2307.10169}}
  (\bibinfo{year}{2023}).
\newblock


\bibitem[\protect\citeauthoryear{Kirchenbauer, Geiping, Wen, Katz, Miers, and
  Goldstein}{Kirchenbauer et~al\mbox{.}}{2023}]%
        {kirchenbauer2023watermark}
\bibfield{author}{\bibinfo{person}{John Kirchenbauer}, \bibinfo{person}{Jonas
  Geiping}, \bibinfo{person}{Yuxin Wen}, \bibinfo{person}{Jonathan Katz},
  \bibinfo{person}{Ian Miers}, {and} \bibinfo{person}{Tom Goldstein}.}
  \bibinfo{year}{2023}\natexlab{}.
\newblock \showarticletitle{A watermark for large language models}.
\newblock \bibinfo{journal}{\emph{arXiv preprint arXiv:2301.10226}}
  (\bibinfo{year}{2023}).
\newblock


\bibitem[\protect\citeauthoryear{Kitchenham and Charters}{Kitchenham and
  Charters}{2007a}]%
        {kitchenham_guidelines_2007}
\bibfield{author}{\bibinfo{person}{Barbara Kitchenham} {and}
  \bibinfo{person}{Stuart Charters}.} \bibinfo{year}{2007}\natexlab{a}.
\newblock \showarticletitle{Guidelines for performing {Systematic} {Literature}
  {Reviews} in {Software} {Engineering}}.
\newblock   \bibinfo{volume}{2} (\bibinfo{date}{Jan.} \bibinfo{year}{2007}).
\newblock


\bibitem[\protect\citeauthoryear{Kitchenham and Charters}{Kitchenham and
  Charters}{2007b}]%
        {kitchenham2007guidelines}
\bibfield{author}{\bibinfo{person}{Barbara Kitchenham} {and}
  \bibinfo{person}{Stuart Charters}.} \bibinfo{year}{2007}\natexlab{b}.
\newblock \showarticletitle{Guidelines for performing systematic literature
  reviews in software engineering}.
\newblock  (\bibinfo{year}{2007}).
\newblock


\bibitem[\protect\citeauthoryear{Kitchenham, Madeyski, and Budgen}{Kitchenham
  et~al\mbox{.}}{2023}]%
        {kitchenham_segress_2023}
\bibfield{author}{\bibinfo{person}{Barbara Kitchenham}, \bibinfo{person}{Lech
  Madeyski}, {and} \bibinfo{person}{David Budgen}.}
  \bibinfo{year}{2023}\natexlab{}.
\newblock \showarticletitle{{SEGRESS}: {Software} {Engineering} {Guidelines}
  for {REporting} {Secondary} {Studies}}.
\newblock \bibinfo{journal}{\emph{IEEE Transactions on Software Engineering}}
  \bibinfo{volume}{49} (\bibinfo{date}{March} \bibinfo{year}{2023}),
  \bibinfo{pages}{1273--1298}.
\newblock
\showISSN{1939-3520}
\urldef\tempurl%
\url{https://doi.org/10.1109/TSE.2022.3174092}
\showDOI{\tempurl}
\newblock
\shownote{Conference Name: IEEE Transactions on Software Engineering.}


\bibitem[\protect\citeauthoryear{Laaber, Basmaci, and Salza}{Laaber
  et~al\mbox{.}}{2021}]%
        {laaber2021predicting}
\bibfield{author}{\bibinfo{person}{Christoph Laaber}, \bibinfo{person}{Mikael
  Basmaci}, {and} \bibinfo{person}{Pasquale Salza}.}
  \bibinfo{year}{2021}\natexlab{}.
\newblock \showarticletitle{Predicting unstable software benchmarks using
  static source code features}.
\newblock \bibinfo{journal}{\emph{Empirical Software Engineering}}
  \bibinfo{volume}{26}, \bibinfo{number}{6} (\bibinfo{year}{2021}),
  \bibinfo{pages}{114}.
\newblock


\bibitem[\protect\citeauthoryear{Lai, Li, Wang, Zhang, Zhong, Zettlemoyer, Yih,
  Fried, yi~Wang, and Yu}{Lai et~al\mbox{.}}{2022}]%
        {Lai2022DS1000AN}
\bibfield{author}{\bibinfo{person}{Yuhang Lai}, \bibinfo{person}{Chengxi Li},
  \bibinfo{person}{Yiming Wang}, \bibinfo{person}{Tianyi Zhang},
  \bibinfo{person}{Ruiqi Zhong}, \bibinfo{person}{Luke Zettlemoyer},
  \bibinfo{person}{Scott Yih}, \bibinfo{person}{Daniel Fried},
  \bibinfo{person}{Si yi Wang}, {and} \bibinfo{person}{Tao Yu}.}
  \bibinfo{year}{2022}\natexlab{}.
\newblock \showarticletitle{DS-1000: A Natural and Reliable Benchmark for Data
  Science Code Generation}.
\newblock \bibinfo{journal}{\emph{ArXiv}}  \bibinfo{volume}{abs/2211.11501}
  (\bibinfo{year}{2022}).
\newblock
\urldef\tempurl%
\url{https://api.semanticscholar.org/CorpusID:253734939}
\showURL{%
\tempurl}


\bibitem[\protect\citeauthoryear{Li, Qi, Liu, Di, Liu, Pei, Yi, and Zhou}{Li
  et~al\mbox{.}}{2023c}]%
        {li2023trustworthy}
\bibfield{author}{\bibinfo{person}{Bo Li}, \bibinfo{person}{Peng Qi},
  \bibinfo{person}{Bo Liu}, \bibinfo{person}{Shuai Di}, \bibinfo{person}{Jingen
  Liu}, \bibinfo{person}{Jiquan Pei}, \bibinfo{person}{Jinfeng Yi}, {and}
  \bibinfo{person}{Bowen Zhou}.} \bibinfo{year}{2023}\natexlab{c}.
\newblock \showarticletitle{Trustworthy AI: From principles to practices}.
\newblock \bibinfo{journal}{\emph{Comput. Surveys}} \bibinfo{volume}{55},
  \bibinfo{number}{9} (\bibinfo{year}{2023}), \bibinfo{pages}{1--46}.
\newblock


\bibitem[\protect\citeauthoryear{Li, Allal, Zi, Muennighoff, Kocetkov, Mou,
  Marone, Akiki, Li, Chim, et~al\mbox{.}}{Li et~al\mbox{.}}{2023a}]%
        {li2023starcoder}
\bibfield{author}{\bibinfo{person}{Raymond Li}, \bibinfo{person}{Loubna~Ben
  Allal}, \bibinfo{person}{Yangtian Zi}, \bibinfo{person}{Niklas Muennighoff},
  \bibinfo{person}{Denis Kocetkov}, \bibinfo{person}{Chenghao Mou},
  \bibinfo{person}{Marc Marone}, \bibinfo{person}{Christopher Akiki},
  \bibinfo{person}{Jia Li}, \bibinfo{person}{Jenny Chim}, {et~al\mbox{.}}}
  \bibinfo{year}{2023}\natexlab{a}.
\newblock \showarticletitle{StarCoder: may the source be with you!}
\newblock \bibinfo{journal}{\emph{arXiv preprint arXiv:2305.06161}}
  (\bibinfo{year}{2023}).
\newblock


\bibitem[\protect\citeauthoryear{Li, Sahu, Talwalkar, and Smith}{Li
  et~al\mbox{.}}{2020}]%
        {li2020federated}
\bibfield{author}{\bibinfo{person}{Tian Li}, \bibinfo{person}{Anit~Kumar Sahu},
  \bibinfo{person}{Ameet Talwalkar}, {and} \bibinfo{person}{Virginia Smith}.}
  \bibinfo{year}{2020}\natexlab{}.
\newblock \showarticletitle{Federated learning: Challenges, methods, and future
  directions}.
\newblock \bibinfo{journal}{\emph{IEEE signal processing magazine}}
  \bibinfo{volume}{37}, \bibinfo{number}{3} (\bibinfo{year}{2020}),
  \bibinfo{pages}{50--60}.
\newblock


\bibitem[\protect\citeauthoryear{Li, Liu, Chen, Xie, Zhang, and Liu}{Li
  et~al\mbox{.}}{2023b}]%
        {li2023multi}
\bibfield{author}{\bibinfo{person}{Yanzhou Li}, \bibinfo{person}{Shangqing
  Liu}, \bibinfo{person}{Kangjie Chen}, \bibinfo{person}{Xiaofei Xie},
  \bibinfo{person}{Tianwei Zhang}, {and} \bibinfo{person}{Yang Liu}.}
  \bibinfo{year}{2023}\natexlab{b}.
\newblock \showarticletitle{Multi-target Backdoor Attacks for Code Pre-trained
  Models}.
\newblock \bibinfo{journal}{\emph{arXiv preprint arXiv:2306.08350}}
  (\bibinfo{year}{2023}).
\newblock


\bibitem[\protect\citeauthoryear{Li, Wang, and Nguyen}{Li
  et~al\mbox{.}}{2021}]%
        {li_vulnerability_2021}
\bibfield{author}{\bibinfo{person}{Yi Li}, \bibinfo{person}{Shaohua Wang},
  {and} \bibinfo{person}{Tien~N. Nguyen}.} \bibinfo{year}{2021}\natexlab{}.
\newblock \showarticletitle{Vulnerability detection with fine-grained
  interpretations}. In \bibinfo{booktitle}{\emph{Proceedings of the 29th {ACM}
  {Joint} {Meeting} on {European} {Software} {Engineering} {Conference} and
  {Symposium} on the {Foundations} of {Software} {Engineering}}}
  \emph{(\bibinfo{series}{{ESEC}/{FSE} 2021})}. \bibinfo{publisher}{Association
  for Computing Machinery}, \bibinfo{address}{New York, NY, USA},
  \bibinfo{pages}{292--303}.
\newblock
\showISBNx{978-1-4503-8562-6}
\urldef\tempurl%
\url{https://doi.org/10.1145/3468264.3468597}
\showDOI{\tempurl}


\bibitem[\protect\citeauthoryear{Li, Pan, Pei, Zhang, Wang, and Li}{Li
  et~al\mbox{.}}{2022}]%
        {li2022robust}
\bibfield{author}{\bibinfo{person}{Zhong Li}, \bibinfo{person}{Minxue Pan},
  \bibinfo{person}{Yu Pei}, \bibinfo{person}{Tian Zhang},
  \bibinfo{person}{Linzhang Wang}, {and} \bibinfo{person}{Xuandong Li}.}
  \bibinfo{year}{2022}\natexlab{}.
\newblock \showarticletitle{Robust Learning of Deep Predictive Models from
  Noisy and Imbalanced Software Engineering Datasets}. In
  \bibinfo{booktitle}{\emph{Proceedings of the 37th IEEE/ACM International
  Conference on Automated Software Engineering}}. \bibinfo{pages}{1--13}.
\newblock


\bibitem[\protect\citeauthoryear{Li, Wang, Liu, Wang, Chen, Wang, and Gao}{Li
  et~al\mbox{.}}{2023d}]%
        {li2023cctest}
\bibfield{author}{\bibinfo{person}{Zongjie Li}, \bibinfo{person}{Chaozheng
  Wang}, \bibinfo{person}{Zhibo Liu}, \bibinfo{person}{Haoxuan Wang},
  \bibinfo{person}{Dong Chen}, \bibinfo{person}{Shuai Wang}, {and}
  \bibinfo{person}{Cuiyun Gao}.} \bibinfo{year}{2023}\natexlab{d}.
\newblock \showarticletitle{Cctest: Testing and repairing code completion
  systems}. In \bibinfo{booktitle}{\emph{2023 IEEE/ACM 45th International
  Conference on Software Engineering (ICSE)}}. IEEE,
  \bibinfo{pages}{1238--1250}.
\newblock


\bibitem[\protect\citeauthoryear{Li, Wang, Ma, Liu, Wang, Wu, and Gao}{Li
  et~al\mbox{.}}{2023e}]%
        {li2023feasibility}
\bibfield{author}{\bibinfo{person}{Zongjie Li}, \bibinfo{person}{Chaozheng
  Wang}, \bibinfo{person}{Pingchuan Ma}, \bibinfo{person}{Chaowei Liu},
  \bibinfo{person}{Shuai Wang}, \bibinfo{person}{Daoyuan Wu}, {and}
  \bibinfo{person}{Cuiyun Gao}.} \bibinfo{year}{2023}\natexlab{e}.
\newblock \showarticletitle{On the feasibility of specialized ability stealing
  for large language code models}.
\newblock \bibinfo{journal}{\emph{arXiv preprint arXiv:2303.03012}}
  (\bibinfo{year}{2023}).
\newblock


\bibitem[\protect\citeauthoryear{Lin, Li, Zhang, Deng, Zeng, Zhang, and
  Wan}{Lin et~al\mbox{.}}{2022}]%
        {lin2022xcode}
\bibfield{author}{\bibinfo{person}{Zehao Lin}, \bibinfo{person}{Guodun Li},
  \bibinfo{person}{Jingfeng Zhang}, \bibinfo{person}{Yue Deng},
  \bibinfo{person}{Xiangji Zeng}, \bibinfo{person}{Yin Zhang}, {and}
  \bibinfo{person}{Yao Wan}.} \bibinfo{year}{2022}\natexlab{}.
\newblock \showarticletitle{XCODE: Towards Cross-Language Code Representation
  with Large-Scale Pre-Training}.
\newblock \bibinfo{journal}{\emph{ACM Transactions on Software Engineering and
  Methodology (TOSEM)}} \bibinfo{volume}{31}, \bibinfo{number}{3}
  (\bibinfo{year}{2022}), \bibinfo{pages}{1--44}.
\newblock


\bibitem[\protect\citeauthoryear{Ling, Blunsom, Grefenstette, Hermann,
  Ko{\v{c}}isk{\`y}, Wang, and Senior}{Ling et~al\mbox{.}}{2016}]%
        {ling2016latent}
\bibfield{author}{\bibinfo{person}{Wang Ling}, \bibinfo{person}{Phil Blunsom},
  \bibinfo{person}{Edward Grefenstette}, \bibinfo{person}{Karl~Moritz Hermann},
  \bibinfo{person}{Tom{\'a}{\v{s}} Ko{\v{c}}isk{\`y}}, \bibinfo{person}{Fumin
  Wang}, {and} \bibinfo{person}{Andrew Senior}.}
  \bibinfo{year}{2016}\natexlab{}.
\newblock \showarticletitle{Latent Predictor Networks for Code Generation}. In
  \bibinfo{booktitle}{\emph{Proceedings of the 54th Annual Meeting of the
  Association for Computational Linguistics (Volume 1: Long Papers)}}.
  \bibinfo{pages}{599--609}.
\newblock


\bibitem[\protect\citeauthoryear{Liu, Gao, Xia, Lo, Grundy, and Yang}{Liu
  et~al\mbox{.}}{2021a}]%
        {liu2021Reproducibility}
\bibfield{author}{\bibinfo{person}{Chao Liu}, \bibinfo{person}{Cuiyun Gao},
  \bibinfo{person}{Xin Xia}, \bibinfo{person}{David Lo}, \bibinfo{person}{John
  Grundy}, {and} \bibinfo{person}{Xiaohu Yang}.}
  \bibinfo{year}{2021}\natexlab{a}.
\newblock \showarticletitle{On the Reproducibility and Replicability of Deep
  Learning in Software Engineering}.
\newblock \bibinfo{journal}{\emph{ACM Trans. Softw. Eng. Methodol.}}
  \bibinfo{volume}{31}, \bibinfo{number}{1}, Article \bibinfo{articleno}{15}
  (\bibinfo{date}{oct} \bibinfo{year}{2021}), \bibinfo{numpages}{46}~pages.
\newblock
\showISSN{1049-331X}
\urldef\tempurl%
\url{https://doi.org/10.1145/3477535}
\showDOI{\tempurl}


\bibitem[\protect\citeauthoryear{Liu, Lin, Lou, Wen, and Zhang}{Liu
  et~al\mbox{.}}{2021b}]%
        {liu_can_2021}
\bibfield{author}{\bibinfo{person}{Chenyao Liu}, \bibinfo{person}{Zeqi Lin},
  \bibinfo{person}{Jian-Guang Lou}, \bibinfo{person}{Lijie Wen}, {and}
  \bibinfo{person}{Dongmei Zhang}.} \bibinfo{year}{2021}\natexlab{b}.
\newblock \showarticletitle{Can {Neural} {Clone} {Detection} {Generalize} to
  {Unseen} {Functionalities}}. In \bibinfo{booktitle}{\emph{2021 36th
  {IEEE}/{ACM} {International} {Conference} on {Automated} {Software}
  {Engineering} ({ASE})}}. \bibinfo{publisher}{IEEE},
  \bibinfo{address}{Melbourne, Australia}, \bibinfo{pages}{617--629}.
\newblock
\showISBNx{978-1-66540-337-5}
\urldef\tempurl%
\url{https://doi.org/10.1109/ASE51524.2021.9678907}
\showDOI{\tempurl}


\bibitem[\protect\citeauthoryear{Liu, Shen, Zhu, Niu, Li, and Zhang}{Liu
  et~al\mbox{.}}{2022b}]%
        {liu_deep_2022}
\bibfield{author}{\bibinfo{person}{Hui Liu}, \bibinfo{person}{Mingzhu Shen},
  \bibinfo{person}{Jiaqi Zhu}, \bibinfo{person}{Nan Niu}, \bibinfo{person}{Ge
  Li}, {and} \bibinfo{person}{Lu Zhang}.} \bibinfo{year}{2022}\natexlab{b}.
\newblock \showarticletitle{Deep {Learning} {Based} {Program} {Generation}
  {From} {Requirements} {Text}: {Are} {We} {There} {Yet}?}
\newblock \bibinfo{journal}{\emph{IIEEE Trans. Software Eng.}}
  \bibinfo{volume}{48}, \bibinfo{number}{4} (\bibinfo{date}{April}
  \bibinfo{year}{2022}), \bibinfo{pages}{1268--1289}.
\newblock
\showISSN{0098-5589, 1939-3520, 2326-3881}
\urldef\tempurl%
\url{https://doi.org/10.1109/TSE.2020.3018481}
\showDOI{\tempurl}


\bibitem[\protect\citeauthoryear{Liu, Lin, Qu, Zhang, De~Vel, Montague, and
  Xiang}{Liu et~al\mbox{.}}{2022a}]%
        {liu_cd-vuld_2022}
\bibfield{author}{\bibinfo{person}{Shigang Liu}, \bibinfo{person}{Guanjun Lin},
  \bibinfo{person}{Lizhen Qu}, \bibinfo{person}{Jun Zhang},
  \bibinfo{person}{Olivier De~Vel}, \bibinfo{person}{Paul Montague}, {and}
  \bibinfo{person}{Yang Xiang}.} \bibinfo{year}{2022}\natexlab{a}.
\newblock \showarticletitle{{CD}-{VulD}: {Cross}-{Domain} {Vulnerability}
  {Discovery} {Based} on {Deep} {Domain} {Adaptation}}.
\newblock \bibinfo{journal}{\emph{IEEE Trans. Dependable and Secure Comput.}}
  \bibinfo{volume}{19}, \bibinfo{number}{1} (\bibinfo{date}{Jan.}
  \bibinfo{year}{2022}), \bibinfo{pages}{438--451}.
\newblock
\showISSN{1545-5971, 1941-0018, 2160-9209}
\urldef\tempurl%
\url{https://doi.org/10.1109/TDSC.2020.2984505}
\showDOI{\tempurl}


\bibitem[\protect\citeauthoryear{Liu, Wu, Xie, Meng, and Liu}{Liu
  et~al\mbox{.}}{2023c}]%
        {liu2023contrabert}
\bibfield{author}{\bibinfo{person}{Shangqing Liu}, \bibinfo{person}{Bozhi Wu},
  \bibinfo{person}{Xiaofei Xie}, \bibinfo{person}{Guozhu Meng}, {and}
  \bibinfo{person}{Yang Liu}.} \bibinfo{year}{2023}\natexlab{c}.
\newblock \showarticletitle{Contrabert: Enhancing code pre-trained models via
  contrastive learning}.
\newblock \bibinfo{journal}{\emph{arXiv preprint arXiv:2301.09072}}
  (\bibinfo{year}{2023}).
\newblock


\bibitem[\protect\citeauthoryear{Liu, Le-Cong, Widyasari, Tantithamthavorn, Li,
  Le, and Lo}{Liu et~al\mbox{.}}{2023a}]%
        {liu2023refining}
\bibfield{author}{\bibinfo{person}{Yue Liu}, \bibinfo{person}{Thanh Le-Cong},
  \bibinfo{person}{Ratnadira Widyasari}, \bibinfo{person}{Chakkrit
  Tantithamthavorn}, \bibinfo{person}{Li Li}, \bibinfo{person}{Xuan-Bach~D Le},
  {and} \bibinfo{person}{David Lo}.} \bibinfo{year}{2023}\natexlab{a}.
\newblock \showarticletitle{Refining ChatGPT-Generated Code: Characterizing and
  Mitigating Code Quality Issues}.
\newblock \bibinfo{journal}{\emph{arXiv preprint arXiv:2307.12596}}
  (\bibinfo{year}{2023}).
\newblock


\bibitem[\protect\citeauthoryear{Liu, Tantithamthavorn, Li, and Liu}{Liu
  et~al\mbox{.}}{2022c}]%
        {liu2022deep}
\bibfield{author}{\bibinfo{person}{Yue Liu}, \bibinfo{person}{Chakkrit
  Tantithamthavorn}, \bibinfo{person}{Li Li}, {and} \bibinfo{person}{Yepang
  Liu}.} \bibinfo{year}{2022}\natexlab{c}.
\newblock \showarticletitle{Deep learning for android malware defenses: a
  systematic literature review}.
\newblock \bibinfo{journal}{\emph{Comput. Surveys}} \bibinfo{volume}{55},
  \bibinfo{number}{8} (\bibinfo{year}{2022}), \bibinfo{pages}{1--36}.
\newblock


\bibitem[\protect\citeauthoryear{Liu, Liu, Xia, and Yang}{Liu
  et~al\mbox{.}}{2023b}]%
        {liu2023towards}
\bibfield{author}{\bibinfo{person}{Zhongxin Liu}, \bibinfo{person}{Kui Liu},
  \bibinfo{person}{Xin Xia}, {and} \bibinfo{person}{Xiaohu Yang}.}
  \bibinfo{year}{2023}\natexlab{b}.
\newblock \showarticletitle{Towards More Realistic Evaluation for Neural Test
  Oracle Generation}.
\newblock \bibinfo{journal}{\emph{arXiv preprint arXiv:2305.17047}}
  (\bibinfo{year}{2023}).
\newblock


\bibitem[\protect\citeauthoryear{Liu, Xia, Hassan, Lo, Xing, and Wang}{Liu
  et~al\mbox{.}}{2018}]%
        {liu2018neural}
\bibfield{author}{\bibinfo{person}{Zhongxin Liu}, \bibinfo{person}{Xin Xia},
  \bibinfo{person}{Ahmed~E Hassan}, \bibinfo{person}{David Lo},
  \bibinfo{person}{Zhenchang Xing}, {and} \bibinfo{person}{Xinyu Wang}.}
  \bibinfo{year}{2018}\natexlab{}.
\newblock \showarticletitle{Neural-machine-translation-based commit message
  generation: how far are we?}. In \bibinfo{booktitle}{\emph{Proceedings of the
  33rd ACM/IEEE International Conference on Automated Software Engineering}}.
  \bibinfo{pages}{373--384}.
\newblock


\bibitem[\protect\citeauthoryear{Lo}{Lo}{2023}]%
        {lo2023trustworthy}
\bibfield{author}{\bibinfo{person}{David Lo}.} \bibinfo{year}{2023}\natexlab{}.
\newblock \showarticletitle{Trustworthy and Synergistic Artificial Intelligence
  for Software Engineering: Vision and Roadmaps}.
\newblock \bibinfo{journal}{\emph{arXiv preprint arXiv:2309.04142}}
  (\bibinfo{year}{2023}).
\newblock


\bibitem[\protect\citeauthoryear{Lu, Yu, Li, Yang, and Zuo}{Lu
  et~al\mbox{.}}{2023}]%
        {lu2023llama}
\bibfield{author}{\bibinfo{person}{Junyi Lu}, \bibinfo{person}{Lei Yu},
  \bibinfo{person}{Xiaojia Li}, \bibinfo{person}{Li Yang}, {and}
  \bibinfo{person}{Chun Zuo}.} \bibinfo{year}{2023}\natexlab{}.
\newblock \showarticletitle{LLaMA-Reviewer: Advancing Code Review Automation
  with Large Language Models through Parameter-Efficient Fine-Tuning (Practical
  Experience Report)}.
\newblock \bibinfo{journal}{\emph{arXiv preprint arXiv:2308.11148}}
  (\bibinfo{year}{2023}).
\newblock


\bibitem[\protect\citeauthoryear{Lu, Guo, Ren, Huang, Svyatkovskiy, Blanco,
  Clement, Drain, Jiang, Tang, et~al\mbox{.}}{Lu et~al\mbox{.}}{2021}]%
        {lu2021codexglue}
\bibfield{author}{\bibinfo{person}{Shuai Lu}, \bibinfo{person}{Daya Guo},
  \bibinfo{person}{Shuo Ren}, \bibinfo{person}{Junjie Huang},
  \bibinfo{person}{Alexey Svyatkovskiy}, \bibinfo{person}{Ambrosio Blanco},
  \bibinfo{person}{Colin Clement}, \bibinfo{person}{Dawn Drain},
  \bibinfo{person}{Daxin Jiang}, \bibinfo{person}{Duyu Tang}, {et~al\mbox{.}}}
  \bibinfo{year}{2021}\natexlab{}.
\newblock \showarticletitle{CodeXGLUE: A Machine Learning Benchmark Dataset for
  Code Understanding and Generation}. In \bibinfo{booktitle}{\emph{Thirty-fifth
  Conference on Neural Information Processing Systems Datasets and Benchmarks
  Track (Round 1)}}.
\newblock


\bibitem[\protect\citeauthoryear{Lukas, Salem, Sim, Tople, Wutschitz, and
  Zanella-B{\'e}guelin}{Lukas et~al\mbox{.}}{2023}]%
        {lukas2023analyzing}
\bibfield{author}{\bibinfo{person}{Nils Lukas}, \bibinfo{person}{Ahmed Salem},
  \bibinfo{person}{Robert Sim}, \bibinfo{person}{Shruti Tople},
  \bibinfo{person}{Lukas Wutschitz}, {and} \bibinfo{person}{Santiago
  Zanella-B{\'e}guelin}.} \bibinfo{year}{2023}\natexlab{}.
\newblock \showarticletitle{Analyzing Leakage of Personally Identifiable
  Information in Language Models}. In \bibinfo{booktitle}{\emph{2023 IEEE
  Symposium on Security and Privacy (SP)}}. IEEE Computer Society,
  \bibinfo{pages}{346--363}.
\newblock


\bibitem[\protect\citeauthoryear{Madsen, Reddy, and Chandar}{Madsen
  et~al\mbox{.}}{2022}]%
        {madsen2022post}
\bibfield{author}{\bibinfo{person}{Andreas Madsen}, \bibinfo{person}{Siva
  Reddy}, {and} \bibinfo{person}{Sarath Chandar}.}
  \bibinfo{year}{2022}\natexlab{}.
\newblock \showarticletitle{Post-hoc interpretability for neural nlp: A
  survey}.
\newblock \bibinfo{journal}{\emph{Comput. Surveys}} \bibinfo{volume}{55},
  \bibinfo{number}{8} (\bibinfo{year}{2022}), \bibinfo{pages}{1--42}.
\newblock


\bibitem[\protect\citeauthoryear{Mastropaolo, Pascarella, Guglielmi, Ciniselli,
  Scalabrino, Oliveto, and Bavota}{Mastropaolo et~al\mbox{.}}{2023}]%
        {mastropaolo2023robustness}
\bibfield{author}{\bibinfo{person}{Antonio Mastropaolo}, \bibinfo{person}{Luca
  Pascarella}, \bibinfo{person}{Emanuela Guglielmi}, \bibinfo{person}{Matteo
  Ciniselli}, \bibinfo{person}{Simone Scalabrino}, \bibinfo{person}{Rocco
  Oliveto}, {and} \bibinfo{person}{Gabriele Bavota}.}
  \bibinfo{year}{2023}\natexlab{}.
\newblock \showarticletitle{On the robustness of code generation techniques: An
  empirical study on github copilot}.
\newblock \bibinfo{journal}{\emph{arXiv preprint arXiv:2302.00438}}
  (\bibinfo{year}{2023}).
\newblock


\bibitem[\protect\citeauthoryear{Meng, Wang, Zhang, Sun, Liu, and Hu}{Meng
  et~al\mbox{.}}{2023}]%
        {meng2023template}
\bibfield{author}{\bibinfo{person}{Xiangxin Meng}, \bibinfo{person}{Xu Wang},
  \bibinfo{person}{Hongyu Zhang}, \bibinfo{person}{Hailong Sun},
  \bibinfo{person}{Xudong Liu}, {and} \bibinfo{person}{Chunming Hu}.}
  \bibinfo{year}{2023}\natexlab{}.
\newblock \showarticletitle{Template-based Neural Program Repair}. In
  \bibinfo{booktitle}{\emph{2023 IEEE/ACM 45th International Conference on
  Software Engineering (ICSE)}}. IEEE, \bibinfo{pages}{1456--1468}.
\newblock


\bibitem[\protect\citeauthoryear{Min, Ross, Sulem, Veyseh, Nguyen, Sainz,
  Agirre, Heintz, and Roth}{Min et~al\mbox{.}}{2021}]%
        {min2021recent}
\bibfield{author}{\bibinfo{person}{Bonan Min}, \bibinfo{person}{Hayley Ross},
  \bibinfo{person}{Elior Sulem}, \bibinfo{person}{Amir Pouran~Ben Veyseh},
  \bibinfo{person}{Thien~Huu Nguyen}, \bibinfo{person}{Oscar Sainz},
  \bibinfo{person}{Eneko Agirre}, \bibinfo{person}{Ilana Heintz}, {and}
  \bibinfo{person}{Dan Roth}.} \bibinfo{year}{2021}\natexlab{}.
\newblock \showarticletitle{Recent advances in natural language processing via
  large pre-trained language models: A survey}.
\newblock \bibinfo{journal}{\emph{Comput. Surveys}} (\bibinfo{year}{2021}).
\newblock


\bibitem[\protect\citeauthoryear{Mozannar, Bansal, Fourney, and
  Horvitz}{Mozannar et~al\mbox{.}}{2022}]%
        {Mozannar2022ReadingBT}
\bibfield{author}{\bibinfo{person}{Hussein Mozannar}, \bibinfo{person}{Gagan
  Bansal}, \bibinfo{person}{Adam Fourney}, {and} \bibinfo{person}{Eric
  Horvitz}.} \bibinfo{year}{2022}\natexlab{}.
\newblock \showarticletitle{Reading Between the Lines: Modeling User Behavior
  and Costs in AI-Assisted Programming}.
\newblock \bibinfo{journal}{\emph{ArXiv}}  \bibinfo{volume}{abs/2210.14306}
  (\bibinfo{year}{2022}).
\newblock
\urldef\tempurl%
\url{https://api.semanticscholar.org/CorpusID:253117056}
\showURL{%
\tempurl}


\bibitem[\protect\citeauthoryear{Mozannar, Bansal, Fourney, and
  Horvitz}{Mozannar et~al\mbox{.}}{2023}]%
        {Mozannar2023WhenTS}
\bibfield{author}{\bibinfo{person}{Hussein Mozannar}, \bibinfo{person}{Gagan
  Bansal}, \bibinfo{person}{Adam Fourney}, {and} \bibinfo{person}{Eric
  Horvitz}.} \bibinfo{year}{2023}\natexlab{}.
\newblock \showarticletitle{When to Show a Suggestion? Integrating Human
  Feedback in AI-Assisted Programming}.
\newblock \bibinfo{journal}{\emph{ArXiv}}  \bibinfo{volume}{abs/2306.04930}
  (\bibinfo{year}{2023}).
\newblock
\urldef\tempurl%
\url{https://api.semanticscholar.org/CorpusID:259108906}
\showURL{%
\tempurl}


\bibitem[\protect\citeauthoryear{Nguyen and Nadi}{Nguyen and Nadi}{2022}]%
        {nguyen2022empirical}
\bibfield{author}{\bibinfo{person}{Nhan Nguyen} {and} \bibinfo{person}{Sarah
  Nadi}.} \bibinfo{year}{2022}\natexlab{}.
\newblock \showarticletitle{An empirical evaluation of GitHub copilot's code
  suggestions}. In \bibinfo{booktitle}{\emph{Proceedings of the 19th
  International Conference on Mining Software Repositories}}.
  \bibinfo{pages}{1--5}.
\newblock


\bibitem[\protect\citeauthoryear{Nie, Li, Wang, Wang, Luo, and Wang}{Nie
  et~al\mbox{.}}{2023}]%
        {nie2023understanding}
\bibfield{author}{\bibinfo{person}{Xu Nie}, \bibinfo{person}{Ningke Li},
  \bibinfo{person}{Kailong Wang}, \bibinfo{person}{Shangguang Wang},
  \bibinfo{person}{Xiapu Luo}, {and} \bibinfo{person}{Haoyu Wang}.}
  \bibinfo{year}{2023}\natexlab{}.
\newblock \showarticletitle{Understanding and Tackling Label Errors in Deep
  Learning-Based Vulnerability Detection (Experience Paper)}. In
  \bibinfo{booktitle}{\emph{Proceedings of the 32nd ACM SIGSOFT International
  Symposium on Software Testing and Analysis}}. \bibinfo{pages}{52--63}.
\newblock


\bibitem[\protect\citeauthoryear{Nikitopoulos, Dritsa, Louridas, and
  Mitropoulos}{Nikitopoulos et~al\mbox{.}}{2021}]%
        {nikitopoulos2021crossvul}
\bibfield{author}{\bibinfo{person}{Georgios Nikitopoulos},
  \bibinfo{person}{Konstantina Dritsa}, \bibinfo{person}{Panos Louridas}, {and}
  \bibinfo{person}{Dimitris Mitropoulos}.} \bibinfo{year}{2021}\natexlab{}.
\newblock \showarticletitle{CrossVul: a cross-language vulnerability dataset
  with commit data}. In \bibinfo{booktitle}{\emph{Proceedings of the 29th ACM
  Joint Meeting on European Software Engineering Conference and Symposium on
  the Foundations of Software Engineering}}. \bibinfo{pages}{1565--1569}.
\newblock


\bibitem[\protect\citeauthoryear{Niu, Li, Ng, Chen, Ge, and Luo}{Niu
  et~al\mbox{.}}{2023b}]%
        {niu2023empirical}
\bibfield{author}{\bibinfo{person}{Changan Niu}, \bibinfo{person}{Chuanyi Li},
  \bibinfo{person}{Vincent Ng}, \bibinfo{person}{Dongxiao Chen},
  \bibinfo{person}{Jidong Ge}, {and} \bibinfo{person}{Bin Luo}.}
  \bibinfo{year}{2023}\natexlab{b}.
\newblock \showarticletitle{An empirical comparison of pre-trained models of
  source code}.
\newblock  (\bibinfo{year}{2023}), \bibinfo{pages}{2136–2148}.
\newblock
\urldef\tempurl%
\url{https://doi.org/10.1109/ICSE48619.2023.00180}
\showDOI{\tempurl}


\bibitem[\protect\citeauthoryear{Niu, Li, Ng, and Luo}{Niu
  et~al\mbox{.}}{2023a}]%
        {niu2023crosscodebench}
\bibfield{author}{\bibinfo{person}{Changan Niu}, \bibinfo{person}{Chuanyi Li},
  \bibinfo{person}{Vincent Ng}, {and} \bibinfo{person}{Bin Luo}.}
  \bibinfo{year}{2023}\natexlab{a}.
\newblock \showarticletitle{CrossCodeBench: Benchmarking Cross-Task
  Generalization of Source Code Models}.
\newblock \bibinfo{journal}{\emph{arXiv preprint arXiv:2302.04030}}
  (\bibinfo{year}{2023}).
\newblock


\bibitem[\protect\citeauthoryear{Niu, Mirza, Maradni, and P{\"o}pper}{Niu
  et~al\mbox{.}}{2023c}]%
        {niu2023codexleaks}
\bibfield{author}{\bibinfo{person}{Liang Niu}, \bibinfo{person}{Shujaat Mirza},
  \bibinfo{person}{Zayd Maradni}, {and} \bibinfo{person}{Christina
  P{\"o}pper}.} \bibinfo{year}{2023}\natexlab{c}.
\newblock \showarticletitle{$\{$CodexLeaks$\}$: Privacy Leaks from Code
  Generation Language Models in $\{$GitHub$\}$ Copilot}. In
  \bibinfo{booktitle}{\emph{32nd USENIX Security Symposium (USENIX Security
  23)}}. \bibinfo{pages}{2133--2150}.
\newblock


\bibitem[\protect\citeauthoryear{Nong, Ou, Pradel, Chen, and Cai}{Nong
  et~al\mbox{.}}{2022}]%
        {nong_generating_2022}
\bibfield{author}{\bibinfo{person}{Yu Nong}, \bibinfo{person}{Yuzhe Ou},
  \bibinfo{person}{Michael Pradel}, \bibinfo{person}{Feng Chen}, {and}
  \bibinfo{person}{Haipeng Cai}.} \bibinfo{year}{2022}\natexlab{}.
\newblock \showarticletitle{Generating realistic vulnerabilities via neural
  code editing: an empirical study}. In \bibinfo{booktitle}{\emph{Proceedings
  of the 30th {ACM} {Joint} {European} {Software} {Engineering} {Conference}
  and {Symposium} on the {Foundations} of {Software} {Engineering}}}.
  \bibinfo{publisher}{ACM}, \bibinfo{address}{Singapore Singapore},
  \bibinfo{pages}{1097--1109}.
\newblock
\showISBNx{978-1-4503-9413-0}
\urldef\tempurl%
\url{https://doi.org/10.1145/3540250.3549128}
\showDOI{\tempurl}


\bibitem[\protect\citeauthoryear{Northcutt, Jiang, and Chuang}{Northcutt
  et~al\mbox{.}}{2021}]%
        {northcutt2021confident}
\bibfield{author}{\bibinfo{person}{Curtis Northcutt}, \bibinfo{person}{Lu
  Jiang}, {and} \bibinfo{person}{Isaac Chuang}.}
  \bibinfo{year}{2021}\natexlab{}.
\newblock \showarticletitle{Confident learning: Estimating uncertainty in
  dataset labels}.
\newblock \bibinfo{journal}{\emph{Journal of Artificial Intelligence Research}}
   \bibinfo{volume}{70} (\bibinfo{year}{2021}), \bibinfo{pages}{1373--1411}.
\newblock


\bibitem[\protect\citeauthoryear{Oda, Fudaba, Neubig, Hata, Sakti, Toda, and
  Nakamura}{Oda et~al\mbox{.}}{2015}]%
        {oda2015learning}
\bibfield{author}{\bibinfo{person}{Yusuke Oda}, \bibinfo{person}{Hiroyuki
  Fudaba}, \bibinfo{person}{Graham Neubig}, \bibinfo{person}{Hideaki Hata},
  \bibinfo{person}{Sakriani Sakti}, \bibinfo{person}{Tomoki Toda}, {and}
  \bibinfo{person}{Satoshi Nakamura}.} \bibinfo{year}{2015}\natexlab{}.
\newblock \showarticletitle{Learning to generate pseudo-code from source code
  using statistical machine translation}. In \bibinfo{booktitle}{\emph{2015
  30th IEEE/ACM International Conference on Automated Software Engineering
  (ASE)}}. IEEE, \bibinfo{pages}{574--584}.
\newblock


\bibitem[\protect\citeauthoryear{Olausson, Inala, Wang, Gao, and
  Solar-Lezama}{Olausson et~al\mbox{.}}{2023}]%
        {olausson2023demystifying}
\bibfield{author}{\bibinfo{person}{Theo~X Olausson},
  \bibinfo{person}{Jeevana~Priya Inala}, \bibinfo{person}{Chenglong Wang},
  \bibinfo{person}{Jianfeng Gao}, {and} \bibinfo{person}{Armando
  Solar-Lezama}.} \bibinfo{year}{2023}\natexlab{}.
\newblock \showarticletitle{Demystifying GPT Self-Repair for Code Generation}.
\newblock \bibinfo{journal}{\emph{arXiv preprint arXiv:2306.09896}}
  (\bibinfo{year}{2023}).
\newblock


\bibitem[\protect\citeauthoryear{Oliynyk, Mayer, and Rauber}{Oliynyk
  et~al\mbox{.}}{2023}]%
        {oliynyk2023know}
\bibfield{author}{\bibinfo{person}{Daryna Oliynyk}, \bibinfo{person}{Rudolf
  Mayer}, {and} \bibinfo{person}{Andreas Rauber}.}
  \bibinfo{year}{2023}\natexlab{}.
\newblock \showarticletitle{I know what you trained last summer: A survey on
  stealing machine learning models and defences}.
\newblock \bibinfo{journal}{\emph{Comput. Surveys}} (\bibinfo{year}{2023}).
\newblock


\bibitem[\protect\citeauthoryear{OpenAI}{OpenAI}{2023}]%
        {openai2023gpt4}
\bibfield{author}{\bibinfo{person}{OpenAI}.} \bibinfo{year}{2023}\natexlab{}.
\newblock \bibinfo{title}{GPT-4 Technical Report}.
\newblock
\newblock
\showeprint[arxiv]{2303.08774}~[cs.CL]


\bibitem[\protect\citeauthoryear{Paltenghi and Pradel}{Paltenghi and
  Pradel}{2021}]%
        {paltenghi_thinking_2021}
\bibfield{author}{\bibinfo{person}{Matteo Paltenghi} {and}
  \bibinfo{person}{Michael Pradel}.} \bibinfo{year}{2021}\natexlab{}.
\newblock \showarticletitle{Thinking {Like} a {Developer}? {Comparing} the
  {Attention} of {Humans} with {Neural} {Models} of {Code}}. In
  \bibinfo{booktitle}{\emph{2021 36th {IEEE}/{ACM} {International} {Conference}
  on {Automated} {Software} {Engineering} ({ASE})}}. \bibinfo{publisher}{IEEE},
  \bibinfo{address}{Melbourne, Australia}, \bibinfo{pages}{867--879}.
\newblock
\showISBNx{978-1-66540-337-5}
\urldef\tempurl%
\url{https://doi.org/10.1109/ASE51524.2021.9678712}
\showDOI{\tempurl}


\bibitem[\protect\citeauthoryear{Pearce, Ahmad, Tan, Dolan-Gavitt, and
  Karri}{Pearce et~al\mbox{.}}{2022}]%
        {pearce_asleep_2022}
\bibfield{author}{\bibinfo{person}{Hammond Pearce}, \bibinfo{person}{Baleegh
  Ahmad}, \bibinfo{person}{Benjamin Tan}, \bibinfo{person}{Brendan
  Dolan-Gavitt}, {and} \bibinfo{person}{Ramesh Karri}.}
  \bibinfo{year}{2022}\natexlab{}.
\newblock \showarticletitle{Asleep at the keyboard? assessing the security of
  github copilot’s code contributions}. In \bibinfo{booktitle}{\emph{2022
  IEEE Symposium on Security and Privacy (SP)}}. IEEE,
  \bibinfo{pages}{754--768}.
\newblock


\bibitem[\protect\citeauthoryear{Pearce, Tan, Ahmad, Karri, and
  Dolan-Gavitt}{Pearce et~al\mbox{.}}{2023}]%
        {pearce_examining_2023}
\bibfield{author}{\bibinfo{person}{Hammond Pearce}, \bibinfo{person}{Benjamin
  Tan}, \bibinfo{person}{Baleegh Ahmad}, \bibinfo{person}{Ramesh Karri}, {and}
  \bibinfo{person}{Brendan Dolan-Gavitt}.} \bibinfo{year}{2023}\natexlab{}.
\newblock \showarticletitle{Examining {Zero}-{Shot} {Vulnerability} {Repair}
  with {Large} {Language} {Models}}. In \bibinfo{booktitle}{\emph{2023 {IEEE}
  {Symposium} on {Security} and {Privacy} ({SP})}}.
  \bibinfo{pages}{2339--2356}.
\newblock
\urldef\tempurl%
\url{https://doi.org/10.1109/SP46215.2023.10179324}
\showDOI{\tempurl}


\bibitem[\protect\citeauthoryear{Pendlebury, Pierazzi, Jordaney, Kinder, and
  Cavallaro}{Pendlebury et~al\mbox{.}}{2019}]%
        {pendlebury2019tesseract}
\bibfield{author}{\bibinfo{person}{Feargus Pendlebury}, \bibinfo{person}{Fabio
  Pierazzi}, \bibinfo{person}{Roberto Jordaney}, \bibinfo{person}{Johannes
  Kinder}, {and} \bibinfo{person}{Lorenzo Cavallaro}.}
  \bibinfo{year}{2019}\natexlab{}.
\newblock \showarticletitle{$\{$TESSERACT$\}$: Eliminating experimental bias in
  malware classification across space and time}. In
  \bibinfo{booktitle}{\emph{28th USENIX Security Symposium (USENIX Security
  19)}}. \bibinfo{pages}{729--746}.
\newblock


\bibitem[\protect\citeauthoryear{Peng, Zhang, Yang, and Stevenson}{Peng
  et~al\mbox{.}}{2022}]%
        {peng2022security}
\bibfield{author}{\bibinfo{person}{Xutan Peng}, \bibinfo{person}{Yipeng Zhang},
  \bibinfo{person}{Jingfeng Yang}, {and} \bibinfo{person}{Mark Stevenson}.}
  \bibinfo{year}{2022}\natexlab{}.
\newblock \showarticletitle{On the Security Vulnerabilities of Text-to-SQL
  Models}.
\newblock \bibinfo{journal}{\emph{arXiv preprint arXiv:2211.15363}}
  (\bibinfo{year}{2022}).
\newblock


\bibitem[\protect\citeauthoryear{Poesia, Polozov, Le, Tiwari, Soares, Meek, and
  Gulwani}{Poesia et~al\mbox{.}}{2022}]%
        {poesia2022synchromesh}
\bibfield{author}{\bibinfo{person}{Gabriel Poesia}, \bibinfo{person}{Oleksandr
  Polozov}, \bibinfo{person}{Vu Le}, \bibinfo{person}{Ashish Tiwari},
  \bibinfo{person}{Gustavo Soares}, \bibinfo{person}{Christopher Meek}, {and}
  \bibinfo{person}{Sumit Gulwani}.} \bibinfo{year}{2022}\natexlab{}.
\newblock \showarticletitle{Synchromesh: Reliable code generation from
  pre-trained language models}.
\newblock \bibinfo{journal}{\emph{arXiv preprint arXiv:2201.11227}}
  (\bibinfo{year}{2022}).
\newblock


\bibitem[\protect\citeauthoryear{Polino, Pascanu, and Alistarh}{Polino
  et~al\mbox{.}}{2018}]%
        {polino2018model}
\bibfield{author}{\bibinfo{person}{Antonio Polino}, \bibinfo{person}{Razvan
  Pascanu}, {and} \bibinfo{person}{Dan Alistarh}.}
  \bibinfo{year}{2018}\natexlab{}.
\newblock \showarticletitle{Model compression via distillation and
  quantization}. In \bibinfo{booktitle}{\emph{International Conference on
  Learning Representations}}.
\newblock


\bibitem[\protect\citeauthoryear{Raatikainen, Tiihonen, and
  M{\"a}nnist{\"o}}{Raatikainen et~al\mbox{.}}{2019}]%
        {raatikainen2019software}
\bibfield{author}{\bibinfo{person}{Mikko Raatikainen}, \bibinfo{person}{Juha
  Tiihonen}, {and} \bibinfo{person}{Tomi M{\"a}nnist{\"o}}.}
  \bibinfo{year}{2019}\natexlab{}.
\newblock \showarticletitle{Software product lines and variability modeling: A
  tertiary study}.
\newblock \bibinfo{journal}{\emph{Journal of Systems and Software}}
  \bibinfo{volume}{149} (\bibinfo{year}{2019}), \bibinfo{pages}{485--510}.
\newblock


\bibitem[\protect\citeauthoryear{Rabin, Bui, Wang, Yu, Jiang, and
  Alipour}{Rabin et~al\mbox{.}}{2021}]%
        {rabin2021generalizability}
\bibfield{author}{\bibinfo{person}{Md~Rafiqul~Islam Rabin},
  \bibinfo{person}{Nghi~DQ Bui}, \bibinfo{person}{Ke Wang},
  \bibinfo{person}{Yijun Yu}, \bibinfo{person}{Lingxiao Jiang}, {and}
  \bibinfo{person}{Mohammad~Amin Alipour}.} \bibinfo{year}{2021}\natexlab{}.
\newblock \showarticletitle{On the generalizability of Neural Program Models
  with respect to semantic-preserving program transformations}.
\newblock \bibinfo{journal}{\emph{Information and Software Technology}}
  \bibinfo{volume}{135} (\bibinfo{year}{2021}), \bibinfo{pages}{106552}.
\newblock


\bibitem[\protect\citeauthoryear{Ren, Guo, Lu, Zhou, Liu, Tang, Sundaresan,
  Zhou, Blanco, and Ma}{Ren et~al\mbox{.}}{2020}]%
        {ren2020codebleu}
\bibfield{author}{\bibinfo{person}{Shuo Ren}, \bibinfo{person}{Daya Guo},
  \bibinfo{person}{Shuai Lu}, \bibinfo{person}{Long Zhou},
  \bibinfo{person}{Shujie Liu}, \bibinfo{person}{Duyu Tang},
  \bibinfo{person}{Neel Sundaresan}, \bibinfo{person}{Ming Zhou},
  \bibinfo{person}{Ambrosio Blanco}, {and} \bibinfo{person}{Shuai Ma}.}
  \bibinfo{year}{2020}\natexlab{}.
\newblock \showarticletitle{Codebleu: a method for automatic evaluation of code
  synthesis}.
\newblock \bibinfo{journal}{\emph{arXiv preprint arXiv:2009.10297}}
  (\bibinfo{year}{2020}).
\newblock


\bibitem[\protect\citeauthoryear{Research and of~Australasia}{Research and
  of~Australasia}{2023}]%
        {aucore2023ranking}
\bibfield{author}{\bibinfo{person}{The~Computing Research} {and}
  \bibinfo{person}{Education~Association of Australasia}.}
  \bibinfo{year}{2023}\natexlab{}.
\newblock \bibinfo{title}{CORE Rankings Portal}.
\newblock \bibinfo{howpublished}{\url{
  https://www.core.edu.au/conference-portal}}.
\newblock


\bibitem[\protect\citeauthoryear{Roy, Fakhoury, and Arnaoudova}{Roy
  et~al\mbox{.}}{2021}]%
        {roy2021reassessing}
\bibfield{author}{\bibinfo{person}{Devjeet Roy}, \bibinfo{person}{Sarah
  Fakhoury}, {and} \bibinfo{person}{Venera Arnaoudova}.}
  \bibinfo{year}{2021}\natexlab{}.
\newblock \showarticletitle{Reassessing automatic evaluation metrics for code
  summarization tasks}. In \bibinfo{booktitle}{\emph{Proceedings of the 29th
  ACM Joint Meeting on European Software Engineering Conference and Symposium
  on the Foundations of Software Engineering}}. \bibinfo{pages}{1105--1116}.
\newblock


\bibitem[\protect\citeauthoryear{Scalabrino, Bavota, Vendome,
  Linares-V{\'a}squez, Poshyvanyk, and Oliveto}{Scalabrino
  et~al\mbox{.}}{2017}]%
        {scalabrino2017automatically}
\bibfield{author}{\bibinfo{person}{Simone Scalabrino},
  \bibinfo{person}{Gabriele Bavota}, \bibinfo{person}{Christopher Vendome},
  \bibinfo{person}{Mario Linares-V{\'a}squez}, \bibinfo{person}{Denys
  Poshyvanyk}, {and} \bibinfo{person}{Rocco Oliveto}.}
  \bibinfo{year}{2017}\natexlab{}.
\newblock \showarticletitle{Automatically assessing code understandability: How
  far are we?}. In \bibinfo{booktitle}{\emph{2017 32nd IEEE/ACM International
  Conference on Automated Software Engineering (ASE)}}. IEEE,
  \bibinfo{pages}{417--427}.
\newblock


\bibitem[\protect\citeauthoryear{Schuster, Song, Tromer, and
  Shmatikov}{Schuster et~al\mbox{.}}{2021}]%
        {schuster2021you}
\bibfield{author}{\bibinfo{person}{Roei Schuster}, \bibinfo{person}{Congzheng
  Song}, \bibinfo{person}{Eran Tromer}, {and} \bibinfo{person}{Vitaly
  Shmatikov}.} \bibinfo{year}{2021}\natexlab{}.
\newblock \showarticletitle{You autocomplete me: Poisoning vulnerabilities in
  neural code completion}. In \bibinfo{booktitle}{\emph{30th USENIX Security
  Symposium (USENIX Security 21)}}. \bibinfo{pages}{1559--1575}.
\newblock


\bibitem[\protect\citeauthoryear{Schwarzkopf, Murray, and Hand}{Schwarzkopf
  et~al\mbox{.}}{2012}]%
        {schwarzkopf2012seven}
\bibfield{author}{\bibinfo{person}{Malte Schwarzkopf}, \bibinfo{person}{Derek~G
  Murray}, {and} \bibinfo{person}{Steven Hand}.}
  \bibinfo{year}{2012}\natexlab{}.
\newblock \showarticletitle{The seven deadly sins of cloud computing research}.
  In \bibinfo{booktitle}{\emph{4th USENIX Workshop on Hot Topics in Cloud
  Computing (HotCloud 12)}}.
\newblock


\bibitem[\protect\citeauthoryear{Shahbazi, Lin, Asudeh, and Jagadish}{Shahbazi
  et~al\mbox{.}}{2023}]%
        {shahbazi2023representation}
\bibfield{author}{\bibinfo{person}{Nima Shahbazi}, \bibinfo{person}{Yin Lin},
  \bibinfo{person}{Abolfazl Asudeh}, {and} \bibinfo{person}{HV Jagadish}.}
  \bibinfo{year}{2023}\natexlab{}.
\newblock \showarticletitle{Representation Bias in Data: A Survey on
  Identification and Resolution Techniques}.
\newblock \bibinfo{journal}{\emph{Comput. Surveys}} (\bibinfo{year}{2023}).
\newblock


\bibitem[\protect\citeauthoryear{Shen, Zhu, Dong, Guo, Zhen, and Li}{Shen
  et~al\mbox{.}}{2022}]%
        {shen_incorporating_2022}
\bibfield{author}{\bibinfo{person}{Sijie Shen}, \bibinfo{person}{Xiang Zhu},
  \bibinfo{person}{Yihong Dong}, \bibinfo{person}{Qizhi Guo},
  \bibinfo{person}{Yankun Zhen}, {and} \bibinfo{person}{Ge Li}.}
  \bibinfo{year}{2022}\natexlab{}.
\newblock \showarticletitle{Incorporating domain knowledge through task
  augmentation for front-end {JavaScript} code generation}. In
  \bibinfo{booktitle}{\emph{Proceedings of the 30th {ACM} {Joint} {European}
  {Software} {Engineering} {Conference} and {Symposium} on the {Foundations} of
  {Software} {Engineering}}} \emph{(\bibinfo{series}{{ESEC}/{FSE} 2022})}.
  \bibinfo{publisher}{Association for Computing Machinery},
  \bibinfo{address}{Singapore Singapore}, \bibinfo{pages}{1533--1543}.
\newblock
\showISBNx{978-1-4503-9413-0}
\urldef\tempurl%
\url{https://doi.org/10.1145/3540250.3558965}
\showDOI{\tempurl}


\bibitem[\protect\citeauthoryear{Shi, Wang, Du, Chen, Han, Zhang, Zhang, and
  Sun}{Shi et~al\mbox{.}}{2022b}]%
        {shi2022evaluation}
\bibfield{author}{\bibinfo{person}{Ensheng Shi}, \bibinfo{person}{Yanlin Wang},
  \bibinfo{person}{Lun Du}, \bibinfo{person}{Junjie Chen}, \bibinfo{person}{Shi
  Han}, \bibinfo{person}{Hongyu Zhang}, \bibinfo{person}{Dongmei Zhang}, {and}
  \bibinfo{person}{Hongbin Sun}.} \bibinfo{year}{2022}\natexlab{b}.
\newblock \showarticletitle{On the evaluation of neural code summarization}. In
  \bibinfo{booktitle}{\emph{Proceedings of the 44th International Conference on
  Software Engineering}}. \bibinfo{pages}{1597--1608}.
\newblock


\bibitem[\protect\citeauthoryear{Shi, Wang, Zhang, Du, Han, Zhang, and Sun}{Shi
  et~al\mbox{.}}{2023a}]%
        {shi2023towards}
\bibfield{author}{\bibinfo{person}{Ensheng Shi}, \bibinfo{person}{Yanlin Wang},
  \bibinfo{person}{Hongyu Zhang}, \bibinfo{person}{Lun Du},
  \bibinfo{person}{Shi Han}, \bibinfo{person}{Dongmei Zhang}, {and}
  \bibinfo{person}{Hongbin Sun}.} \bibinfo{year}{2023}\natexlab{a}.
\newblock \showarticletitle{Towards Efficient Fine-tuning of Pre-trained Code
  Models: An Experimental Study and Beyond}.
\newblock \bibinfo{journal}{\emph{arXiv preprint arXiv:2304.05216}}
  (\bibinfo{year}{2023}).
\newblock


\bibitem[\protect\citeauthoryear{Shi, Yang, Kang, Xu, He, and Lo}{Shi
  et~al\mbox{.}}{2023b}]%
        {shi2023smaller}
\bibfield{author}{\bibinfo{person}{Jieke Shi}, \bibinfo{person}{Zhou Yang},
  \bibinfo{person}{Hong~Jin Kang}, \bibinfo{person}{Bowen Xu},
  \bibinfo{person}{Junda He}, {and} \bibinfo{person}{David Lo}.}
  \bibinfo{year}{2023}\natexlab{b}.
\newblock \showarticletitle{Smaller, Faster, Greener: Compressing Pre-trained
  Code Models via Surrogate-Assisted Optimization}.
\newblock \bibinfo{journal}{\emph{arXiv preprint arXiv:2309.04076}}
  (\bibinfo{year}{2023}).
\newblock


\bibitem[\protect\citeauthoryear{Shi, Yang, Xu, Kang, and Lo}{Shi
  et~al\mbox{.}}{2022c}]%
        {shi2022compressing}
\bibfield{author}{\bibinfo{person}{Jieke Shi}, \bibinfo{person}{Zhou Yang},
  \bibinfo{person}{Bowen Xu}, \bibinfo{person}{Hong~Jin Kang}, {and}
  \bibinfo{person}{David Lo}.} \bibinfo{year}{2022}\natexlab{c}.
\newblock \showarticletitle{Compressing pre-trained models of code into 3 mb}.
  In \bibinfo{booktitle}{\emph{Proceedings of the 37th IEEE/ACM International
  Conference on Automated Software Engineering}}. \bibinfo{pages}{1--12}.
\newblock


\bibitem[\protect\citeauthoryear{Shi, Mu, Chen, Wang, Wang, Yang, Li, Xia, and
  Wang}{Shi et~al\mbox{.}}{2022a}]%
        {shi2022we}
\bibfield{author}{\bibinfo{person}{Lin Shi}, \bibinfo{person}{Fangwen Mu},
  \bibinfo{person}{Xiao Chen}, \bibinfo{person}{Song Wang},
  \bibinfo{person}{Junjie Wang}, \bibinfo{person}{Ye Yang}, \bibinfo{person}{Ge
  Li}, \bibinfo{person}{Xin Xia}, {and} \bibinfo{person}{Qing Wang}.}
  \bibinfo{year}{2022}\natexlab{a}.
\newblock \showarticletitle{Are we building on the rock? on the importance of
  data preprocessing for code summarization}. In
  \bibinfo{booktitle}{\emph{Proceedings of the 30th ACM Joint European Software
  Engineering Conference and Symposium on the Foundations of Software
  Engineering}}. \bibinfo{pages}{107--119}.
\newblock


\bibitem[\protect\citeauthoryear{Steenhoek, Rahman, Jiles, and Le}{Steenhoek
  et~al\mbox{.}}{2023}]%
        {steenhoek2023empirical}
\bibfield{author}{\bibinfo{person}{Benjamin Steenhoek},
  \bibinfo{person}{Md~Mahbubur Rahman}, \bibinfo{person}{Richard Jiles}, {and}
  \bibinfo{person}{Wei Le}.} \bibinfo{year}{2023}\natexlab{}.
\newblock \showarticletitle{An empirical study of deep learning models for
  vulnerability detection}. In \bibinfo{booktitle}{\emph{2023 IEEE/ACM 45th
  International Conference on Software Engineering (ICSE)}}. IEEE,
  \bibinfo{pages}{2237--2248}.
\newblock


\bibitem[\protect\citeauthoryear{Sun, Chen, Tao, Fang, Zhang, Zhang, and
  Luo}{Sun et~al\mbox{.}}{2023a}]%
        {sun2023backdooring}
\bibfield{author}{\bibinfo{person}{Weisong Sun}, \bibinfo{person}{Yuchen Chen},
  \bibinfo{person}{Guanhong Tao}, \bibinfo{person}{Chunrong Fang},
  \bibinfo{person}{Xiangyu Zhang}, \bibinfo{person}{Quanjun Zhang}, {and}
  \bibinfo{person}{Bin Luo}.} \bibinfo{year}{2023}\natexlab{a}.
\newblock \showarticletitle{Backdooring Neural Code Search}.
\newblock \bibinfo{journal}{\emph{arXiv preprint arXiv:2305.17506}}
  (\bibinfo{year}{2023}).
\newblock


\bibitem[\protect\citeauthoryear{Sun, Du, Song, and Li}{Sun
  et~al\mbox{.}}{2023b}]%
        {sun2023codemark}
\bibfield{author}{\bibinfo{person}{Zhensu Sun}, \bibinfo{person}{Xiaoning Du},
  \bibinfo{person}{Fu Song}, {and} \bibinfo{person}{Li Li}.}
  \bibinfo{year}{2023}\natexlab{b}.
\newblock \showarticletitle{CodeMark: Imperceptible Watermarking for Code
  Datasets against Neural Code Completion Models}.
\newblock \bibinfo{journal}{\emph{arXiv preprint arXiv:2308.14401}}
  (\bibinfo{year}{2023}).
\newblock


\bibitem[\protect\citeauthoryear{Sun, Du, Song, Ni, and Li}{Sun
  et~al\mbox{.}}{2022a}]%
        {sun2022coprotector}
\bibfield{author}{\bibinfo{person}{Zhensu Sun}, \bibinfo{person}{Xiaoning Du},
  \bibinfo{person}{Fu Song}, \bibinfo{person}{Mingze Ni}, {and}
  \bibinfo{person}{Li Li}.} \bibinfo{year}{2022}\natexlab{a}.
\newblock \showarticletitle{Coprotector: Protect open-source code against
  unauthorized training usage with data poisoning}. In
  \bibinfo{booktitle}{\emph{Proceedings of the ACM Web Conference 2022}}.
  \bibinfo{pages}{652--660}.
\newblock


\bibitem[\protect\citeauthoryear{Sun, Li, Liu, Du, and Li}{Sun
  et~al\mbox{.}}{2022b}]%
        {sun_importance_2022}
\bibfield{author}{\bibinfo{person}{Zhensu Sun}, \bibinfo{person}{Li Li},
  \bibinfo{person}{Yan Liu}, \bibinfo{person}{Xiaoning Du}, {and}
  \bibinfo{person}{Li Li}.} \bibinfo{year}{2022}\natexlab{b}.
\newblock \showarticletitle{On the importance of building high-quality training
  datasets for neural code search}. In \bibinfo{booktitle}{\emph{Proceedings of
  the 44th {International} {Conference} on {Software} {Engineering}}}.
  \bibinfo{publisher}{ACM}, \bibinfo{address}{Pittsburgh Pennsylvania},
  \bibinfo{pages}{1609--1620}.
\newblock
\showISBNx{978-1-4503-9221-1}
\urldef\tempurl%
\url{https://doi.org/10.1145/3510003.3510160}
\showDOI{\tempurl}


\bibitem[\protect\citeauthoryear{Svyatkovskiy, Deng, Fu, and
  Sundaresan}{Svyatkovskiy et~al\mbox{.}}{2020}]%
        {svyatkovskiy2020intellicode}
\bibfield{author}{\bibinfo{person}{Alexey Svyatkovskiy},
  \bibinfo{person}{Shao~Kun Deng}, \bibinfo{person}{Shengyu Fu}, {and}
  \bibinfo{person}{Neel Sundaresan}.} \bibinfo{year}{2020}\natexlab{}.
\newblock \showarticletitle{Intellicode compose: Code generation using
  transformer}. In \bibinfo{booktitle}{\emph{Proceedings of the 28th ACM Joint
  Meeting on European Software Engineering Conference and Symposium on the
  Foundations of Software Engineering}}. \bibinfo{pages}{1433--1443}.
\newblock


\bibitem[\protect\citeauthoryear{Svyatkovskiy, Lee, Hadjitofi, Riechert,
  Franco, and Allamanis}{Svyatkovskiy et~al\mbox{.}}{2021}]%
        {svyatkovskiy2021fast}
\bibfield{author}{\bibinfo{person}{Alexey Svyatkovskiy},
  \bibinfo{person}{Sebastian Lee}, \bibinfo{person}{Anna Hadjitofi},
  \bibinfo{person}{Maik Riechert}, \bibinfo{person}{Juliana~Vicente Franco},
  {and} \bibinfo{person}{Miltiadis Allamanis}.}
  \bibinfo{year}{2021}\natexlab{}.
\newblock \showarticletitle{Fast and memory-efficient neural code completion}.
  In \bibinfo{booktitle}{\emph{2021 IEEE/ACM 18th International Conference on
  Mining Software Repositories (MSR)}}. IEEE, \bibinfo{pages}{329--340}.
\newblock


\bibitem[\protect\citeauthoryear{Tantithamthavorn, Cito, Hemmati, and
  Chandra}{Tantithamthavorn et~al\mbox{.}}{2023}]%
        {10109341}
\bibfield{author}{\bibinfo{person}{Chakkrit Tantithamthavorn},
  \bibinfo{person}{Jürgen Cito}, \bibinfo{person}{Hadi Hemmati}, {and}
  \bibinfo{person}{Satish Chandra}.} \bibinfo{year}{2023}\natexlab{}.
\newblock \showarticletitle{Explainable AI for SE: Challenges and Future
  Directions}.
\newblock \bibinfo{journal}{\emph{IEEE Software}} \bibinfo{volume}{40},
  \bibinfo{number}{3} (\bibinfo{year}{2023}), \bibinfo{pages}{29--33}.
\newblock
\urldef\tempurl%
\url{https://doi.org/10.1109/MS.2023.3246686}
\showDOI{\tempurl}


\bibitem[\protect\citeauthoryear{Tantithamthavorn and Hassan}{Tantithamthavorn
  and Hassan}{2018}]%
        {DBLP:conf/icse/Tantithamthavorn18}
\bibfield{author}{\bibinfo{person}{Chakkrit Tantithamthavorn} {and}
  \bibinfo{person}{Ahmed~E. Hassan}.} \bibinfo{year}{2018}\natexlab{}.
\newblock \showarticletitle{An experience report on defect modelling in
  practice: pitfalls and challenges}. In \bibinfo{booktitle}{\emph{Proceedings
  of the 40th International Conference on Software Engineering: Software
  Engineering in Practice, {ICSE} {(SEIP)} 2018, Gothenburg, Sweden, May 27 -
  June 03, 2018}}, \bibfield{editor}{\bibinfo{person}{Frances Paulisch} {and}
  \bibinfo{person}{Jan Bosch}} (Eds.). \bibinfo{publisher}{{ACM}},
  \bibinfo{pages}{286--295}.
\newblock
\urldef\tempurl%
\url{https://doi.org/10.1145/3183519.3183547}
\showDOI{\tempurl}


\bibitem[\protect\citeauthoryear{Tantithamthavorn and
  Jiarpakdee}{Tantithamthavorn and Jiarpakdee}{2021}]%
        {DBLP:conf/kbse/Tantithamthavorn21}
\bibfield{author}{\bibinfo{person}{Chakkrit Tantithamthavorn} {and}
  \bibinfo{person}{Jirayus Jiarpakdee}.} \bibinfo{year}{2021}\natexlab{}.
\newblock \showarticletitle{Explainable {AI} for Software Engineering}. In
  \bibinfo{booktitle}{\emph{36th {IEEE/ACM} International Conference on
  Automated Software Engineering, {ASE} 2021, Melbourne, Australia, November
  15-19, 2021}}. \bibinfo{publisher}{{IEEE}}, \bibinfo{pages}{1--2}.
\newblock
\urldef\tempurl%
\url{https://doi.org/10.1109/ASE51524.2021.9678580}
\showDOI{\tempurl}


\bibitem[\protect\citeauthoryear{Tantithamthavorn, McIntosh, Hassan, Ihara, and
  Matsumoto}{Tantithamthavorn et~al\mbox{.}}{2015}]%
        {tantithamthavorn2015impact}
\bibfield{author}{\bibinfo{person}{Chakkrit Tantithamthavorn},
  \bibinfo{person}{Shane McIntosh}, \bibinfo{person}{Ahmed~E Hassan},
  \bibinfo{person}{Akinori Ihara}, {and} \bibinfo{person}{Kenichi Matsumoto}.}
  \bibinfo{year}{2015}\natexlab{}.
\newblock \showarticletitle{The impact of mislabelling on the performance and
  interpretation of defect prediction models}. In
  \bibinfo{booktitle}{\emph{2015 IEEE/ACM 37th IEEE International Conference on
  Software Engineering}}, Vol.~\bibinfo{volume}{1}. IEEE,
  \bibinfo{pages}{812--823}.
\newblock


\bibitem[\protect\citeauthoryear{Thongtanunam, Pornprasit, and
  Tantithamthavorn}{Thongtanunam et~al\mbox{.}}{2022}]%
        {thongtanunam_autotransform_2022}
\bibfield{author}{\bibinfo{person}{Patanamon Thongtanunam},
  \bibinfo{person}{Chanathip Pornprasit}, {and} \bibinfo{person}{Chakkrit
  Tantithamthavorn}.} \bibinfo{year}{2022}\natexlab{}.
\newblock \showarticletitle{{AutoTransform}: automated code transformation to
  support modern code review process}. In \bibinfo{booktitle}{\emph{Proceedings
  of the 44th {International} {Conference} on {Software} {Engineering}}}
  \emph{(\bibinfo{series}{{ICSE} '22})}. \bibinfo{publisher}{Association for
  Computing Machinery}, \bibinfo{address}{Pittsburgh Pennsylvania},
  \bibinfo{pages}{237--248}.
\newblock
\showISBNx{978-1-4503-9221-1}
\urldef\tempurl%
\url{https://doi.org/10.1145/3510003.3510067}
\showDOI{\tempurl}


\bibitem[\protect\citeauthoryear{Tian, Cui, Liang, and Yu}{Tian
  et~al\mbox{.}}{2022}]%
        {tian2022comprehensive}
\bibfield{author}{\bibinfo{person}{Zhiyi Tian}, \bibinfo{person}{Lei Cui},
  \bibinfo{person}{Jie Liang}, {and} \bibinfo{person}{Shui Yu}.}
  \bibinfo{year}{2022}\natexlab{}.
\newblock \showarticletitle{A comprehensive survey on poisoning attacks and
  countermeasures in machine learning}.
\newblock \bibinfo{journal}{\emph{Comput. Surveys}} \bibinfo{volume}{55},
  \bibinfo{number}{8} (\bibinfo{year}{2022}), \bibinfo{pages}{1--35}.
\newblock


\bibitem[\protect\citeauthoryear{van~der Kouwe, Heiser, Andriesse, Bos, and
  Giuffrida}{van~der Kouwe et~al\mbox{.}}{2019}]%
        {van2019sok}
\bibfield{author}{\bibinfo{person}{Erik van~der Kouwe}, \bibinfo{person}{Gernot
  Heiser}, \bibinfo{person}{Dennis Andriesse}, \bibinfo{person}{Herbert Bos},
  {and} \bibinfo{person}{Cristiano Giuffrida}.}
  \bibinfo{year}{2019}\natexlab{}.
\newblock \showarticletitle{SoK: Benchmarking flaws in systems security}. In
  \bibinfo{booktitle}{\emph{2019 IEEE European Symposium on Security and
  Privacy (EuroS\&P)}}. IEEE, \bibinfo{pages}{310--325}.
\newblock


\bibitem[\protect\citeauthoryear{Vaswani, Shazeer, Parmar, Uszkoreit, Jones,
  Gomez, Kaiser, and Polosukhin}{Vaswani et~al\mbox{.}}{2017}]%
        {vaswani2017attention}
\bibfield{author}{\bibinfo{person}{Ashish Vaswani}, \bibinfo{person}{Noam
  Shazeer}, \bibinfo{person}{Niki Parmar}, \bibinfo{person}{Jakob Uszkoreit},
  \bibinfo{person}{Llion Jones}, \bibinfo{person}{Aidan~N Gomez},
  \bibinfo{person}{{\L}ukasz Kaiser}, {and} \bibinfo{person}{Illia
  Polosukhin}.} \bibinfo{year}{2017}\natexlab{}.
\newblock \showarticletitle{Attention is all you need}.
\newblock \bibinfo{journal}{\emph{Advances in neural informfation processing
  systems}}  \bibinfo{volume}{30} (\bibinfo{year}{2017}).
\newblock


\bibitem[\protect\citeauthoryear{Wan, Zhang, Zhang, Sui, Xu, Yao, Jin, and
  Sun}{Wan et~al\mbox{.}}{2022a}]%
        {wan_you_2022}
\bibfield{author}{\bibinfo{person}{Yao Wan}, \bibinfo{person}{Shijie Zhang},
  \bibinfo{person}{Hongyu Zhang}, \bibinfo{person}{Yulei Sui},
  \bibinfo{person}{Guandong Xu}, \bibinfo{person}{Dezhong Yao},
  \bibinfo{person}{Hai Jin}, {and} \bibinfo{person}{Lichao Sun}.}
  \bibinfo{year}{2022}\natexlab{a}.
\newblock \showarticletitle{You see what {I} want you to see: poisoning
  vulnerabilities in neural code search}. In
  \bibinfo{booktitle}{\emph{Proceedings of the 30th {ACM} {Joint} {European}
  {Software} {Engineering} {Conference} and {Symposium} on the {Foundations} of
  {Software} {Engineering}}} \emph{(\bibinfo{series}{{ESEC}/{FSE} 2022})}.
  \bibinfo{publisher}{Association for Computing Machinery},
  \bibinfo{address}{Singapore Singapore}, \bibinfo{pages}{1233--1245}.
\newblock
\showISBNx{978-1-4503-9413-0}
\urldef\tempurl%
\url{https://doi.org/10.1145/3540250.3549153}
\showDOI{\tempurl}


\bibitem[\protect\citeauthoryear{Wan, Zhao, Zhang, Sui, Xu, and Jin}{Wan
  et~al\mbox{.}}{2022b}]%
        {wan_what_2022}
\bibfield{author}{\bibinfo{person}{Yao Wan}, \bibinfo{person}{Wei Zhao},
  \bibinfo{person}{Hongyu Zhang}, \bibinfo{person}{Yulei Sui},
  \bibinfo{person}{Guandong Xu}, {and} \bibinfo{person}{Hai Jin}.}
  \bibinfo{year}{2022}\natexlab{b}.
\newblock \showarticletitle{What do they capture?: a structural analysis of
  pre-trained language models for source code}. In
  \bibinfo{booktitle}{\emph{Proceedings of the 44th {International}
  {Conference} on {Software} {Engineering}}} \emph{(\bibinfo{series}{{ICSE}
  '22})}. \bibinfo{publisher}{Association for Computing Machinery},
  \bibinfo{address}{Pittsburgh Pennsylvania}, \bibinfo{pages}{2377--2388}.
\newblock
\showISBNx{978-1-4503-9221-1}
\urldef\tempurl%
\url{https://doi.org/10.1145/3510003.3510050}
\showDOI{\tempurl}


\bibitem[\protect\citeauthoryear{Wan, Zhao, Yang, Xu, Ying, Wu, and Yu}{Wan
  et~al\mbox{.}}{2018}]%
        {wan_improving_2018}
\bibfield{author}{\bibinfo{person}{Yao Wan}, \bibinfo{person}{Zhou Zhao},
  \bibinfo{person}{Min Yang}, \bibinfo{person}{Guandong Xu},
  \bibinfo{person}{Haochao Ying}, \bibinfo{person}{Jian Wu}, {and}
  \bibinfo{person}{Philip~S. Yu}.} \bibinfo{year}{2018}\natexlab{}.
\newblock \showarticletitle{Improving automatic source code summarization via
  deep reinforcement learning}. In \bibinfo{booktitle}{\emph{Proceedings of the
  33rd {ACM}/{IEEE} {International} {Conference} on {Automated} {Software}
  {Engineering}}} \emph{(\bibinfo{series}{{ASE} '18})}.
  \bibinfo{publisher}{Association for Computing Machinery},
  \bibinfo{address}{Montpellier France}, \bibinfo{pages}{397--407}.
\newblock
\showISBNx{978-1-4503-5937-5}
\urldef\tempurl%
\url{https://doi.org/10.1145/3238147.3238206}
\showDOI{\tempurl}


\bibitem[\protect\citeauthoryear{Wang, Jia, Li, Yu, Xiong, Dong, and Liao}{Wang
  et~al\mbox{.}}{2022b}]%
        {wang2022bridging}
\bibfield{author}{\bibinfo{person}{Deze Wang}, \bibinfo{person}{Zhouyang Jia},
  \bibinfo{person}{Shanshan Li}, \bibinfo{person}{Yue Yu}, \bibinfo{person}{Yun
  Xiong}, \bibinfo{person}{Wei Dong}, {and} \bibinfo{person}{Xiangke Liao}.}
  \bibinfo{year}{2022}\natexlab{b}.
\newblock \showarticletitle{Bridging pre-trained models and downstream tasks
  for source code understanding}. In \bibinfo{booktitle}{\emph{Proceedings of
  the 44th International Conference on Software Engineering}}.
  \bibinfo{pages}{287--298}.
\newblock


\bibitem[\protect\citeauthoryear{Wang, Ma, Yuan, Liu, Wang, Tang, Nie, and
  Wu}{Wang et~al\mbox{.}}{2023d}]%
        {wang_enhancing_2023}
\bibfield{author}{\bibinfo{person}{Huaijin Wang}, \bibinfo{person}{Pingchuan
  Ma}, \bibinfo{person}{Yuanyuan Yuan}, \bibinfo{person}{Zhibo Liu},
  \bibinfo{person}{Shuai Wang}, \bibinfo{person}{Qiyi Tang},
  \bibinfo{person}{Sen Nie}, {and} \bibinfo{person}{Shi Wu}.}
  \bibinfo{year}{2023}\natexlab{d}.
\newblock \showarticletitle{Enhancing {DNN}-{Based} {Binary} {Code} {Function}
  {Search} {With} {Low}-{Cost} {Equivalence} {Checking}}.
\newblock \bibinfo{journal}{\emph{IIEEE Trans. Software Eng.}}
  \bibinfo{volume}{49}, \bibinfo{number}{1} (\bibinfo{date}{Jan.}
  \bibinfo{year}{2023}), \bibinfo{pages}{226--250}.
\newblock
\showISSN{0098-5589, 1939-3520, 2326-3881}
\urldef\tempurl%
\url{https://doi.org/10.1109/TSE.2022.3149240}
\showDOI{\tempurl}


\bibitem[\protect\citeauthoryear{Wang, Hu, Hou, Chen, Zheng, Wang, Yang, Huang,
  Ye, Geng, et~al\mbox{.}}{Wang et~al\mbox{.}}{2023a}]%
        {wang2023robustness}
\bibfield{author}{\bibinfo{person}{Jindong Wang}, \bibinfo{person}{Xixu Hu},
  \bibinfo{person}{Wenxin Hou}, \bibinfo{person}{Hao Chen},
  \bibinfo{person}{Runkai Zheng}, \bibinfo{person}{Yidong Wang},
  \bibinfo{person}{Linyi Yang}, \bibinfo{person}{Haojun Huang},
  \bibinfo{person}{Wei Ye}, \bibinfo{person}{Xiubo Geng}, {et~al\mbox{.}}}
  \bibinfo{year}{2023}\natexlab{a}.
\newblock \showarticletitle{On the robustness of chatgpt: An adversarial and
  out-of-distribution perspective}.
\newblock \bibinfo{journal}{\emph{arXiv preprint arXiv:2302.12095}}
  (\bibinfo{year}{2023}).
\newblock


\bibitem[\protect\citeauthoryear{Wang, Huang, Chen, Liu, Wang, and Wang}{Wang
  et~al\mbox{.}}{2023b}]%
        {wang2023software}
\bibfield{author}{\bibinfo{person}{Junjie Wang}, \bibinfo{person}{Yuchao
  Huang}, \bibinfo{person}{Chunyang Chen}, \bibinfo{person}{Zhe Liu},
  \bibinfo{person}{Song Wang}, {and} \bibinfo{person}{Qing Wang}.}
  \bibinfo{year}{2023}\natexlab{b}.
\newblock \showarticletitle{Software Testing with Large Language Model: Survey,
  Landscape, and Vision}.
\newblock \bibinfo{journal}{\emph{arXiv preprint arXiv:2307.07221}}
  (\bibinfo{year}{2023}).
\newblock


\bibitem[\protect\citeauthoryear{Wang, Huang, Gao, Ge, Zhang, Feng, Satyarth,
  Li, Zhang, and Ng}{Wang et~al\mbox{.}}{2022a}]%
        {wang2022machine}
\bibfield{author}{\bibinfo{person}{Simin Wang}, \bibinfo{person}{Liguo Huang},
  \bibinfo{person}{Amiao Gao}, \bibinfo{person}{Jidong Ge},
  \bibinfo{person}{Tengfei Zhang}, \bibinfo{person}{Haitao Feng},
  \bibinfo{person}{Ishna Satyarth}, \bibinfo{person}{Ming Li},
  \bibinfo{person}{He Zhang}, {and} \bibinfo{person}{Vincent Ng}.}
  \bibinfo{year}{2022}\natexlab{a}.
\newblock \showarticletitle{Machine/deep learning for software engineering: A
  systematic literature review}.
\newblock \bibinfo{journal}{\emph{IEEE Transactions on Software Engineering}}
  \bibinfo{volume}{49}, \bibinfo{number}{3} (\bibinfo{year}{2022}),
  \bibinfo{pages}{1188--1231}.
\newblock


\bibitem[\protect\citeauthoryear{Wang, Huang, Gao, Ge, Zhang, Feng, Satyarth,
  Li, Zhang, and Ng}{Wang et~al\mbox{.}}{2023c}]%
        {wang_machinedeep_2023}
\bibfield{author}{\bibinfo{person}{Simin Wang}, \bibinfo{person}{Liguo Huang},
  \bibinfo{person}{Amiao Gao}, \bibinfo{person}{Jidong Ge},
  \bibinfo{person}{Tengfei Zhang}, \bibinfo{person}{Haitao Feng},
  \bibinfo{person}{Ishna Satyarth}, \bibinfo{person}{Ming Li},
  \bibinfo{person}{He Zhang}, {and} \bibinfo{person}{Vincent Ng}.}
  \bibinfo{year}{2023}\natexlab{c}.
\newblock \showarticletitle{Machine/{Deep} {Learning} for {Software}
  {Engineering}: {A} {Systematic} {Literature} {Review}}.
\newblock \bibinfo{journal}{\emph{IEEE Transactions on Software Engineering}}
  \bibinfo{volume}{49}, \bibinfo{number}{3} (\bibinfo{date}{March}
  \bibinfo{year}{2023}), \bibinfo{pages}{1188--1231}.
\newblock
\showISSN{1939-3520}
\urldef\tempurl%
\url{https://doi.org/10.1109/TSE.2022.3173346}
\showDOI{\tempurl}


\bibitem[\protect\citeauthoryear{Wang, Li, Qian, Yang, Wang, Shang, Kumar, Tan,
  Ray, Bhatia, et~al\mbox{.}}{Wang et~al\mbox{.}}{2022c}]%
        {wang2022recode}
\bibfield{author}{\bibinfo{person}{Shiqi Wang}, \bibinfo{person}{Zheng Li},
  \bibinfo{person}{Haifeng Qian}, \bibinfo{person}{Chenghao Yang},
  \bibinfo{person}{Zijian Wang}, \bibinfo{person}{Mingyue Shang},
  \bibinfo{person}{Varun Kumar}, \bibinfo{person}{Samson Tan},
  \bibinfo{person}{Baishakhi Ray}, \bibinfo{person}{Parminder Bhatia},
  {et~al\mbox{.}}} \bibinfo{year}{2022}\natexlab{c}.
\newblock \showarticletitle{ReCode: Robustness Evaluation of Code Generation
  Models}.
\newblock \bibinfo{journal}{\emph{arXiv preprint arXiv:2212.10264}}
  (\bibinfo{year}{2022}).
\newblock


\bibitem[\protect\citeauthoryear{Wang, Zhang, Sui, Wan, Zhao, Wu, Yu, and
  Xu}{Wang et~al\mbox{.}}{2022d}]%
        {wang_reinforcement-learning-guided_2022}
\bibfield{author}{\bibinfo{person}{Wenhua Wang}, \bibinfo{person}{Yuqun Zhang},
  \bibinfo{person}{Yulei Sui}, \bibinfo{person}{Yao Wan}, \bibinfo{person}{Zhou
  Zhao}, \bibinfo{person}{Jian Wu}, \bibinfo{person}{Philip~S. Yu}, {and}
  \bibinfo{person}{Guandong Xu}.} \bibinfo{year}{2022}\natexlab{d}.
\newblock \showarticletitle{Reinforcement-{Learning}-{Guided} {Source} {Code}
  {Summarization} {Using} {Hierarchical} {Attention}}.
\newblock \bibinfo{journal}{\emph{IIEEE Trans. Software Eng.}}
  \bibinfo{volume}{48}, \bibinfo{number}{1} (\bibinfo{date}{Jan.}
  \bibinfo{year}{2022}), \bibinfo{pages}{102--119}.
\newblock
\showISSN{0098-5589, 1939-3520, 2326-3881}
\urldef\tempurl%
\url{https://doi.org/10.1109/TSE.2020.2979701}
\showDOI{\tempurl}


\bibitem[\protect\citeauthoryear{Wang, Wang, Joty, and Hoi}{Wang
  et~al\mbox{.}}{2021}]%
        {wang2021codet5}
\bibfield{author}{\bibinfo{person}{Yue Wang}, \bibinfo{person}{Weishi Wang},
  \bibinfo{person}{Shafiq Joty}, {and} \bibinfo{person}{Steven~CH Hoi}.}
  \bibinfo{year}{2021}\natexlab{}.
\newblock \showarticletitle{Codet5: Identifier-aware unified pre-trained
  encoder-decoder models for code understanding and generation}.
\newblock \bibinfo{journal}{\emph{arXiv preprint arXiv:2109.00859}}
  (\bibinfo{year}{2021}).
\newblock


\bibitem[\protect\citeauthoryear{Watson, Cooper, Palacio, Moran, and
  Poshyvanyk}{Watson et~al\mbox{.}}{2022}]%
        {watson2022systematic}
\bibfield{author}{\bibinfo{person}{Cody Watson}, \bibinfo{person}{Nathan
  Cooper}, \bibinfo{person}{David~Nader Palacio}, \bibinfo{person}{Kevin
  Moran}, {and} \bibinfo{person}{Denys Poshyvanyk}.}
  \bibinfo{year}{2022}\natexlab{}.
\newblock \showarticletitle{A systematic literature review on the use of deep
  learning in software engineering research}.
\newblock \bibinfo{journal}{\emph{ACM Transactions on Software Engineering and
  Methodology (TOSEM)}} \bibinfo{volume}{31}, \bibinfo{number}{2}
  (\bibinfo{year}{2022}), \bibinfo{pages}{1--58}.
\newblock


\bibitem[\protect\citeauthoryear{Wei, Xia, and Zhang}{Wei
  et~al\mbox{.}}{2023}]%
        {wei2023copiloting}
\bibfield{author}{\bibinfo{person}{Yuxiang Wei},
  \bibinfo{person}{Chunqiu~Steven Xia}, {and} \bibinfo{person}{Lingming
  Zhang}.} \bibinfo{year}{2023}\natexlab{}.
\newblock \showarticletitle{Copiloting the Copilots: Fusing Large Language
  Models with Completion Engines for Automated Program Repair}.
\newblock \bibinfo{journal}{\emph{arXiv preprint arXiv:2309.00608}}
  (\bibinfo{year}{2023}).
\newblock


\bibitem[\protect\citeauthoryear{Weyssow, Zhou, Kim, Lo, and Sahraoui}{Weyssow
  et~al\mbox{.}}{2023}]%
        {weyssow2023exploring}
\bibfield{author}{\bibinfo{person}{Martin Weyssow}, \bibinfo{person}{Xin Zhou},
  \bibinfo{person}{Kisub Kim}, \bibinfo{person}{David Lo}, {and}
  \bibinfo{person}{Houari Sahraoui}.} \bibinfo{year}{2023}\natexlab{}.
\newblock \showarticletitle{Exploring Parameter-Efficient Fine-Tuning
  Techniques for Code Generation with Large Language Models}.
\newblock \bibinfo{journal}{\emph{arXiv preprint arXiv:2308.10462}}
  (\bibinfo{year}{2023}).
\newblock


\bibitem[\protect\citeauthoryear{Wu, Jiang, Pham, Lutellier, Davis, Tan,
  Babkin, and Shah}{Wu et~al\mbox{.}}{2023}]%
        {wu2023effective}
\bibfield{author}{\bibinfo{person}{Yi Wu}, \bibinfo{person}{Nan Jiang},
  \bibinfo{person}{Hung~Viet Pham}, \bibinfo{person}{Thibaud Lutellier},
  \bibinfo{person}{Jordan Davis}, \bibinfo{person}{Lin Tan},
  \bibinfo{person}{Petr Babkin}, {and} \bibinfo{person}{Sameena Shah}.}
  \bibinfo{year}{2023}\natexlab{}.
\newblock \showarticletitle{How Effective Are Neural Networks for Fixing
  Security Vulnerabilities}.
\newblock \bibinfo{journal}{\emph{arXiv preprint arXiv:2305.18607}}
  (\bibinfo{year}{2023}).
\newblock


\bibitem[\protect\citeauthoryear{Xia, Wei, and Zhang}{Xia
  et~al\mbox{.}}{2023}]%
        {xia2023automated}
\bibfield{author}{\bibinfo{person}{Chunqiu~Steven Xia},
  \bibinfo{person}{Yuxiang Wei}, {and} \bibinfo{person}{Lingming Zhang}.}
  \bibinfo{year}{2023}\natexlab{}.
\newblock \showarticletitle{Automated program repair in the era of large
  pre-trained language models}. In \bibinfo{booktitle}{\emph{Proceedings of the
  45th International Conference on Software Engineering (ICSE 2023).
  Association for Computing Machinery}}.
\newblock


\bibitem[\protect\citeauthoryear{Xu, Alon, Neubig, and Hellendoorn}{Xu
  et~al\mbox{.}}{2022}]%
        {xu2022systematic}
\bibfield{author}{\bibinfo{person}{Frank~F Xu}, \bibinfo{person}{Uri Alon},
  \bibinfo{person}{Graham Neubig}, {and} \bibinfo{person}{Vincent~Josua
  Hellendoorn}.} \bibinfo{year}{2022}\natexlab{}.
\newblock \showarticletitle{A systematic evaluation of large language models of
  code}. In \bibinfo{booktitle}{\emph{Proceedings of the 6th ACM SIGPLAN
  International Symposium on Machine Programming}}. \bibinfo{pages}{1--10}.
\newblock


\bibitem[\protect\citeauthoryear{Xu, Li, and Deng}{Xu et~al\mbox{.}}{2021}]%
        {Xu2021DifferentialTA}
\bibfield{author}{\bibinfo{person}{Jiayun Xu}, \bibinfo{person}{Yingjiu Li},
  {and} \bibinfo{person}{Robert~H. Deng}.} \bibinfo{year}{2021}\natexlab{}.
\newblock \showarticletitle{Differential Training: A Generic Framework to
  Reduce Label Noises for Android Malware Detection}. In
  \bibinfo{booktitle}{\emph{Network and Distributed System Security
  Symposium}}.
\newblock
\urldef\tempurl%
\url{https://api.semanticscholar.org/CorpusID:231879075}
\showURL{%
\tempurl}


\bibitem[\protect\citeauthoryear{Yang, Wang, Li, and Wang}{Yang
  et~al\mbox{.}}{2023a}]%
        {yang2023does}
\bibfield{author}{\bibinfo{person}{Xu Yang}, \bibinfo{person}{Shaowei Wang},
  \bibinfo{person}{Yi Li}, {and} \bibinfo{person}{Shaohua Wang}.}
  \bibinfo{year}{2023}\natexlab{a}.
\newblock \showarticletitle{Does data sampling improve deep learning-based
  vulnerability detection? Yeas! and Nays!}. In \bibinfo{booktitle}{\emph{2023
  IEEE/ACM 45th International Conference on Software Engineering (ICSE)}}.
  IEEE, \bibinfo{pages}{2287--2298}.
\newblock


\bibitem[\protect\citeauthoryear{Yang, Xia, Lo, and Grundy}{Yang
  et~al\mbox{.}}{2022b}]%
        {yang2022survey}
\bibfield{author}{\bibinfo{person}{Yanming Yang}, \bibinfo{person}{Xin Xia},
  \bibinfo{person}{David Lo}, {and} \bibinfo{person}{John Grundy}.}
  \bibinfo{year}{2022}\natexlab{b}.
\newblock \showarticletitle{A survey on deep learning for software
  engineering}.
\newblock \bibinfo{journal}{\emph{ACM Computing Surveys (CSUR)}}
  \bibinfo{volume}{54}, \bibinfo{number}{10s} (\bibinfo{year}{2022}),
  \bibinfo{pages}{1--73}.
\newblock


\bibitem[\protect\citeauthoryear{Yang, Shi, He, and Lo}{Yang
  et~al\mbox{.}}{2022a}]%
        {yang2022natural}
\bibfield{author}{\bibinfo{person}{Zhou Yang}, \bibinfo{person}{Jieke Shi},
  \bibinfo{person}{Junda He}, {and} \bibinfo{person}{David Lo}.}
  \bibinfo{year}{2022}\natexlab{a}.
\newblock \showarticletitle{Natural attack for pre-trained models of code}. In
  \bibinfo{booktitle}{\emph{Proceedings of the 44th International Conference on
  Software Engineering}}. \bibinfo{pages}{1482--1493}.
\newblock


\bibitem[\protect\citeauthoryear{Yang, Zhao, Wang, Shi, Kim, Han, and Lo}{Yang
  et~al\mbox{.}}{2023b}]%
        {yang2023code}
\bibfield{author}{\bibinfo{person}{Zhou Yang}, \bibinfo{person}{Zhipeng Zhao},
  \bibinfo{person}{Chenyu Wang}, \bibinfo{person}{Jieke Shi},
  \bibinfo{person}{Dongsun Kim}, \bibinfo{person}{DongGyun Han}, {and}
  \bibinfo{person}{David Lo}.} \bibinfo{year}{2023}\natexlab{b}.
\newblock \bibinfo{title}{What Do Code Models Memorize? An Empirical Study on
  Large Language Models of Code}.
\newblock
\newblock
\showeprint[arxiv]{2308.09932}~[cs.SE]


\bibitem[\protect\citeauthoryear{Yatish, Jiarpakdee, Thongtanunam, and
  Tantithamthavorn}{Yatish et~al\mbox{.}}{2019}]%
        {yatish2019mining}
\bibfield{author}{\bibinfo{person}{Suraj Yatish}, \bibinfo{person}{Jirayus
  Jiarpakdee}, \bibinfo{person}{Patanamon Thongtanunam}, {and}
  \bibinfo{person}{Chakkrit Tantithamthavorn}.}
  \bibinfo{year}{2019}\natexlab{}.
\newblock \showarticletitle{Mining software defects: Should we consider
  affected releases?}. In \bibinfo{booktitle}{\emph{2019 IEEE/ACM 41st
  International Conference on Software Engineering (ICSE)}}. IEEE,
  \bibinfo{pages}{654--665}.
\newblock


\bibitem[\protect\citeauthoryear{Yedida and Menzies}{Yedida and
  Menzies}{2021}]%
        {yedida2021value}
\bibfield{author}{\bibinfo{person}{Rahul Yedida} {and} \bibinfo{person}{Tim
  Menzies}.} \bibinfo{year}{2021}\natexlab{}.
\newblock \showarticletitle{On the value of oversampling for deep learning in
  software defect prediction}.
\newblock \bibinfo{journal}{\emph{IEEE Transactions on Software Engineering}}
  \bibinfo{volume}{48}, \bibinfo{number}{8} (\bibinfo{year}{2021}),
  \bibinfo{pages}{3103--3116}.
\newblock


\bibitem[\protect\citeauthoryear{Yin, Deng, Chen, Vasilescu, and Neubig}{Yin
  et~al\mbox{.}}{2018}]%
        {yin2018learning}
\bibfield{author}{\bibinfo{person}{Pengcheng Yin}, \bibinfo{person}{Bowen
  Deng}, \bibinfo{person}{Edgar Chen}, \bibinfo{person}{Bogdan Vasilescu},
  {and} \bibinfo{person}{Graham Neubig}.} \bibinfo{year}{2018}\natexlab{}.
\newblock \showarticletitle{Learning to mine aligned code and natural language
  pairs from stack overflow}. In \bibinfo{booktitle}{\emph{Proceedings of the
  15th international conference on mining software repositories}}.
  \bibinfo{pages}{476--486}.
\newblock


\bibitem[\protect\citeauthoryear{Zeng, Tan, Zhang, Li, Zhang, and Zhang}{Zeng
  et~al\mbox{.}}{2022}]%
        {zeng_extensive_2022}
\bibfield{author}{\bibinfo{person}{Zhengran Zeng}, \bibinfo{person}{Hanzhuo
  Tan}, \bibinfo{person}{Haotian Zhang}, \bibinfo{person}{Jing Li},
  \bibinfo{person}{Yuqun Zhang}, {and} \bibinfo{person}{Lingming Zhang}.}
  \bibinfo{year}{2022}\natexlab{}.
\newblock \showarticletitle{An extensive study on pre-trained models for
  program understanding and generation}. In
  \bibinfo{booktitle}{\emph{Proceedings of the 31st {ACM} {SIGSOFT}
  {International} {Symposium} on {Software} {Testing} and {Analysis}}}
  \emph{(\bibinfo{series}{{ISSTA} 2022})}. \bibinfo{publisher}{Association for
  Computing Machinery}, \bibinfo{address}{Virtual South Korea},
  \bibinfo{pages}{39--51}.
\newblock
\showISBNx{978-1-4503-9379-9}
\urldef\tempurl%
\url{https://doi.org/10.1145/3533767.3534390}
\showDOI{\tempurl}


\bibitem[\protect\citeauthoryear{Zeng, Zhang, Zhang, and Zhang}{Zeng
  et~al\mbox{.}}{2021}]%
        {zeng_deep_2021}
\bibfield{author}{\bibinfo{person}{Zhengran Zeng}, \bibinfo{person}{Yuqun
  Zhang}, \bibinfo{person}{Haotian Zhang}, {and} \bibinfo{person}{Lingming
  Zhang}.} \bibinfo{year}{2021}\natexlab{}.
\newblock \showarticletitle{Deep just-in-time defect prediction: how far are
  we?}. In \bibinfo{booktitle}{\emph{Proceedings of the 30th {ACM} {SIGSOFT}
  {International} {Symposium} on {Software} {Testing} and {Analysis}}}
  \emph{(\bibinfo{series}{{ISSTA} 2021})}. \bibinfo{publisher}{Association for
  Computing Machinery}, \bibinfo{address}{New York, NY, USA},
  \bibinfo{pages}{427--438}.
\newblock
\showISBNx{978-1-4503-8459-9}
\urldef\tempurl%
\url{https://doi.org/10.1145/3460319.3464819}
\showDOI{\tempurl}


\bibitem[\protect\citeauthoryear{Zhang, Chen, Zhao, and Peng}{Zhang
  et~al\mbox{.}}{2023a}]%
        {zhang2023slice}
\bibfield{author}{\bibinfo{person}{Fengyi Zhang}, \bibinfo{person}{Bihuan
  Chen}, \bibinfo{person}{Yufei Zhao}, {and} \bibinfo{person}{Xin Peng}.}
  \bibinfo{year}{2023}\natexlab{a}.
\newblock \showarticletitle{Slice-Based Code Change Representation Learning}.
  In \bibinfo{booktitle}{\emph{2023 IEEE International Conference on Software
  Analysis, Evolution and Reengineering (SANER)}}. IEEE,
  \bibinfo{pages}{319--330}.
\newblock


\bibitem[\protect\citeauthoryear{Zhang, Babar, and Tell}{Zhang
  et~al\mbox{.}}{2011}]%
        {zhang2011identifying}
\bibfield{author}{\bibinfo{person}{He Zhang}, \bibinfo{person}{Muhammad~Ali
  Babar}, {and} \bibinfo{person}{Paolo Tell}.} \bibinfo{year}{2011}\natexlab{}.
\newblock \showarticletitle{Identifying relevant studies in software
  engineering}.
\newblock \bibinfo{journal}{\emph{Information and Software Technology}}
  \bibinfo{volume}{53}, \bibinfo{number}{6} (\bibinfo{year}{2011}),
  \bibinfo{pages}{625--637}.
\newblock


\bibitem[\protect\citeauthoryear{Zhang, Fu, Li, Ma, Zhao, Yang, Sun, Liu, and
  Jin}{Zhang et~al\mbox{.}}{2022a}]%
        {zhang_towards_2022}
\bibfield{author}{\bibinfo{person}{Huangzhao Zhang}, \bibinfo{person}{Zhiyi
  Fu}, \bibinfo{person}{Ge Li}, \bibinfo{person}{Lei Ma},
  \bibinfo{person}{Zhehao Zhao}, \bibinfo{person}{Hua’an Yang},
  \bibinfo{person}{Yizhe Sun}, \bibinfo{person}{Yang Liu}, {and}
  \bibinfo{person}{Zhi Jin}.} \bibinfo{year}{2022}\natexlab{a}.
\newblock \showarticletitle{Towards {Robustness} of {Deep} {Program}
  {Processing} {Models} {Detection}, {Estimation}, and {Enhancement}}.
\newblock \bibinfo{journal}{\emph{ACM Trans. Softw. Eng. Methodol.}}
  \bibinfo{volume}{31}, \bibinfo{number}{3} (\bibinfo{date}{July}
  \bibinfo{year}{2022}), \bibinfo{pages}{50:1--50:40}.
\newblock
\showISSN{1049-331X, 1557-7392}
\urldef\tempurl%
\url{https://doi.org/10.1145/3511887}
\showDOI{\tempurl}


\bibitem[\protect\citeauthoryear{Zhang, Li, Li, Ma, Liu, and Jin}{Zhang
  et~al\mbox{.}}{2020}]%
        {zhang_generating_2020}
\bibfield{author}{\bibinfo{person}{Huangzhao Zhang}, \bibinfo{person}{Zhuo Li},
  \bibinfo{person}{Ge Li}, \bibinfo{person}{Lei Ma}, \bibinfo{person}{Yang
  Liu}, {and} \bibinfo{person}{Zhi Jin}.} \bibinfo{year}{2020}\natexlab{}.
\newblock \showarticletitle{Generating {Adversarial} {Examples} for {Holding}
  {Robustness} of {Source} {Code} {Processing} {Models}}.
\newblock \bibinfo{journal}{\emph{AAAI}} \bibinfo{volume}{34},
  \bibinfo{number}{01} (\bibinfo{date}{April} \bibinfo{year}{2020}),
  \bibinfo{pages}{1169--1176}.
\newblock
\showISSN{2374-3468, 2159-5399}
\urldef\tempurl%
\url{https://doi.org/10.1609/aaai.v34i01.5469}
\showDOI{\tempurl}


\bibitem[\protect\citeauthoryear{Zhang, Guo, Zhang, Sui, Xue, and Xu}{Zhang
  et~al\mbox{.}}{2023b}]%
        {zhang_challenging_2023}
\bibfield{author}{\bibinfo{person}{Weiwei Zhang}, \bibinfo{person}{Shengjian
  Guo}, \bibinfo{person}{Hongyu Zhang}, \bibinfo{person}{Yulei Sui},
  \bibinfo{person}{Yinxing Xue}, {and} \bibinfo{person}{Yun Xu}.}
  \bibinfo{year}{2023}\natexlab{b}.
\newblock \showarticletitle{Challenging {Machine} {Learning}-{Based} {Clone}
  {Detectors} via {Semantic}-{Preserving} {Code} {Transformations}}.
\newblock \bibinfo{journal}{\emph{IIEEE Trans. Software Eng.}}
  \bibinfo{volume}{49}, \bibinfo{number}{5} (\bibinfo{date}{May}
  \bibinfo{year}{2023}), \bibinfo{pages}{3052--3070}.
\newblock
\showISSN{0098-5589, 1939-3520, 2326-3881}
\urldef\tempurl%
\url{https://doi.org/10.1109/TSE.2023.3240118}
\showDOI{\tempurl}


\bibitem[\protect\citeauthoryear{Zhang, Zhang, Shen, and Gu}{Zhang
  et~al\mbox{.}}{2022b}]%
        {zhang_diet_2022}
\bibfield{author}{\bibinfo{person}{Zhaowei Zhang}, \bibinfo{person}{Hongyu
  Zhang}, \bibinfo{person}{Beijun Shen}, {and} \bibinfo{person}{Xiaodong Gu}.}
  \bibinfo{year}{2022}\natexlab{b}.
\newblock \showarticletitle{Diet code is healthy: simplifying programs for
  pre-trained models of code}. In \bibinfo{booktitle}{\emph{Proceedings of the
  30th {ACM} {Joint} {European} {Software} {Engineering} {Conference} and
  {Symposium} on the {Foundations} of {Software} {Engineering}}}
  \emph{(\bibinfo{series}{{ESEC}/{FSE} 2022})}. \bibinfo{publisher}{Association
  for Computing Machinery}, \bibinfo{address}{Singapore Singapore},
  \bibinfo{pages}{1073--1084}.
\newblock
\showISBNx{978-1-4503-9413-0}
\urldef\tempurl%
\url{https://doi.org/10.1145/3540250.3549094}
\showDOI{\tempurl}


\bibitem[\protect\citeauthoryear{Zheng, Pujar, Lewis, Buratti, Epstein, Yang,
  Laredo, Morari, and Su}{Zheng et~al\mbox{.}}{2021}]%
        {zheng2021d2a}
\bibfield{author}{\bibinfo{person}{Yunhui Zheng}, \bibinfo{person}{Saurabh
  Pujar}, \bibinfo{person}{Burn Lewis}, \bibinfo{person}{Luca Buratti},
  \bibinfo{person}{Edward Epstein}, \bibinfo{person}{Bo Yang},
  \bibinfo{person}{Jim Laredo}, \bibinfo{person}{Alessandro Morari}, {and}
  \bibinfo{person}{Zhong Su}.} \bibinfo{year}{2021}\natexlab{}.
\newblock \showarticletitle{D2a: A dataset built for ai-based vulnerability
  detection methods using differential analysis}. In
  \bibinfo{booktitle}{\emph{2021 IEEE/ACM 43rd International Conference on
  Software Engineering: Software Engineering in Practice (ICSE-SEIP)}}. IEEE,
  \bibinfo{pages}{111--120}.
\newblock


\bibitem[\protect\citeauthoryear{Zhou, Gao, Wu, Grundy, Chen, Chen, and
  Li}{Zhou et~al\mbox{.}}{2023}]%
        {zhou2023modelobfuscator}
\bibfield{author}{\bibinfo{person}{Mingyi Zhou}, \bibinfo{person}{Xiang Gao},
  \bibinfo{person}{Jing Wu}, \bibinfo{person}{John Grundy},
  \bibinfo{person}{Xiao Chen}, \bibinfo{person}{Chunyang Chen}, {and}
  \bibinfo{person}{Li Li}.} \bibinfo{year}{2023}\natexlab{}.
\newblock \showarticletitle{ModelObfuscator: Obfuscating Model Information to
  Protect Deployed ML-Based Systems}.
\newblock  (\bibinfo{year}{2023}), \bibinfo{pages}{1005–1017}.
\newblock
\showISBNx{9798400702211}
\urldef\tempurl%
\url{https://doi.org/10.1145/3597926.3598113}
\showDOI{\tempurl}


\bibitem[\protect\citeauthoryear{Zhou, Zhang, Shen, Han, Chen, and Gall}{Zhou
  et~al\mbox{.}}{2022}]%
        {zhou_adversarial_2022}
\bibfield{author}{\bibinfo{person}{Yu Zhou}, \bibinfo{person}{Xiaoqing Zhang},
  \bibinfo{person}{Juanjuan Shen}, \bibinfo{person}{Tingting Han},
  \bibinfo{person}{Taolue Chen}, {and} \bibinfo{person}{Harald Gall}.}
  \bibinfo{year}{2022}\natexlab{}.
\newblock \showarticletitle{Adversarial {Robustness} of {Deep} {Code} {Comment}
  {Generation}}.
\newblock \bibinfo{journal}{\emph{ACM Trans. Softw. Eng. Methodol.}}
  \bibinfo{volume}{31}, \bibinfo{number}{4} (\bibinfo{date}{July}
  \bibinfo{year}{2022}), \bibinfo{pages}{60:1--60:30}.
\newblock
\showISSN{1049-331X, 1557-7392}
\urldef\tempurl%
\url{https://doi.org/10.1145/3501256}
\showDOI{\tempurl}


\bibitem[\protect\citeauthoryear{Zini and Awad}{Zini and Awad}{2022}]%
        {zini2022explainability}
\bibfield{author}{\bibinfo{person}{Julia~El Zini} {and}
  \bibinfo{person}{Mariette Awad}.} \bibinfo{year}{2022}\natexlab{}.
\newblock \showarticletitle{On the explainability of natural language
  processing deep models}.
\newblock \bibinfo{journal}{\emph{Comput. Surveys}} \bibinfo{volume}{55},
  \bibinfo{number}{5} (\bibinfo{year}{2022}), \bibinfo{pages}{1--31}.
\newblock


\bibitem[\protect\citeauthoryear{Zou, Zhu, Xu, Li, Jin, and Ye}{Zou
  et~al\mbox{.}}{2021}]%
        {zou_interpreting_2021}
\bibfield{author}{\bibinfo{person}{Deqing Zou}, \bibinfo{person}{Yawei Zhu},
  \bibinfo{person}{Shouhuai Xu}, \bibinfo{person}{Zhen Li},
  \bibinfo{person}{Hai Jin}, {and} \bibinfo{person}{Hengkai Ye}.}
  \bibinfo{year}{2021}\natexlab{}.
\newblock \showarticletitle{Interpreting {Deep} {Learning}-based
  {Vulnerability} {Detector} {Predictions} {Based} on {Heuristic} {Searching}}.
\newblock \bibinfo{journal}{\emph{ACM Trans. Softw. Eng. Methodol.}}
  \bibinfo{volume}{30}, \bibinfo{number}{2} (\bibinfo{date}{March}
  \bibinfo{year}{2021}), \bibinfo{pages}{23:1--23:31}.
\newblock
\showISSN{1049-331X, 1557-7392}
\urldef\tempurl%
\url{https://doi.org/10.1145/3429444}
\showDOI{\tempurl}


\end{thebibliography}

\appendix
\section{Details for Paper Collection and Selection}
\label{sec:approach}
\begin{figure}
    \centering
    \includegraphics[width=\linewidth]{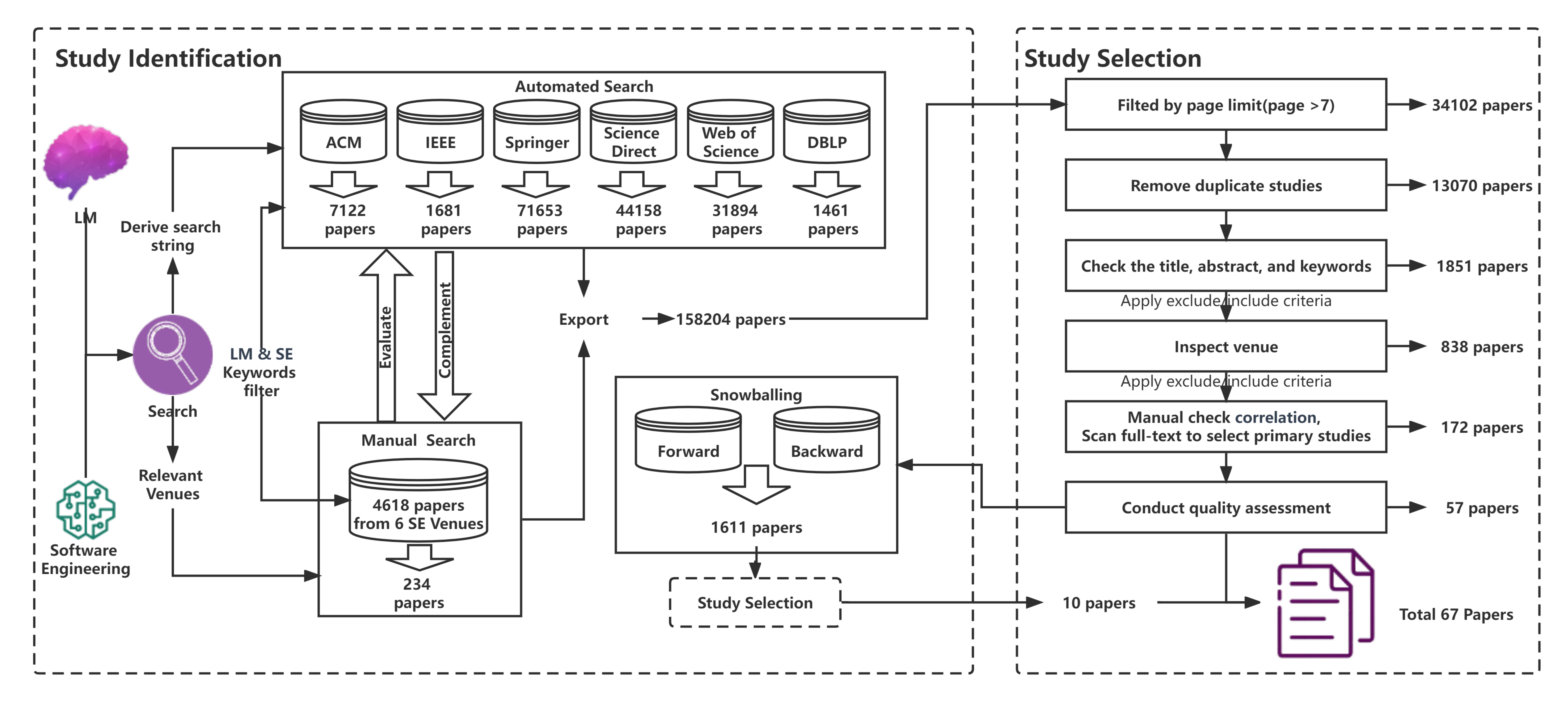}
    \caption{The overview of the review process.}
    \label{fig:Study_identification_and_selection}
\end{figure}

Since we explore the state-of-the-art research of the pitfalls within LM4Code, we first outline our methodology for identifying relevant research studies, and then provide an overview of our collected literature.
We follow the rigorous methodology proposed by Kitchenham~\ea~\cite{kitchenham_guidelines_2007,kitchenham_segress_2023} and Zhang~\ea~\cite{zhang2011identifying} to perform our lightweight Systematic Literature Review (SLR).
The primary steps in our systematic literature review can be summarized as follows: (1) planning the review and formulating a review protocol, (2) proposing research questions, (3) designing search strategies and proposing inclusion/exclusion criteria, (4) conducting a lightweight snowballing, (5) data extraction, and (6) data synthesis. 
The outline for our review can be represented as presented in Figure~\ref{fig:Study_identification_and_selection}.


\subsection{Research Questions and Motivations}
In recent research, language models trained for code intelligence have shown promising performance~\cite{hou2023large, wang2023software, zeng_extensive_2022}.
However, an increasing number of literature~\cite{sun_importance_2022, shi2022we, nie2023understanding} has highlighted the existence of pitfalls in LM4Code that can skew their realistic performance, leading to either substantial overestimation or underestimation of their effectiveness.
The aim of conducting this systematic review is to gain an in-depth understanding of the pitfalls present in language models tailored for code intelligence.
Ensuring the robustness, reliability, and ethical deployment of such models is important for their effective integration into the software development lifecycle.
Consequently, it is crucial to discern the nature of these pitfalls, comprehend their implications, and examine existing solutions.
Thus, we aim to answer the following research questions:

\begin{itemize}
    \item \textbf{RQ1: What types of pitfalls are prevalent in language models for code intelligence?} This research question aims to identify the prevalent pitfalls in LM4Code systems, exploring how they could affect various stages of the learning-based system lifecycle.
    \item \textbf{RQ2: What are the implications of these pitfalls?} This research question investigates the implications of the identified pitfalls, specifically focusing on their impacts on the effectiveness, reliability, and ethical considerations of automated code intelligence systems.
    \item \textbf{RQ3: What solutions have been proposed to address these biases and pitfalls?} This research question reviews the existing body of literature to pinpoint proposed approaches for mitigating the identified pitfalls.

\end{itemize}

\subsection{Search Strategy}
To find out all potentially relevant research papers, we utilized the “Quasi-Gold Standard” (QGS)~\cite{zhang2011identifying} approach, which combines both manual and automated search strategies. 
Using QGS offers an optimal balance between efficiency and research coverage, as evidenced in several previous studies\cite{wang2022machine, durelli2019machine}. 
As illustrated in Figure~\ref{fig:Study_identification_and_selection}, our search strategy involved the following sequential steps:

\begin{table}[t]
  \centering
  \resizebox{\linewidth}{!}{
    \begin{tabular}{rl}
    \toprule
   \textbf{Acronym} & \multicolumn{1}{c}{\textbf{Venues}} \\
    \hline
    ASE   & International Conference on Automated Software Engineering  \\
    ESEC/FSE & Joint European Software Engineering Conference and Symposium on the Foundations of Software Engineering \\
    ICSE  & International Conference on Software Engineering  \\
    ISSTA  & International Symposium on Software Testing and Analysis  \\
    TOSEM  & Transactions on Software Engineering and Methodology  \\
    TSE   & Transactions on Software Engineering \\
    \bottomrule
    \end{tabular}%
    }
\caption{Publication venues for manual search}
  \label{tab:manual_search_venues}%
  
\end{table}%

\begin{enumerate}
    \item {Select appropriate publication venues for manual search and select digital databases for automated search that encompass all the chosen venues.} 
    \item {Establish QGS: Screen all papers for manual search and filter by inclusion/exclusion criteria(defined in Table \ref{tab:manual_search_venues}).}
    \item {Define search string based on domain knowledge from Language Models (LM) and Software Engineering (SE).}
    \item {Conduct an automated search using the search keywords defined in Step 3.}
    \item {Evaluate the quality of included studies through QGS.}
\end{enumerate}

In our research approach, we integrated both manual and keyword-based search methodologies to identify relevant papers. 
During the manual approach, we focused on six top-tier SE conferences and journals, namely ICSE, ESEC/FSE, ASE, ISSTA, TOSEM, and TSE, all of which are CCF~\cite{ccf2023ranking} A-ranked venues in the software engineering domains (as indicated in Table~\ref{tab:manual_search_venues}).
We systematically crawled a list comprising 4,618 published papers from the top venues.
After automating scanning with scripts, we thoroughly verified and discovered 234 papers relating to LM4Code.
These 234 relevant papers served as the foundation for developing the Quasi-Gold Standard (QGS) for the following automated search. 
Our search string should include two sets of keywords:  one for LMs and another for code intelligence. 
Only if the paper contains both types of keywords does it have a higher probability of being the one we require. 
The full list of search keywords is as follows:

\begin{itemize}
    \item Keywords related to LMs: "LLM" OR "Large Language Model*" OR "Language Model*" OR "LM" OR "PLM" OR "Pre-trained" OR "Pre-training" OR "Natural Language Processing" OR "NLP" OR "Machine Learning" OR "ML" OR "Deep Learning" OR "DL" OR "Artificial Intelligence" OR "AI" OR "Transformer" OR "BERT" OR "CODEX" OR "GPT" OR "T5" OR "Sequence Model*" OR "Attention Model*" OR "Transfer Learning" OR "Neural Network*" OR "ChatGPT" OR "GPT-*" OR "Deep neural network*" OR "DNN*"
    
    \item Keywords  related to code intelligence tasks: "Software Engineering" OR "Software Development" OR "Program*" OR "Software Testing" OR "Software Mainten*" OR "SE" OR "Software Lifecycle" OR "Software Design*" OR "Code representation" OR "Code generation" OR "Code comment generation" OR "Code search" OR "Code localization" OR "Code completion" OR "Code summarization" OR "Method name generation" OR "Bug detection" OR "Bug localization" OR "Vulnerability detection" OR "Testing techniques" OR "Test case generation" OR "Program analysis" OR "Bug classification" OR "Defect prediction" OR "Program repair" OR "Code clone detection" OR "Bug report" OR "Software quality evaluation" OR "SATD detection" OR "Code smell detection" OR "Compiled-related" OR "Code review" OR "Software classification" OR "Code classification" OR "Code change" OR "Incident detection" OR "Requirement extraction" OR "Requirement traceability" OR "Requirement validation" OR "Effort cost prediction" OR "Mining GitHub/Github mining" OR "Mining SO (StackOverflow)/SO mining" OR "Mining app/App mining" OR "Mining tag/Tag mining" OR "Developer-based mining"
\end{itemize}

It's important to highlight that our list of keywords is specific to LM4Code, and we intentionally omitted keywords that are related to pitfalls during the paper search process.
The reason behind this derives from the ambiguity around the term ``pitfalls'', which is open to various interpretations.
We decided to rely on our rigorous inclusion/exclusion criteria because it is difficult to precisely categorize the types of pitfalls that exist within LM4Code, which is also the main motivation behind this taxonomy study. 
We were able to include relevant papers with this methodology, even if they didn't explicitly mention ``pitfalls'' in their content.

After establishing the search string, we proceeded to conduct an automated search across six widely used databases to ensure comprehensive coverage of all relevant published papers. 
Specifically, our search spanned four major academic publishers: ACM Digital Library, IEEE Xplorer, Springer, and Science Direct. Additionally, we included two renowned indexing databases: DBLP and Web of Science, which indexes several other smaller academic databases. 
Similarly to previous studies~\cite{croft2022data, wang2022machine}, we did not use other search engines like Google Scholar due to the existence of excessive irrelevant information in the search results and the requirement for subjective criteria to choose when to stop the search process. 
Finally, we obtained 7,122 papers from the ACM Digital Library, 1,681 papers from IEEE Xplore, 71,653 papers from Springer, 44,158 papers from ScienceDirect, 44,158 papers from Web of Science, and 1,461 papers from DBLP. 

\begin{table}[t]
\resizebox{0.75\linewidth}{!}{
\begin{tabular}{lp{10cm}}
\toprule
\multicolumn{2}{l}{\textbf{Inclusion Criteria}} \\
\midrule
I1) & Studies explicitly utilizing LMs. \\
I2) & Studies claim that the study involves code-related tasks. \\
I3) & Studies highlighting pitfalls, particularly emphasizing unrealistic performance evaluation or factors that negatively influence performance in LM4Code. \\
\midrule
\multicolumn{2}{l}{\textbf{Exclusion Criteria}} \\
\midrule
E1) & Studies whose full-text is inaccessible. \\
E2) & The study whose number of pages is less than 8. \\
E3) & Redundant or nearly identical studies from the same authors. \\
E4) & Papers not written in English. \\
E5) & Systematic literature reviews, reviews, or surveys. \\
E6) & Studies from workshops, doctoral symposiums, books, theses, monographs, keynotes,  or panels. \\
E7) & Non peer-reviewed academic literature. \\
E8) & Studies not related to language models or code intelligence. \\
E9) & Studies emphasizing LM4Code without discussing pitfalls related to realistic performance. \\
E10) & Studies with a primary focus on cyberattacks or Operating Systems. \\
E11) & Studies that are published in a journal or conference with a CORE~\cite{aucore2023ranking} ranking of less than A. \\
\bottomrule
\end{tabular}
}
\caption{Inclusion and Exclusion Criteria}
\label{tab:inclusion_criteria}
\end{table}

\subsection{Study Selection}
From among the papers collected by the paper search process, we attempted to select any research study that focused on the pitfalls of LM4Code.
As we discussed before, we define these ``pitfalls'' as any significant issues or constraints present within the datasets, model architectures, experimental designs, or even model deployment that could potentially undermine the reliability or realistic performance of the proposed LM4Code systems.
The inclusion/exclusion criteria we adopted, shown in Table ~\ref{tab:inclusion_criteria}, were inspired by similar studies~\cite{liu2022deep, croft2022data, raatikainen2019software}.
It is noted that to maintain a reliable taxonomy of the pitfalls of LM4Code based on high-quality research studies, we removed the studies published in low-quality venues: venue ranking below A using the CORE ranking system~\cite{aucore2023ranking}.

To determine whether the studies met the inclusion requirements, we combined thorough manual assessment with automated script filtering.
The study selection process involves six distinct stages, as illustrated in Figure~\ref{fig:Study_identification_and_selection}. 
The first two stages (filtering and deduplication) used automated scripts, substantially reducing the initial set to 13,070 papers. 
Subsequently, in stages three and four, we applied the inclusion/exclusion criteria, delving into the titles, abstracts, keywords, and publication venues of each paper.
It led to a sharp reduction in the number of remaining papers, leaving us with 838. 
The primary reason for exclusion was the lack of keywords correlating with both language models and code intelligence. 
Furthermore, we dismissed 890 papers classified as grey literature or misaligned with our main focus, as well as 105 systematic literature review articles to maintain our emphasis on the pitfalls associated with LM4Code tasks.
In the fifth and sixth stages, every paper underwent a manual evaluation for relevance and Quality Assessment (QA), which led to the final selection of 57 papers. 

For SLRs, it is important to analyze the quality of collected studies to ensure that we form an accurate and unbiased representation of the actual research~\cite{kitchenham_guidelines_2007}.
Thus, we undertook the quality assessment process using a pre-defined quality checklist. 
We established four quality assurance criteria (QA1 to QA4) to reassess all selected papers.
\begin{itemize}
    \item \textbf{QA1.} Are the pitfalls of LM4Code clearly described?
    \item \textbf{QA2.} Are the implications of the LM4Code pitfalls clearly stated or demonstrated?
    \item \textbf{QA3.} Is there a robust evaluation or solution for the identified LM4Code pitfalls in the proposed methodology?
    \item \textbf{QA4.} Is the contribution of the research clearly stated?
\end{itemize}
We critically evaluate and rate each QA on a scale of 0 to 4 (with 4 denoting ``high'', while 0 denoting ``low''). 
Then, we calculate the average quality score based on four quality criteria.
We set the threshold equal to 2.5 (50 percent of the percentage score), which means if the average quality score of the research isn't larger than 2.5, the study would be excluded.

\subsection{Snowballing}
In order not to miss some important work, we conducted a lightweight snowballing \cite{kitchenham2007guidelines}, which is a commonly used search approach to complement automated queries. 
We executed both forward and backward snowballing, incorporating references and citations into consideration. 
Specifically, each primary study was examined for its references (backward snowballing) and its subsequent citations using Google Scholar (forward snowballing).
The set of 57 papers from the prior step served as the initial set. 
From the forward and backward snowballing, we garnered 1003 and 1819 papers respectively. 
Following the processes of merging, deduplication, and the removal of articles already uncovered in our automated and manual searching, we ended up with a pool of 1611 papers. 
These papers underwent the same study selection process, culminating in the identification of an extra 10 papers. 
As a result, we collected a total of 67 papers focusing on LM4Code pitfalls.

\begin{figure}[t]
\centering
  \begin{subfigure}{0.5\linewidth}
      \centering
    \includegraphics[height=4cm, keepaspectratio]{images/fig1_time_distribution.pdf}
    \caption{Distribution of papers over years}
    \label{fig:annual-count}
  \end{subfigure}%
  \begin{subfigure}{0.5\linewidth}
      \centering
    \includegraphics[height=4cm, keepaspectratio]{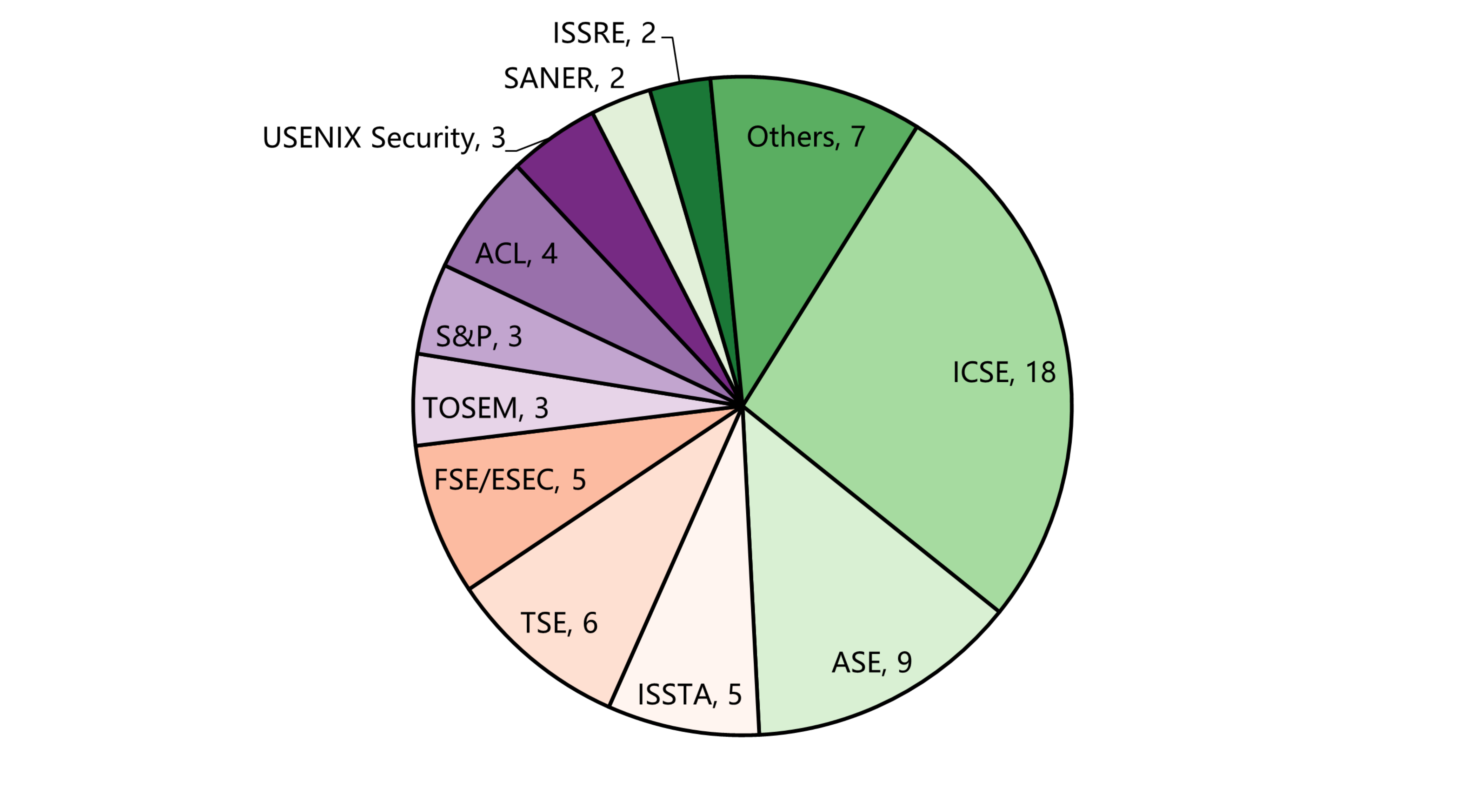}
    \caption{Distribution of papers across venues}
    \label{fig:venue-distr}
  \end{subfigure}
  \caption{Overview of papers}
  \label{fig:overview_paper}
\end{figure}

\subsection{Statistics of Collected Publications}
Figure~\ref{fig:Study_identification_and_selection} concludes the statistics of literature during our paper collection and selection process, and we finally obtained 67 research studies related to our research focus. 
Figure~\ref{fig:overview_paper} displays the distribution of the collected research studies across the published year and published venues.
From Figure~\ref{fig:annual-count}, we have noted that there is a significant increase in the number of relevant research studies published annually from 2021.
Prior studies, such as Watson~\ea~\cite{watson2022systematic} and Yang~\ea~\cite{yang2022survey}, have acknowledged the prevalence of language models in code-related tasks between the years 2014 to 2020.
However, our results indicate that there has been limited attention given to the identification and analysis of pitfalls in LM4Code.
Nevertheless, Figure~\ref{fig:annual-count} shows a rising trend over the last three years, indicating an increasing research interest within the research community about the potential pitfalls in LM4Code. 
Figure~\ref{fig:venue-distr} presents the distribution of the conferences and journals where the collected papers have been published, spanning the Software Engineering (SE), Security (SEC), and Artificial Intelligence (AI) domains.
Some venues displayed in Figure~\ref{fig:annual-count} that aren't listed in Table~\ref{tab:manual_search_venues} include: S\&P (IEEE Symposium on Security and Privacy), ACL (The Annual Meeting of the Association for Computational Linguistics), Usenix Security (USENIX Security Symposium), SANER (IEEE International Conference on Software Analysis, Evolution, and Reengineering), and ISSRE (The International Symposium on Software Reliability Engineering). 
It is noted that most of our papers are from the SE domain, particularly from top-tier venues such as ICSE and ASE.
Also, we collected six papers from the top-tier SEC conferences: three from S\&P and another three from Usenix Security.

\begin{figure}[t]
  \centering
  \begin{minipage}{0.5\linewidth} 
    \centering
    \includegraphics[width=\linewidth]{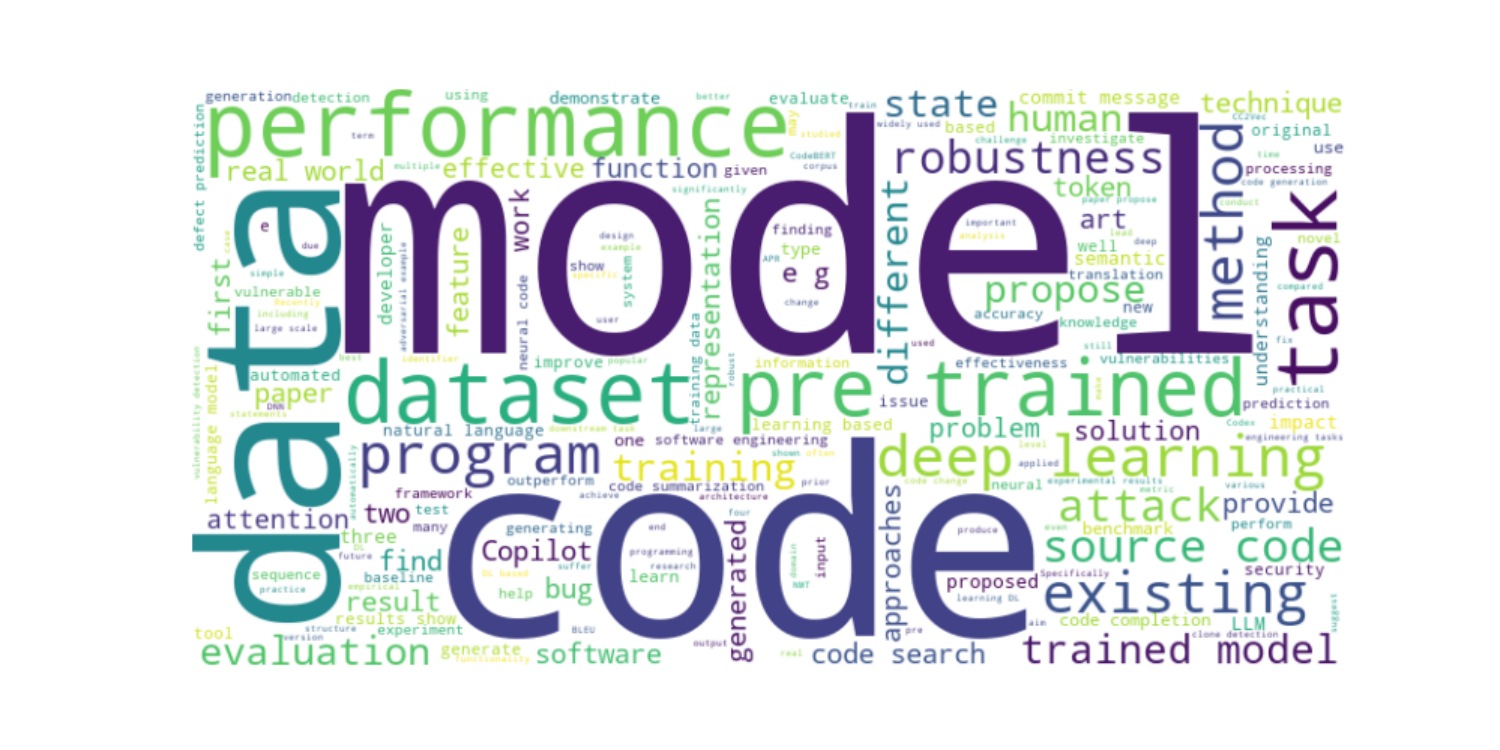} 
    \caption{WordCloud of paper}
    \label{fig:wordcloud}
  \end{minipage}%
  \begin{minipage}{0.5\linewidth} 
    \centering
    \includegraphics[width=\linewidth]{images/fig3_lm_models_distribution.pdf} 
    \caption{Distribution of papers across LMs}
    \label{fig:lm_ditribution}
  \end{minipage}
\end{figure}

\begin{table}[t]
  \centering
  \footnotesize
    \begin{tabular}{lllr}
    \toprule
    \textbf{SE activities} & \multicolumn{2}{c}{\textbf{SE tasks}} & \textbf{Total} \\
    \midrule
    \multicolumn{1}{c}{\multirow{5}[2]{*}{\textbf{Software development}}} & Code Generation(14) & Code Classification(3) & \multirow{5}[2]{*}{52} \\
      & Code Summarization(12) & Code Representation(2) &  \\
      & Code Search(9) & Code Comment Generation(1) &  \\
      & Code Completion(5) & Authorship Attribution(1) &  \\
      & Code Translation(4) & Named Entity Recognition(1) &  \\
    \midrule
    \textbf{Software quality assurance} & Vulnerability Detection(12) & Test generation(1) & 13 \\
    \midrule
    \multirow{4}[2]{*}{\textbf{Software maintainance}} & Clone Detection(10) & Duplicate Bug Report Detection(1) & \multirow{4}[2]{*}{28} \\
      & Program Repair(7) & Bug Report Summarization(1) &  \\
      & Defect Prediction(5) & Bug-Fix Commit Identification(1) &  \\
      & Commit Message Generation(2) & Bug Report Classification(1) &  \\
    \bottomrule
    \end{tabular}%
\caption{{Distribution of papers across SE activities}}
  \label{tab:SE_tasks}%
\end{table}%

Figure~\ref{fig:wordcloud} shows a word cloud generated from the abstracts of the collected papers, highlighting that most of the prominent terms are associated with LM4Code.
Figure~\ref{fig:lm_ditribution} further presents the distribution of language modes used in the collected paper.
It is important to note that while both LSTM and GRU are types of RNN, in this study, papers that only specify the use of RNN without further detail are categorized under ``General RNN''.
Similarly, despite observing the utilization of several popular transformer-based architectures such as CodeBERT, CodeX, and CodeT5, papers that merely claim the use of a self-defined or custom-designed transformer are classified as ``General Transformer'' in subsequent sections.
As depicted in Figure~\ref{fig:lm_ditribution}, it is evident that the LSTM model has a higher prevalence compared to other types. In the past two years, there has been a significant increase in research inquiries focused on transformer-based language models, specifically targeting pre-trained models like CodeBERT and Codex.
These observations are consistent with previous survey studies related to learning-based software engineering~\cite{watson2022systematic, wang2023software, hou2023large}.
Table~\ref{tab:SE_tasks} presents a summary of the distribution of SE tasks that the collated papers address.
Notably, tasks such as code generation, code summarization, code search, vulnerability detection, and clone detection emerge as the dominant scenarios for investigating the pitfalls in LM4Code. These primarily encompass classification tasks, as well as code or text generation challenges in LM4Code.

Overall, our analysis underscores that LM4Code has been an increasing area of interest.
Particularly, since 2021, there has been an increasing amount of literature emphasizing the pitfalls of LM4Code.
These pitfalls may hinder realistic performance and can affect how reliable and practical LM4Code systems are in real-world situations.
Our collected papers cover a wide range of language models and many different software engineering tasks.
As we delve deeper into our primary topic in the subsequent sections, this preliminary overview sets the foundation for a comprehensive analysis of the challenges and opportunities in LM4Code.

\end{document}